\documentclass[usenatbib]{mn2e}

\usepackage{graphicx}

\newcommand{\arepo}{{\sc arepo}}
\newcommand{\gadget}{{\sc gadget}}
\newcommand{\ramses}{{\sc ramses}}
\newcommand{\subfind}{{\sc subfind}}
\newcommand{\kpc}{{\rm kpc}}
\newcommand{\hi} {{\rm H}\,{\small\rm I}}
\newcommand{\kms} {{\rm km~s}^{-1}}

\newcommand{\pc} {{\rm pc}}

\newcommand{\mo}{{\rm M}_\odot}

\newcommand{\Myr}{{\rm Myr}}
\newcommand{\Gyr}{{\rm Gyr}}
\newcommand{\K}{{\rm K}}

\newcommand{\mokpc}{M_{\odot} {\rm kpc}^{-2}}

\newcommand{\gsim}{\lower.7ex\hbox{$\;\stackrel{\textstyle>}{\sim}\;$}}
\newcommand{\lsim}{\lower.7ex\hbox{$\;\stackrel{\textstyle<}{\sim}\;$}}
\newcommand{\eqref}[1]{eq.~(\ref{#1})}
\newcommand{\Sersic}{S{\'e}rsic}

\newcommand{\FM}[1]{#1}

\title[Disc galaxies in moving-mesh simulations]
{The formation of disc galaxies in high resolution 
moving-mesh cosmological simulations} 
\author[F. Marinacci, R.~Pakmor and V.~Springel]
{Federico Marinacci$^{1,2}$, R\"udiger~Pakmor$^1$ and Volker~Springel$^{1,2}$\vspace*{0.2cm}\\
  $^1$Heidelberger Institut f\"{u}r Theoretische Studien,
  Schloss-Wolfsbrunnenweg 35, D-69118 Heidelberg, Germany\\
  $^2$Zentrum f\"ur Astronomie der Universit\"at Heidelberg,
  Astronomisches Recheninstitut, M\"{o}nchhofstr. 12-14, D-69120
  Heidelberg, Germany}

\date{Accepted 2013 October 17.  Received 2013 October 16; in original form 2013 May 22}

\begin{document}

\pagerange{\pageref{firstpage}--\pageref{lastpage}}
\pubyear{2013}

\maketitle

\label{firstpage}

\begin{abstract}
  We present cosmological hydrodynamical simulations of eight Milky
  Way-sized haloes that have been previously studied with dark matter
  only in the Aquarius project. For the first time, we employ the
  moving-mesh code \arepo\ in zoom simulations combined with a
  comprehensive model for galaxy formation physics designed for large
  cosmological simulations.  Our simulations form in most of the eight
  haloes strongly disc-dominated systems with realistic rotation
  curves, close to exponential surface density profiles, a
  stellar-mass to halo-mass ratio that matches expectations from
  abundance matching techniques, and galaxy sizes and ages consistent
  with expectations from large galaxy surveys in the local
  Universe. There is no evidence for any dark matter core formation in
  our simulations, even so they include repeated baryonic outflows by
  supernova-driven winds and black hole quasar feedback.  For one of
  our haloes, the object studied in the recent `Aquila' code comparison
  project, we carried out a resolution study with our techniques,
  covering a dynamic range of 64 in mass resolution.  Without any
  change in our feedback parameters, the final galaxy properties are
  reassuringly similar, in contrast to other modelling techniques used
  in the field that are inherently resolution dependent. This success
  in producing realistic disc galaxies is reached, \FM{in the context
    of our interstellar medium treatment,} without resorting to a high density
  threshold for star formation, a low star formation efficiency, or
  early stellar feedback, factors deemed crucial for disc formation by
  other recent numerical studies.
\end{abstract}

\begin{keywords}
  methods: numerical -- galaxies: evolution -- galaxies: formation -- galaxies: spiral
\end{keywords}

\section{Introduction}

Forming realistic disc galaxies in self-consistent hydrodynamical
simulations of the $\Lambda$ cold dark matter ($\Lambda$CDM) cosmology has been a nagging problem
for more than two decades \citep[after pioneering work
  by][]{Katz1991}.  In stark contrast to the successes reached with
dark matter only simulations of cosmic large-scale structure
\citep[e.g.][]{Davis1985, Springel2006} and with semi-analytic galaxy
formation models \citep[e.g.][]{Guo2011}, making {\em realistic}
spiral galaxies on the computer has emerged as an unexpectedly
difficult endeavour that has withstood countless attempts at solving it
over the years.  Instead, the simulated galaxies were often too small
due to an angular momentum deficit \citep[e.g.][]{Navarro2000}, they
featured at best anaemic discs and almost universally too concentrated
and massive bulges \citep[e.g.][]{Scannapieco2009}, their rotation
curves had unrealistic shapes \citep[e.g.][]{Hummels2012}, and in
general they were far too luminous \citep[e.g.][]{Martig2012} as a
result of the ``overcooling catastrophe'' \citep{Balogh2001}.

Recently, however, the situation has profoundly changed, and there are
now several studies that obtained disc galaxies in quite reasonable
agreement with key observables \citep{Governato2010, Agertz2011,
 Brooks2011, Guedes2011, Aumer2013b, Stinson2013a}. In particular, for
the first time, we have seen simulations that produce reasonably small
bulges and a dominant disc, combined with realistic rotation curves,
roughly correct sizes, and low enough stellar masses to be compatible
with abundance matching expectations \citep{Aumer2013b, Stinson2013a}.
This progress raises the question of what cut the Gordian knot that
had allowed only incremental advances for many years
\citep{Governato2004, Governato2007, Robertson2004, Scannapieco2008,
  Scannapieco2009, Scannapieco2011, Sales2009, Sales2010, Stinson2010,
  Piontek2011}.

There are different claims in the literature about the key remedy for
the impasse.  Some studies have argued that very high numerical
resolution is a central and potentially sufficient requirement
\citep[e.g.][]{Governato2007, Kaufmann2007}, whereas other works
emphasized that the degree of success critically depends on the
modelling of the physics of star formation and feedback
\citep[e.g.][]{Okamoto2005, Scannapieco2008, Sales2010}.  Recently,
some authors suggested that a high star formation threshold is a key
factor in allowing the successful formation of a late-type spiral
galaxy like the Milky Way \citep{Guedes2011}.  On the other hand,
\citet{Agertz2011} find that a low star-formation efficiency,
particularly at high redshift, is important, whereas an opposite
conclusion was reached by \cite{Sommer2003} and \cite{Sales2010}, who
favoured a high star formation efficiency instead.  Other studies
pointed out that additional feedback channels such as cosmic rays
\citep[e.g.][]{Uhlig2012}, or stellar evolutionary processes 
in the form of mass return \citep{Leitner2011} need to be considered.

Which of these factors reflects essential physics important for disc
formation rather than numerics or the particularities of a specific
modelling technique is far from clear. A recurrent theme, though, is
that in all the recent successful simulations of disc galaxies a much
stronger feedback than employed in previous calculations is invoked.
In particular, the advances in reproducing disc galaxies by
\citet{Stinson2013a} and \citet{Aumer2013b} are attributed to `early
stellar feedback' introduced by the authors. In the actual numerical
implementation of \citet{Stinson2013a}, this \FM{form of feedback} is injecting 
considerably more energy than from supernovae alone and allows the simulations to
finally curtail the overproduction of stars at high redshift that
invariably led to an excessively massive central bulge component later
on.  Whether the adopted subgrid modelling of the physics of radiation
pressure of newly born stars is realistic has to be seen, but
optimistic assumptions about the efficiency of feedback are needed by
all authors to reduce early star formation and delay disc formation to
sufficiently late times.

Indeed, perhaps the most general lesson of recent simulation work on
galaxy formation is that the importance of feedback for the outcome of
hydrodynamical cosmological simulations can hardly be overstated. The
Aquila comparison project \citep{Scannapieco2012} has shown that
different numerical codes can give widely different outcomes for the
same initial conditions. Even the same code can give substantially
different answers if small details in the implementation of feedback
and star formation physics are changed. Many feedback implementations
in current use are not robust to resolution changes and require
`retuning' of the free parameters of the model to obtain the same or a
similar result when the resolution is changed -- if at all
possible. The differences can be as extreme as those reported in
\citet{Okamoto2005}, where a galaxy's morphology simulated at
different resolution varied over the entire range of Hubble types.
Part of this sensitivity to small model and simulation details can be
attributed to the highly non-linear nature of the feedback loops that
need to drastically reduce star formation both in small and large
haloes. This is already a significant complication for strong
supernova-driven winds but is especially evident for active galactic nuclei (AGN) feedback
\citep{Springel2005a}. For some periods of time, \FM{black holes (BHs)} are expected to
grow exponentially, a process that can hugely amplify any tiny
numerical differences of the conditions around the BH that set the
accretion rate.

However, part of the lack of robustness in simulation outcomes
certainly also needs to be blamed on numerical models that are
essentially ill posed, in the sense that the feedback prescriptions
used are often heuristic, and not derived rigorously as a
discretization of some well-defined partial differential equations
that approximate the physics. As a result, the response of the models
to resolution changes is poorly defined and sometimes not well
understood, as some studies candidly concede
\citep[e.g.][]{Stinson2006,Aumer2013b}. This often reflects a fuzzy
notion of how `subgrid' physics (which invariably plays an important
role in this problem) should be treated, or sometimes an ignorance of
this unavoidable limitation altogether. We argue in this paper that
resolution-dependent feedback implementations make it hard to separate
physics from numerical effects, and we hence advocate the use of more
explicit subgrid models.

Even at the level of the ordinary hydrodynamical equations that
describe an ideal gas, simulation results can be strongly affected by
the numerical scheme employed for solving hydrodynamics, as emphasized
recently \citep{Keres2012, Sijacki2012, Torrey2012, Vogelsberger2012}. 
These studies have shown that accuracy differences
between smoothed particle hydrodynamics (SPH, as implemented in the
\gadget\ code) and the moving-mesh approach of the \arepo\ code
directly translate into sizable changes of predicted galaxy
properties.  In fact, there is an artificial numerical quenching of the
cooling rate in large haloes in SPH, caused by viscosity and noise
effects \citep{Bauer2012}.  The standard formulation of SPH also
creates spurious dense gas clumps orbiting in haloes. This greatly
modifies how galaxies acquire their gas in large haloes, suppressing
the relative importance of hot mode gas accretion in these systems
\citep{Nelson2013}.

Unfortunately, the size of these numerical uncertainties is so large
that they can mask important physical processes and induce incorrect
calibrations of the required feedback strength. It is therefore
important to use an as accurate numerical technique for hydrodynamics
as is available.  Similarly, sensitive dependences on fine details of
feedback implementations, especially with respect to numerical
resolution, are highly undesirable as this will add to the difficulty
of separating physics from numerics, and ultimately compromise the
predictive power of the simulations. Hence, we argue that a crucial
requirement for the current generation of cosmological simulations of
galaxy formation is that their numerical models should be sufficiently
well posed to yield results approximately invariant with numerical
resolution, at least over a reasonable range where crucial physics
remains subgrid and can only be treated in a phenomenological way. To
our knowledge, this requirement is not yet fulfilled by the reported
successful simulations of disc galaxy formation in the recent
literature.

In this paper, we study the problem by applying a newly developed
numerical methodology for cosmological galaxy formation  to
`zoom' simulations of Milky Way-sized galaxies. The objects we study
are taken from the Aquarius project \citep{Springel2008}, where they
have been examined in great detail with dark matter only
simulations. We added two further haloes to the Aquarius set which
were not run at high resolution in the original project but were still
part of its target list.  This same extended set of Milky Way-sized
haloes has been previously studied with hydrodynamics by
\citet{Scannapieco2009} using SPH, which hence serves as an
interesting comparison for our results.  One of the Aquarius haloes,
the `Aq-C' system, has been the object selected for the Aquila
code comparison project \citep{Scannapieco2012}, yielding another
reference for direct comparisons. Finally, a subset of the Aquarius
systems has been simulated very recently by \citet{Aumer2013b} with an
updated SPH code \citep[based on][]{Scannapieco2009}, yielding
considerably improved results, in particular with respect to the
disc-to-bulge (D/B) ratio and the total stellar mass.

The novel simulation methodology we apply to all eight haloes consists
of our moving-mesh code \arepo\ \citep{Arepo} combined with a
comprehensive model for the galaxy formation physics, as described in
full detail in \citet{Vogelsberger2013}.  We also include a resolution
study by considering both a higher and a lower resolution run by a
factor of 8 in mass around the nominal resolutions of \citet{Scannapieco2009}
and \citet{Aumer2013b}, which is equal to the default resolution that we have
picked here. Our primary goal is to investigate whether our new
numerical treatments yield reasonable galaxy morphologies and
properties in these systems, despite the fact that we do not use a
high density threshold for star formation, a low star formation
efficiency, or early stellar feedback -- or in other words, some of
the ingredients that have been deemed essential by other studies for
successfully forming disc galaxies.

This paper is structured as follows. In Section~\ref{SecMethods}, we
briefly summarize the numerical methodology used in our moving-mesh
simulations, and we detail the simulation set that we examine. In
Section~\ref{SecStructure}, we analyse the present-day structures of
the galaxies that we obtain in the Aquarius haloes, including an analysis
of their gas content. In Section~\ref{SecHistory}, we turn to an
analysis of the formation history of the galaxies, both in terms of
their stars and their embedded supermassive BHs. A brief
analysis of the halo mass structure and the impact of baryonic physics
on the dark matter distribution is given in
Section~\ref{SecBaryonImpact}, followed by results of a resolution
study in Section~\ref{SecResolution}. Finally, we discuss our findings 
and present our conclusions in Section~\ref{SecDiscussion}.

\begin{table*}
\centering
\begin{tabular}{lrrrrrrrrrrrr}
\hline
Run & 
$R_{\rm vir}$  &
$M_{\rm tot}$  &
$M_{\rm gas}$  &
$M_\star$      &
$M_{\rm dm}$   &
$N_{\rm cells}$ &
$N_\star$      &
$N_{\rm dm}$  &
$m_{\rm gas}$ &
$m_{\rm dm}$  &
$f_{\rm b}$ \\
 &
$(\kpc)$   & 
$(10^{10}\mo)$ & 
$(10^{10}\mo)$ & 
$(10^{10}\mo)$ & 
$(10^{10}\mo)$ & 
&
&
&
$(10^5\mo)$ & 
$(10^5\mo)$ & 
 \\
\hline
Aq-A-5 & 239.0 & 169.13 & 11.21 &  4.95 & 152.95 &  203822 &  152476 &  579342 &  5.03 &  26.40  & 0.55\\
Aq-B-5 & 183.0 &  75.93 &  4.08 &  4.88 &  66.97 &  108806 &  234310 &  444557 &  3.35 &  17.59  & 0.70 \\
Aq-C-5 & 234.5 & 159.74 &  7.09 &  7.00 & 145.64 &  163726 &  273124 &  674547 &  4.11 &  21.59  & 0.51 \\
Aq-D-5 & 240.2 & 171.67 &  7.59 & 12.10 & 151.97 &  159591 &  442966 &  657760 &  4.40 &  23.10  & 0.68 \\
Aq-E-5 & 206.3 & 108.74 &  3.58 &  8.75 &  96.39 &  101041 &  431167 &  550757 &  3.33 &  17.50  & 0.67 \\
Aq-F-5 & 209.0 & 113.05 &  8.65 &  8.86 &  95.51 &  331692 &  620784 &  791829 &  2.30 &  12.06  & 0.96 \\
Aq-G-5 & 204.4 & 105.83 & 11.43 &  6.00 &  88.40 &  346061 &  328784 &  708979 &  2.83 &  14.88  & 1.03 \\
Aq-H-5 & 183.1 &  76.06 &  2.95 &  5.01 &  68.10 &   91792 &  273228 &  525235 &  2.96 &  15.56  & 0.61 \\
\hline
Aq-C-6 & 235.5 & 161.82 &  9.86 &  5.95 & 146.00 &   28702 &   26803 &   84525 & 32.90 & 172.73  & 0.57 \\
Aq-C-4 & 234.4 & 159.48 &  8.39 &  5.31 & 145.71 & 1526514 & 1637981 & 5399079 &  0.51 &   2.70  & 0.49 \\
\hline
\end{tabular}
\caption{Primary numerical parameters of the simulated haloes at $z = 0$. We list
  the virial radius defined as a sphere enclosing an overdensity of
  200 with respect to the critical density. The further columns give
  total mass, gas mass, stellar mass and dark matter particle mass
  inside the virial radius. The corresponding numbers of gaseous cells,
  star particles, and dark matter particles are given next, followed
  by the gas mass and dark matter resolutions in the high-resolution
  region. Finally, the last column gives the baryon fraction, $f_{\rm
    b} \equiv ({\Omega_{\rm dm}}/{\Omega_{\rm b}}) ({M_{\rm
      gas}+M_\star + M_{\rm bh}})/{M_{\rm dm}}$ relative to the
  cosmological mean. In all the runs, the gravitational softening has
  been kept fixed in comoving units at $z\ge 1$ and in physical units
  ($680\,{\rm pc}$ for level 5 runs) at $0 \le z<1$.}
\label{tab:properties}
\end{table*}

\section{Numerical methodology and simulation set} \label{SecMethods}

\subsection{Initial conditions}

We use initial conditions from the Aquarius suite of high-resolution
dark matter simulations of Milky Way-sized haloes
\citep{Springel2008}.  The simulated volume is a periodic cube with a
side length of $100\,h^{-1}{\rm Mpc}$. The adopted $\Lambda$CDM
cosmology uses the parameters $\Omega_{\rm m} = \Omega_{\rm dm} +
\Omega_{\rm b} = 0.25$, $\Omega_{\rm b} = 0.04$,
$\Omega_\Lambda=0.75$, $\sigma_8=0.9$, $n_s=1$, and a Hubble constant
of $H_0 =100\,h\,{\rm km\,s^{-1}\,Mpc^{-1}} = 73\,{\rm
  km\,s^{-1}\,Mpc^{-1}}$. These cosmological parameters are the same
as in the Millennium and Millennium-II simulations \citep{Millenium,
  MilleniumII}. While they are now in tension with the diminished
error bars of the latest cosmological constraints from the 
\textit{Wilkinson Microwave Anisotropy Probe} and
\textit{Planck} satellites, this is of no relevance for this study.  In
order to achieve the high resolution needed to resolve the formation
of a Milky Way-like galaxy, our initial conditions utilize the
`zoom-in technique', i.e.~the Lagrangian region from which the main
galaxy forms is sampled with a high number of low-mass particles
whereas the rest of the simulation volume is filled with progressively
higher mass particles whose mass grows with distance from the target
galaxy. This saves computational time while still ensuring the correct
cosmological tidal field and mass infall rate for the forming target
galaxy.

The target galaxies themselves have been selected randomly from a
small mass interval around $10^{12}\,{\rm M}_\odot$, applying only a
rather mild isolation criterion that excluded objects close to another
massive galaxy. Specifically, haloes with mass greater than half that
of the candidate object were required to be at least $1.37\,{\rm Mpc}$
away from the candidate halo at $z = 0$.  This weak criterion is not
very restrictive; it was not met by about 22\% of the haloes from the
purely mass-selected sample, but it still helps to preferentially
select galaxies with quiet merger histories which are expected to be
favourable sites for producing late-type galaxies. A detailed analysis
of the formation history of the selected haloes and the scatter among
the set can be found in \citet{Boylan-Kolchin2010}.

We note that recent estimates of the mass of the Galaxy's dark matter
halo range from $1$ to $3 \times 10^{12}$ \citep{Wilkinson1999,
  Sakamoto2003, Battaglia2005, Dehnen2006, Li2008, Xue2008}. In order
to alleviate tensions due to the `missing massive satellites
problem' pointed out by \cite{Boylan-Kolchin2011,
  Boylan-Kolchin2012}, a recent analysis by \citet{Wang2012} favours
masses at the lower end of this interval, at around $10^{12}\,{\rm
  M}_\odot$, which is also the centre of the narrow mass interval that
we study.  On the other hand, using the space motion of the Leo~I
dwarf spheroidal galaxy, \citet{Boylan-Kolchin2012b} put a strong
lower limit on the Milky Way's mass, finding a median mass of $1.6
\times 10^{12}\,{\rm M}_\odot$, rather similar to several of our
candidate systems.

Following the nomenclature used in the Aquarius project, the primary
simulations that we carry out are hydrodynamical versions of Aq-A-5,
Aq-B-5, Aq-C-5, Aq-D-5, Aq-E-5, Aq-F-5, Aq-G-5, and Aq-H-5.  Here the
`5' in the name refers to the resolution level, corresponding to a
baryonic mass resolution of $\sim 4.1 \times 10^5\,{\rm M}_\odot$ and
a dark matter mass resolution of $\sim 2.2 \times 10^6\,{\rm M}_\odot$
(for Aq-C-5).  At $z>1$, we keep the gravitational softening length of
all mass components in the high-resolution region constant in comoving
units, growing the physical gravitational softening length to a
maximum of $680 \, \mathrm{pc}$, which was then held constant for
$z\le 1$.  In our resolution study of the Aq-C halo (which is the
object studied in the Aquila comparison project), we also consider
simulations adopting a baryonic mass resolution of $3.2 \times
10^6\,{\rm M}_\odot$ with a gravitational softening of $1.36 \;
\mathrm{kpc}$, as well as $5 \times 10^4\,{\rm M}_\odot$ with a
gravitational softening length of $340 \; \mathrm{pc}$, which
correspond to mass resolutions of a factor of 8 better or worse
(equivalent to levels 4 and 6 in the Aquarius project) than our
default, respectively. We note that these softening values follow the
optimum choices derived by \citet{Power2003}. Smaller softening values
would lead to significant two-body effects and a spurious heating of
the gas, particularly at high redshift, and are hence not well
justified.

In Table~\ref{tab:properties}, we list the principal numerical
parameters of our full simulation set\footnote{Omitting the dark
  matter only simulations considered in section \ref{SecBaryonImpact}
  for the sake of brevity.}. The present-day virial
masses\footnote{Following standard procedure, we define the virial
  mass as the mass contained within a sphere that encloses a mean
  matter density 200 times the critical density for closure,
  $\rho_{\rm crit} = 3 H^2(z) / (8\pi G)$.} that we quote are for the
evolved hydrodynamical haloes. Note that in a pure dark matter
simulation, the corresponding masses will be slightly larger because
the reduction of the baryon content below the universal mean through
non-gravitational feedback slows the mass growth of the haloes.  The
baryon fraction $f_{\rm b}$ of our haloes relative to the cosmological
mean is reduced typically by $30\%$ at $z=0$, with some systems having
lost up to half their baryons (Aq-A and Aq-C), and others essentially
none (Aq-F and Aq-G).

The original Aquarius initial conditions contained only dark matter
particles. For our simulations, we add gas by splitting each dark
matter particle into a pair of one dark matter and one gaseous cell,
with their masses set according to the cosmological baryon mass
fraction, and a separation equal to half the original mean
interparticle spacing, keeping the centre of mass and centre-of-mass
velocity of each pair fixed. In this way, two interleaved grids (or
actually `glasses', in the case of our high-resolution region) of dark
matter particles and gaseous cells are formed. We note that we split
{\em all} the particles, regardless of whether they are part of the
high-resolution region or the surrounding low-resolution volume, such
that the whole volume is filled with gas. There is hence no pressure
discontinuity at the boundary of the high-resolution region. The
evolved haloes at $z=0$ show zero contamination of the virialized
regions by low resolution dark matter particles, a tribute to the high
quality of the initial conditions (which were created by Adrian
Jenkins for the Aquarius project).

\subsection{Simulation code}

In the following, we briefly describe our simulation code and the most
important parameter settings used in this work. In the interest of
brevity, we only discuss the most important code characteristics and
refer, for further details, to the code paper of \arepo\
\citep{Arepo} and the application tests discussed in
\citet{Vogelsberger2012} and \citet{Sijacki2012}.

The moving-mesh code \arepo\ employs a dynamic Voronoi mesh for a
finite-volume discretization of the Euler equations.  The fluxes
between the individual Voronoi cells are calculated using a
second-order Godunov scheme together with an exact Riemann solver.
This approach is akin to ordinary grid-based Eulerian schemes for
hydrodynamics, except that an unstructured mesh is used that is
generated as the Voronoi tessellation of a set of mesh-generating
points. In addition, these mesh-generating points may be moved freely,
inducing a dynamical and continuous transformation of the mesh without
the occurrence of pathological mesh distortions. The most interesting
way to exploit this freedom of a dynamic mesh is to move the
mesh-generating points with the local flow velocity. In this default
mode of operation, a pseudo-Lagrangian method results where the mass
per cell is kept approximately constant and a Galilean-invariant
numerical method is obtained.

The automatic adaptivity of \arepo\ is thus similar to that of
SPH, but the mass per cell is not
forced to stay strictly constant. Instead, local variations in the gas
mass per cell may occur, but in case the mass deviates by more than a
factor of $2$ from the target gas mass resolution, we either split the
cell into two, or dissolve it \citep[as in][]{Vogelsberger2012}, which
is very similar to a Lagrangian refinement criterion in adaptive mesh
refinement (AMR) codes. But thanks to the adaptive nature of the
dynamic mesh, such refinement and derefinement operations are needed
much less frequently. Perhaps the most important advantage of
\arepo\ compared to traditional mesh codes with a static mesh is a
significant reduction of advection errors, which becomes particularly
relevant for highly supersonic motions. Compared to SPH, the most
important advantages are the absence of an artificial viscosity, a
reduced sampling noise, a higher accuracy of gradient estimates, and a
faster convergence rate in multi-dimensional flow.

As far as the gravity solver and collisionless dynamics are concerned,
\arepo\ applies the same techniques as the TreePM code
\gadget\ \citep{Springel2005b}. This makes the two codes particularly
well suited for a code comparison that focuses on an analysis of the
differences induced by the hydrodynamic treatment alone, and/or
differences due to feedback implementations, as differences originating
in the treatment of gravity can be largely excluded.  In previous
work, we have carried out such comparisons
\citep[e.g.][]{Scannapieco2012, Sijacki2012, Vogelsberger2012} for
models with identical (minimal) feedback physics. The simulations
presented in this paper use a new strong feedback model (see below)
and can be directly compared with the \gadget\ results obtained by
\citet{Scannapieco2009} and \citet{Aumer2013b} for matching haloes
from the Aquarius sample, as well as with the results reported in the
Aquila project \citep{Scannapieco2012} for a multitude of codes and
feedback models applied to the Aq-C halo.

\subsection{Physical model for galaxy formation} \label{sec:GFM}

We here employ a novel implementation of the most important physical
processes of galaxy formation in \arepo, presented in detail in
\citet{Vogelsberger2013}. For the sake of brevity, we list here only
the most important characteristics and refer to the papers of
\citet{Vogelsberger2013} and \citet{Torrey2013}  for full details
and cosmological tests. The model includes the following.

\begin{enumerate}
\item
 Primordial and metal-line cooling with self-shielding
corrections.
\item
A simple subresolution model for the
interstellar medium (ISM), which pictures the ISM as a two-phase
medium that is predominantly composed of cold clouds embedded in a
tenuous, supernova-heated phase \citep{SFR_paper}.
\item
 Stellar evolution, gas recycling and chemical enrichment.  The
 chemical enrichment follows nine elements (H, He, C, N, O, Ne, Mg,
 Si and Fe) independently, and tracks the overall metallicity and the
 total mass return from stars to gas.
\item
 Stellar feedback realized through a kinetic wind scheme in which
 the wind velocity is scaled with the local halo size \citep[similar
   to][]{Puchwein2013}, which in turn is estimated by the dark matter
 velocity dispersion. The adopted scaling of the mass loading of winds
 corresponds to energy-driven winds.
\item
 A metal loading of outflows that is determined independently of the mass
 loading of the winds. This is required to simultaneously reproduce
 the stellar mass content of low mass haloes and their gas oxygen
 abundances.
\item
 BH seeding, BH accretion and BH merging procedures based
 on an updated version of the model described in
 \citet{Springel2005a}. The BH growth distinguishes between quasar-
 and radio-mode feedback. In addition, a novel prescription for
 radiative feedback from AGN is included that
 modifies the ionization state and hence the cooling rate nearby to an
 active BH. This implementation assumes an average spectral energy
 distribution and a luminosity-dependent scaling of obscuration
 effects.
\item A spatially uniform UV background following the model of
  \citet{FaucherGiguere2009}, which completes \hi\ reionization at a
  redshift of $z\simeq 6$.
\item A new Lagrangian tracer particle formalism introduced by
  \citet{Genel2013} that follows the flow faithfully with a 
  Monte Carlo-based approach.
\end{enumerate}

We set the free parameters of the model to identical values 
  (modulo a minor change, as described below) as identified by
\citet{Vogelsberger2013} for their best match model in cosmological
simulations of galaxy formation. These fiducial settings produce a
good match to the stellar mass to halo mass function, the galaxy
luminosity functions, the history of the cosmic star formation rate
(SFR) density and to several other key observables. These parameters hence
represent a good candidate for testing the model at higher resolution
than possible in simulations of uniformly sampled cosmological
volumes.  We only deviate from \citet{Vogelsberger2013} with respect
to two minor points. As their radio-mode AGN feedback in large haloes
is based on the stochastic triggering of hot bubbles in halo
atmospheres, numerical convergence with varying resolution can not
necessarily be expected for individual objects, but only for the
population mean. Because this could spoil our convergence study, we
replaced the bubble heating with a much gentler halo heating model
where more bubbles of individually much weaker strength are
created. We note however that this feedback channel is almost
unimportant for our galaxies because of their moderate halo mass. The
other small change that we made concerns the galactic winds, which we
opted to endow with some amount of thermal energy in order to make
 them `hot' rather than 'cold' when they are launched. 
Our tests have shown that providing some amount of thermal
energy to galactic winds makes the gas flows in the haloes 
  smoother and more regular without changing the stellar mass of the
galaxies in any significant way. In contrast, if the wind is
  ejected cold, we sometimes see cold string-like gas features in the
  circumgalactic medium, which appear to be an artefact of this wind
  model. To preclude the possibility of in situ star formation in
  these features, we adopted a modification of the wind model where
  the energy given to the wind is split into equal parts into thermal
  and kinetic form, instead of assigning it all as kinetic energy and
  leaving it fully up to the hydrodynamical interactions to dissipate
  it to heat.  Since all the other wind parameters (in particular the wind 
velocity and the wind energy flux) are fixed as in
\citet{Vogelsberger2013}, this reduces the wind mass loading slightly
because the wind now also carries away some energy in thermal
form.

\begin{figure*}
\resizebox{17.8cm}{!}{\includegraphics{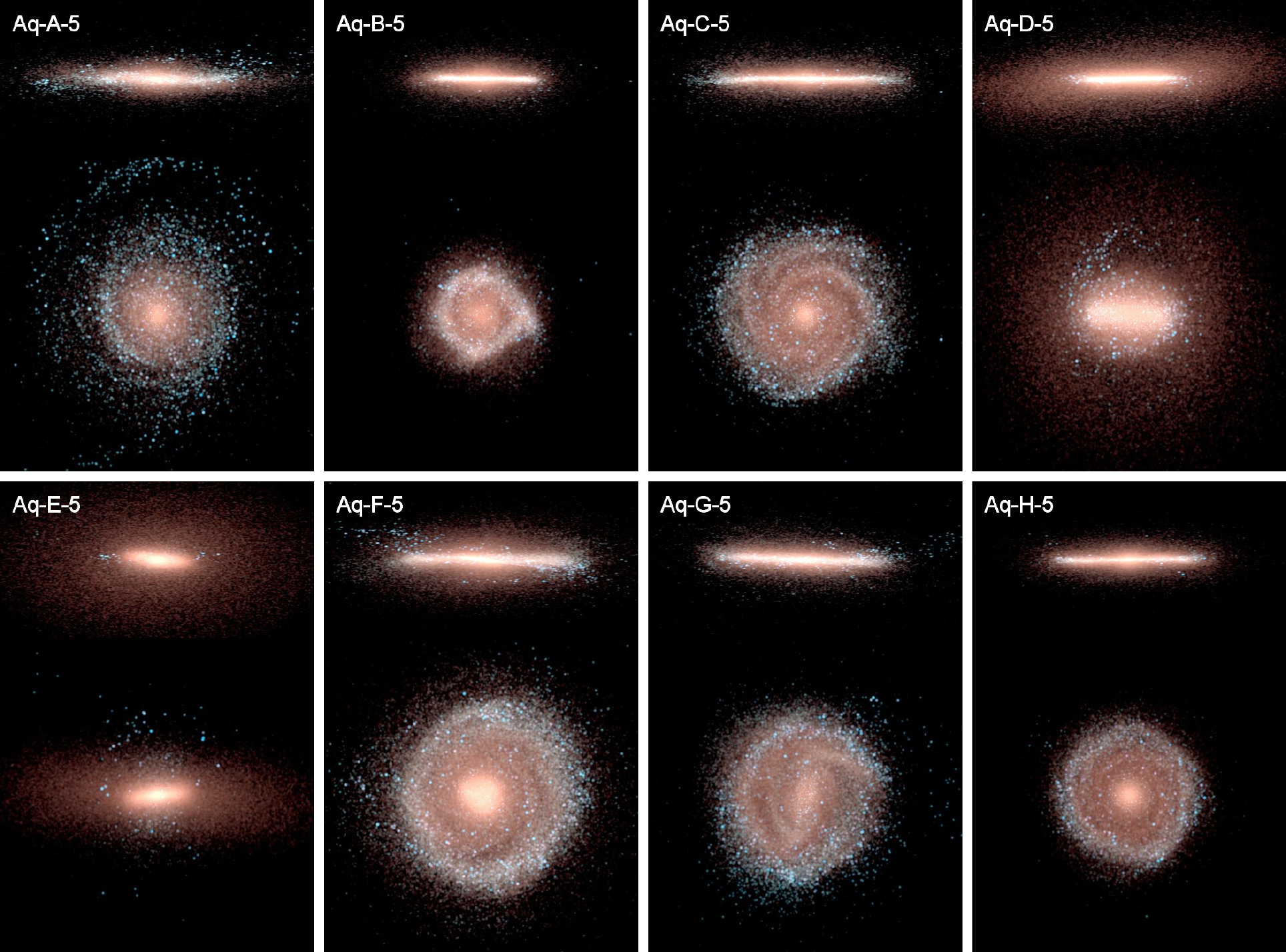}}
\caption{Projected stellar density for the eight simulated haloes at
  $z=0$. The chosen projection box is $50~\kpc$ in all directions and
  is centred on the halo potential minimum.  Edge-on (top portion of
  each panel) and face-on views (bottom portion of each panel) are
  displayed. A stellar disc is detectable in all the simulated
  haloes. The images are obtained by logarithmically mapping the $K$-, $B$-
  and $U$-band luminosity of the stars to the red, green and blue colour channels in order to
  have a visual impression of the age of the different stellar
  populations contained in the final galaxy. As a result, very young
  stars show up blue while older stars appear progressively redder.}
\label{fig:stellardisks}
\end{figure*}

It is worthwhile to emphasize the relation of our simulations to
  the runs carried out for the Aq-C halo in the Aquila comparison
  project \citep{Scannapieco2012}.  There, \arepo\ had been run
  without an explicit feedback model (except for the supernova
  feedback implicit in the ISM pressurization model), and without
  metal cooling. Interestingly, simulations with comparably weak
  feedback done with the \ramses\ code cooled a very similar
  amount of gas as \arepo, which is a reassuring agreement and
  suggests that systematic differences between AMR and our moving-mesh
  technique are not significant for determining how much gas becomes
  available for star formation in the first place (we note that this
  does not preclude, however, that there are other systematic
  differences elsewhere). There has been no change in the basic
  methods for calculating gravity and hydrodynamics in the \arepo\ 
  code since the Aquila project; hence, the improvements in
  the results we report in this paper can be entirely attributed to
  the new physics models for star formation, BH growth and
  associated feedback channels that we adopt here. In particular, the
  key new process invoked in the present simulations is an explicit
  modelling of galactic winds, which is the primary process
  responsible for the differences with respect to the \arepo\ runs
  in the Aquila project. The winds curtail excessive high-redshift
  star formation, thereby reducing the size of the central spheroids
  while leaving enough gas around to form a rotationally supported
  disc at late times. This allows our new simulations to follow the
  path for successful disc formation identified in the Aquila
  project.

\begin{figure}
\begin{center}
\resizebox{6.5cm}{!}{\includegraphics{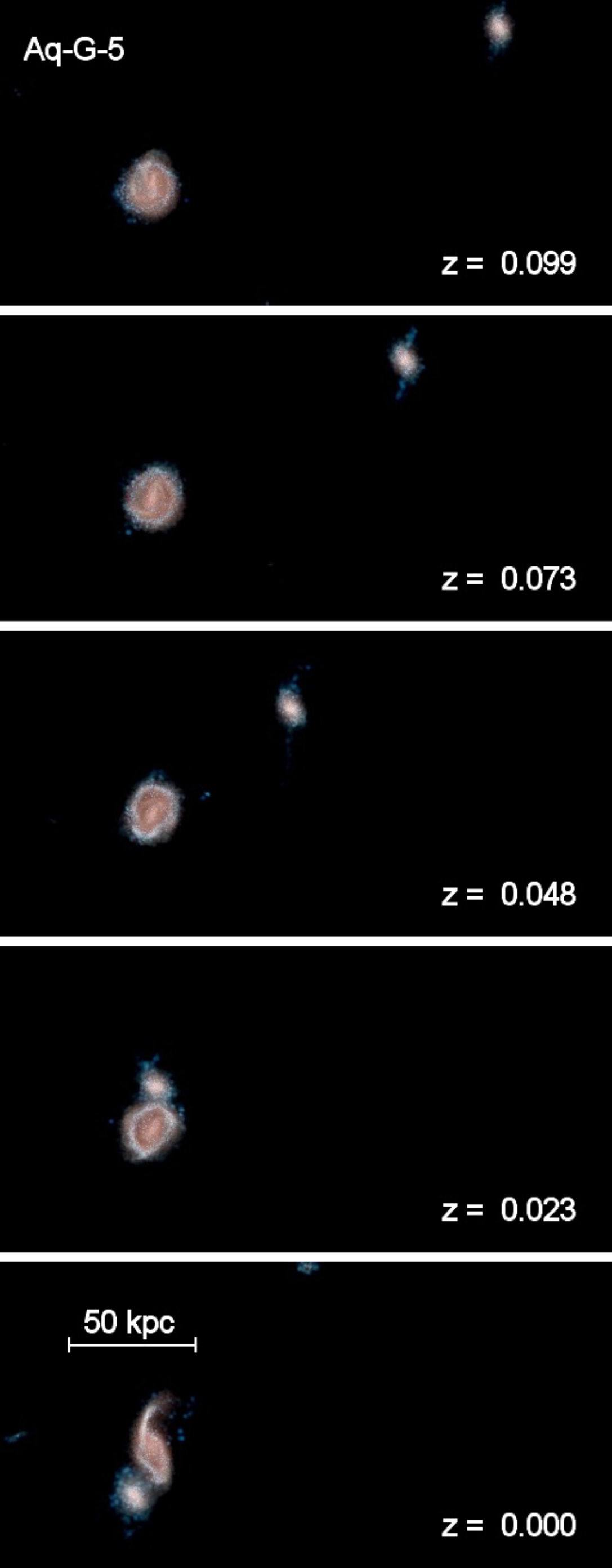}}
\end{center}
\caption{Time evolution of the Aq-G-5 simulation directly before $z=0$.}
\label{fig:g5merger}
\end{figure}

\section{Present day galaxy structure}  \label{SecStructure}

\subsection{Stellar discs} \label{sec:stellardics}

In Fig.~\ref{fig:stellardisks}, we show the stellar mass
distributions of all of our eight simulated haloes at $z=0$, both in
face-on and edge-on projections. The images were constructed by
mapping the $K$-, $B$- and $U$-band luminosities to the red, green and blue
channels of a full colour composite image. Young stellar populations
hence appear blue, old stellar components appear red. All images use
the same logarithmic mapping of stellar luminosity to image intensity
and display the same physical extension of $50\,{\rm kpc}$ on a
side. The face-on orientation used for the projections was defined
through the angular momentum of the cold galactic gas. Using instead
the major axis of the moment-of-inertia tensor of the stars or the
stellar angular momentum vector yields essentially the same
directions, and represents an equally well working choice for these
galaxies. For definiteness, we define an aligned coordinate system
$(x'y'z')$ for each galaxy where the $z'$-axis points along the
angular momentum of the cold gas, while the $y'$-axis points along the
intersection of the $z'=0$ plane with the $z=0$ plane of the
simulation's original coordinate system. This leaves two possible
directions for the $x'$-axis: we pick the one with the smaller angle
between the positive $x'$ and $x$ axes.

Clearly evident in Fig.~\ref{fig:stellardisks} is the pronounced disc
morphology of almost all of the systems. The one exception is the
galaxy Aq-E-5, which is the reddest among the set. Most of its stars
appear to lie in an elongated spheroid that shows substantially
flattening. There is also a feeble disc of young stars misaligned with
the old flattened stellar distribution.  The disc of galaxy Aq-D-5 is
dominated by a prominent bar, and also is comparatively red.  The
other six galaxies feature nicely symmetric, thin and extended discs,
with some indication of a red bulge in the centre, which is however
not prominent enough to be readily apparent in the edge-on
projections.  Interestingly, these well-defined discs show blue outer
rings and some spiral features that indicate substantial star
formation in these regions. \FM{The blue/red appearance of the discs
  shows a qualitative resemblance with some observed galaxies, such as
  NGC 7217\footnote{see
    e.g. http://skycenter.arizona.edu/gallery/Galaxies/NGC7217} which
  compares quite nicely with Aq-H-5, for example.}

Actually, we have shown the galaxy Aq-G-5 at $z=0.07$ rather than
at $z=0$ in Fig.~\ref{fig:stellardisks}. This is done because Aq-G-5
happens to just undergo an encounter with a massive satellite galaxy
at $z=0$, inducing a significant tidal perturbation that started to
deform its disc and created a prominent tidal arm to the North. This
is seen explicitly in Fig.~\ref{fig:g5merger}, which displays
snapshots of the time evolution of the Aq-G-5 system from $z=0.1$ to
$0$. In order to avoid that our results are distorted by the strong
encounter right at $z=0$, we use the $z=0.07$ output of Aq-G-5 for all
subsequent analysis.

\begin{figure*}
\hspace*{-0.2cm}\resizebox{18.2cm}{!}{\includegraphics{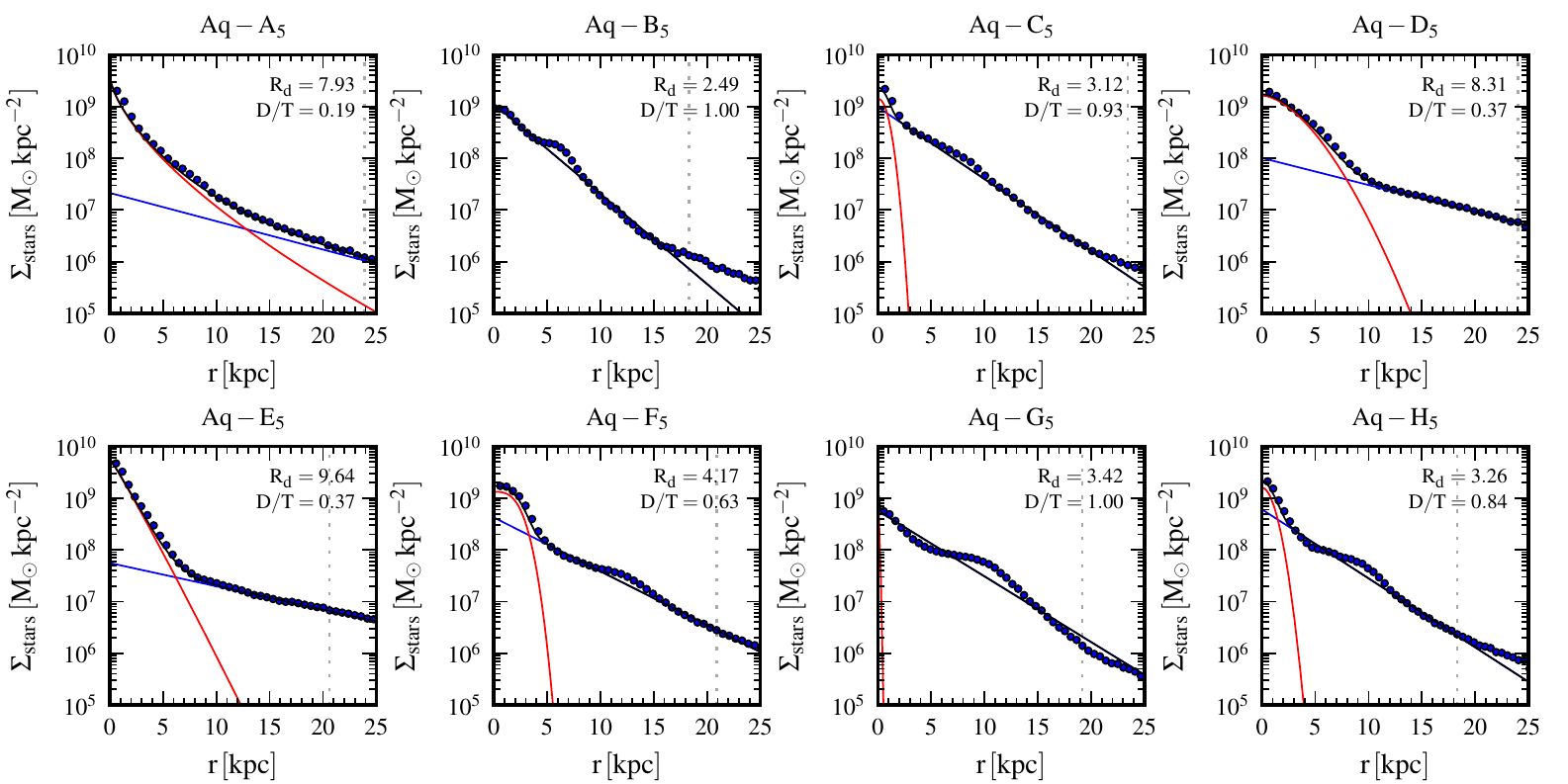}}
\caption{Stellar surface density profiles of the simulated galaxies,
  seen face-on.  Decompositions of the total profile into an
  exponential disc and a \Sersic\ profile are also shown (provided the
  \Sersic\ fit is sufficiently well defined).  The fits are carried
  out up to the vertical dotted line located at $0.1\times R_{\rm
    vir}$.  The resulting disc scalelength and D/T mass
  ratio are indicated in each panel.  A comprehensive list of the
  structural parameters derived from the disc-bulge decomposition can
  be found in Table~\ref{tab:surfbright}. }
\label{fig:surfacebrightness}
\end{figure*}

In Fig.~\ref{fig:surfacebrightness}, we show azimuthally averaged
face-on stellar surface density profiles of our eight galaxies (black
circles). The profiles are obtained by projecting on the $x'y'$ plane
(see above) the mass of all stellar particles contained in a  box
centred on the halo potential minimum, with a total height equal to
$0.1\times R_{\rm vir}$. Stars are binned in circular annuli in the
projection plane, and the total stellar mass of each bin is then
divided by its surface area to get the surface density. 

Interestingly, most of the galaxies show surface density profiles that
present, in their outer parts, the characteristic exponentially
declining trend observed in many spiral galaxies. In some of the
simulated galaxies a clear central excess can be seen, signifying the
presence of a central bulge. To quantify the importance of the two
contributions we have carried out two-component fits of the surface
density profiles with a \cite{Sersic1963} profile (red lines) for the
central part, and an exponential profile (blue lines) for the outer
parts. The fits are performed by \FM{simultaneously varying the
parameters of the exponential and \Sersic\ profiles. 
Finding first the optimal parameters of the exponential profile 
and subsequently modelling  the central residual
excess with the \Sersic\ profile, as done in \cite{Scannapieco2011}, does not 
lead to significant differences in the results. 
The best-fitting parameters are listed in Table~\ref{tab:surfbright}.}

We find that these fits work quite well in most of the cases, and
hence yield a tentative estimate of the bulge and disc stellar masses,
and of the disc-to-total (D/T) ratio. These quantities are also reported in
Table~\ref{tab:surfbright}. We note however that such profile fits,
which form the basis of photometric determinations of the
bulge-to-disc ratio, are often not sufficiently unique to allow a
robust decomposition.  In particular, fits with very similar quality
parameter $Q$ can be obtained for quite different values of $r_{\rm
  cut}$, and sometimes this affects the structural parameters of the
disc and bulge strongly even though the total recovered profile does
not show any appreciable difference. This technique is also known to
typically overestimate the D/B mass ratio. The
essentially {\em pure discs} we obtain for Aq-B, Aq-C and Aq-G
therefore need to be taken with a grain of salt.

\begin{table*}
\centering
\begin{tabular}{ccccccccccc}
\hline
   Run & $\log_{10}\Sigma_{\rm d}$ & $R_{\rm d}$ & $\log_{10}\Sigma_{\rm eff}$ & $r_{\rm eff}$ & $n$ & Disc mass & Bulge mass & $D/T$ & Total mass & Fit Mass \\ 
       & $(\mokpc)$ & $(\kpc)$ & $(\mokpc)$ & $(\kpc)$ &  & $(10^{10}\mo)$ & $(10^{10}\mo)$ & & $(10^{10}\mo)$ & $(10^{10}\mo)$ \\ 
\hline

Aq-A-5 & 7.329 & 7.929 & 6.831 & 3.148 & 1.551 & 0.843 & 3.632       & 0.19 & 4.363 & 4.475 \\ 
Aq-B-5 & 9.052 & 2.493 & 7.614 & 0.021 & 0.693 & 4.399 & $< 10^{-4}$ & 1.00 & 4.691 & 4.399 \\ 
Aq-C-5 & 8.988 & 3.124 & 8.560 & 0.835 & 0.446 & 5.961 & 0.474       & 0.93 & 6.617 & 6.435 \\ 
Aq-D-5 & 8.004 & 8.306 & 8.339 & 3.353 & 0.593 & 4.379 & 7.391       & 0.37 & 10.78 & 11.77 \\ 
Aq-E-5 & 7.747 & 9.641 & 8.204 & 1.971 & 0.939 & 3.264 & 5.484       & 0.37 & 7.919 & 8.748 \\ 
Aq-F-5 & 8.628 & 4.170 & 8.765 & 1.924 & 0.323 & 4.637 & 2.697       & 0.63 & 7.256 & 7.334 \\ 
Aq-G-5 & 8.758 & 3.420 & 8.013 & 0.120 & 0.664 & 4.207 & 0.006       & 1.00 & 4.229 & 4.213 \\ 
Aq-H-5 & 8.784 & 3.255 & 8.500 & 1.046 & 0.504 & 4.045 & 0.784       & 0.84 & 4.865 & 4.830 \\ 
\hline
\end{tabular}
\caption{Parameters of the surface density profile decomposition. For
  each run the columns give (from left to right): the logarithm of the central surface
  density of the disc, the disc scale-length, the logarithm of the
  bulge surface density at the effective radius,
  the bulge effective radius (defined as the radius enclosing half of
  the bulge mass), the \Sersic\ index of the bulge, the inferred disc
  mass, the inferred bulge mass, the D/T mass ratio, the
  total mass of the system as computed by the simulation output and the
  total mass of the system as derived from the fit.
}
\label{tab:surfbright}
\end{table*}

\begin{figure*}
\hspace*{-0.3cm}\resizebox{18.7cm}{!}{\includegraphics{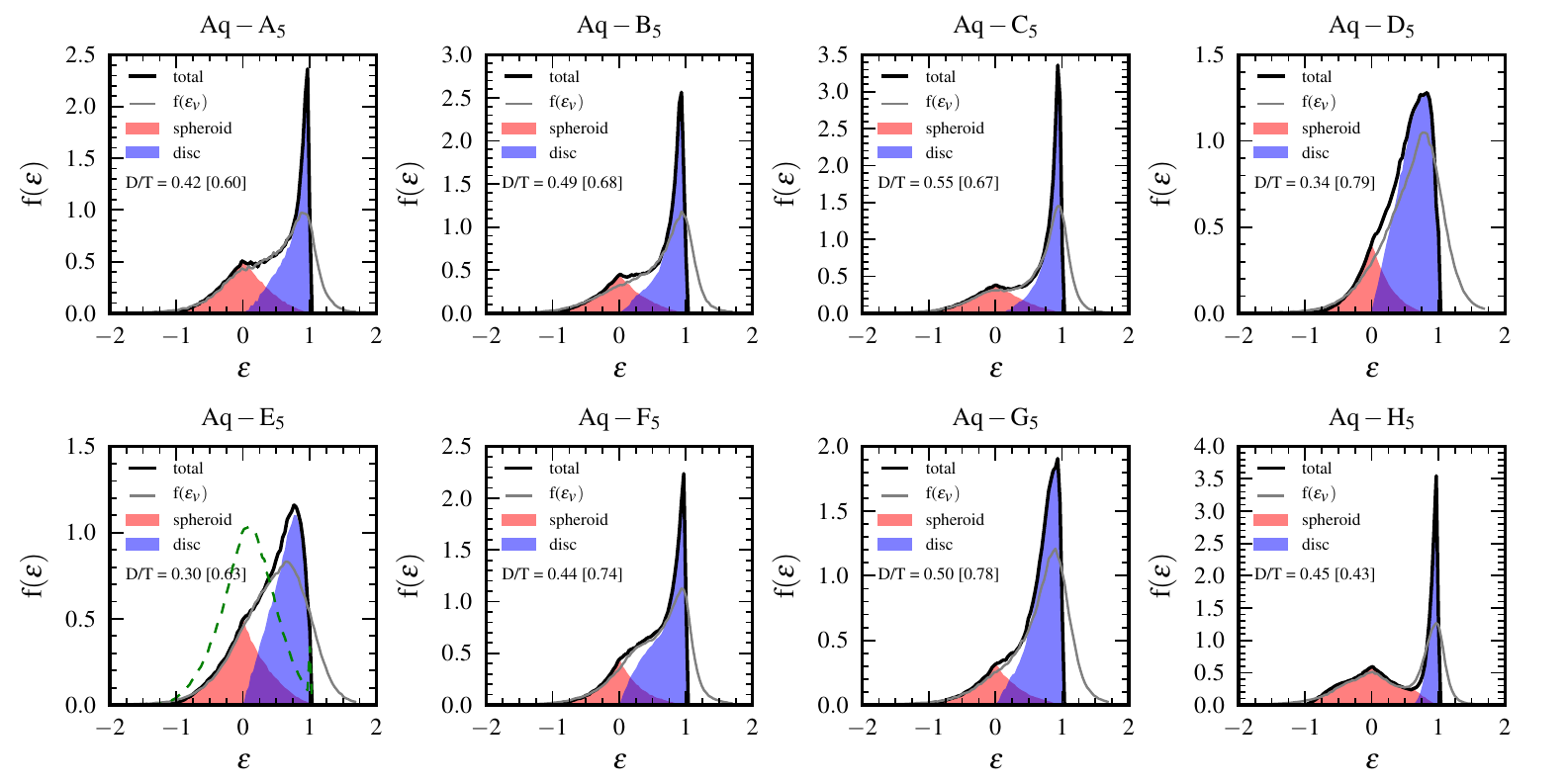}}
\caption{Distribution of the mass-weighted stellar circularities
  $\epsilon$ for the eight Aquarius haloes at $z = 0$. The plots are
  obtained by considering only stars with $r < 0.1~R_{\rm vir}$. The
  solid black lines show the $\epsilon$ distributions obtained by
  following the definition of \citet{Abadi2}. The distributions are
  further subdivided into a bulge (red) and a disc
  (blue) component by mirroring around zero the fraction of 
  stars with negative $\epsilon$ and considering the resulting distribution 
  as making up the bulge. The D/T mass ratios
  derived from this kinematic decomposition are reported in the legend
  of each panel in square brackets, while the other values are obtained
  from the fraction of stars with $\epsilon > 0.7$. 
  The thin grey lines show the
  distribution of stellar circularities if the definition of
  $\epsilon_V$ adopted in the Aquila project is used. For the Aq-E-5
  system, an additional dashed line is shown that gives the
  distribution of $\epsilon$ if the face-on projection is aligned with
  the disc of young stars forming in this system, which is roughly
  perpendicular to the rotation direction of the old stars. }
\label{fig:circularityplot}
\end{figure*}

This is corroborated by a much more reliable kinematic bulge-to-disc
decomposition which we consider in
Fig.~\ref{fig:circularityplot}. Here we follow \citet{Abadi2} and
define for every star with specific angular moment $J_z$ around the
symmetry axis a circularity parameter
\begin{equation}
\epsilon = \frac{J_z}{J(E)},
\end{equation}
where $J(E)$ is the maximum specific angular momentum possible at the
specific binding energy $E$ of the star.  We note that another
definition of circularity that is sometimes used in the literature
\citep[e.g.][]{Scannapieco2009, Scannapieco2012} is to replace $J(E)$
with the angular momentum $r\,v_c(r)$ of a star in circular motion at
the star radial distance $r$, where $v_c(r) = \sqrt {GM(<r)/r}$ is
the circular velocity, yielding
\begin{equation}
\epsilon_{V} = \frac{J_z}{r v_c(r)}.
\end{equation}
To allow an easy comparison with both literature conventions, we
include results for both definitions in our mass-weighted circularity
distributions $f(\epsilon)$. We note that the distributions presented
in Fig.~\ref{fig:circularityplot} are normalized such that $\int
f(\epsilon)\,{\rm d}\epsilon = 1$.

Perfectly cold stellar discs should show up as a narrow distribution
around $\epsilon \sim 1$. As we see from
Fig.~\ref{fig:circularityplot}, there are massive discs in all of
our systems, but with varying contributions to the total stellar
mass. The best characterized disc -- considering both the prominence
of the peak at $\epsilon \sim 1$ and the resulting mass fraction -- is
actually found in halo Aq-C, consistently with the original motivation
for picking this halo for the Aquila comparison project, which was
simply the desire to select a system that is most likely to make a
disc. This has, among other factors, to do with the quiet formation
history of Aq-C. Its halo forms comparatively early for systems of
this mass \citep{Boylan-Kolchin2010}, thus favouring the unperturbed
growth of a nice disc over an extended period of time.

\begin{figure}
\includegraphics{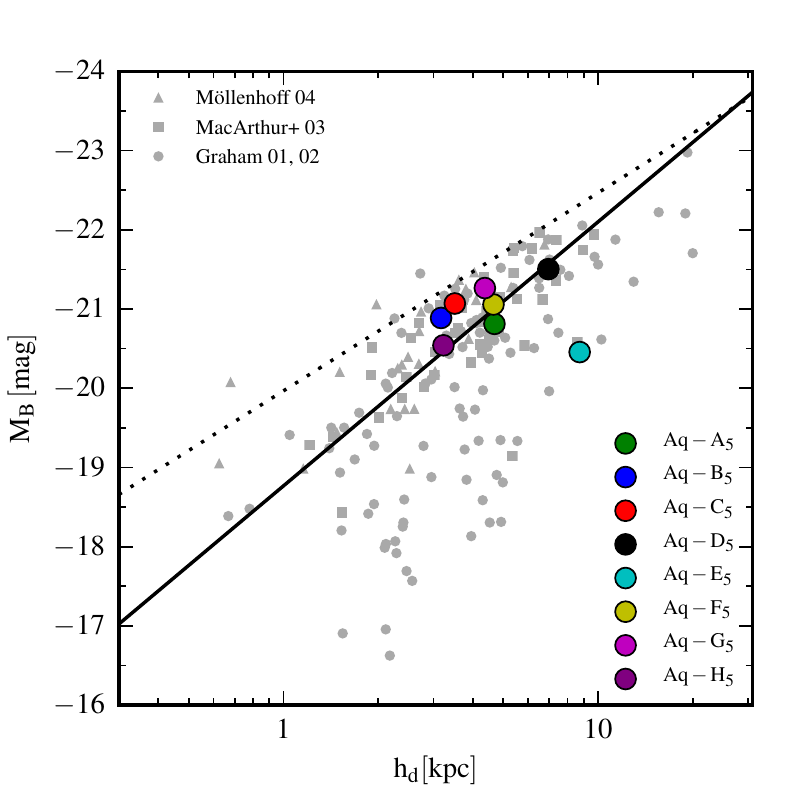}
\caption{$B$-band luminosity versus disc scale-length for the eight
  simulated haloes (coloured circles) compared with the compilation of
  data sets (grey symbols) considered in \citet{Graham2008}. The disc
  scale-lengths of the simulations come from \FM{fits to the surface brightness profiles}, 
  and for the determination of the $B$-band luminosity no dust attenuation was taken into account. 
  Data points are corrected for inclination and dust
  effects as in \citet{Graham2008}. From the same reference we also
  plot the best-fitting relation (solid line) and the observed upper
  boundary to the surface brightness (dotted line).}
\label{fig:luminositysize}
\end{figure}

The D/T ratios obtained from the kinematic
decomposition are included in the individual panels of
Fig.~\ref{fig:circularityplot}, based on the fraction of stars with
$\epsilon > 0.7$ \citep[][find this measure to be roughly equivalent
  to the fraction $\epsilon_V > 0.8$]{Aumer2013b}. Yet another
kinematic measure of the disc fraction is based on the original
approach of \citet{Abadi2}, who defined as bulge component twice the
fraction of stars with $\epsilon < 0$. This gives the highest
kinematic estimates of D/T ratios for the systems and can be seen as a
plausible upper limit on the disc fraction. We also include these
values in square brackets in the individual panels of
Fig.~\ref{fig:circularityplot}. The highest D/T ratios we obtain are
about $\sim 0.5$ based on the $\epsilon > 0.7$ definition, and even
reaching up to 0.8 when the measure of \citet{Abadi2} is adopted.  Our
discs are equally prominent as the best cases in \citet{Aumer2013b},
but not significantly stronger. They are better defined and feature a
higher D/B ratio than most of the other successful disc
formations in the recent literature, which were in part reporting D/T
values based on photometric decompositions. Note that with the latter
method, we would conclude to have obtained pure discs in some of the
systems.

\begin{figure}
\includegraphics{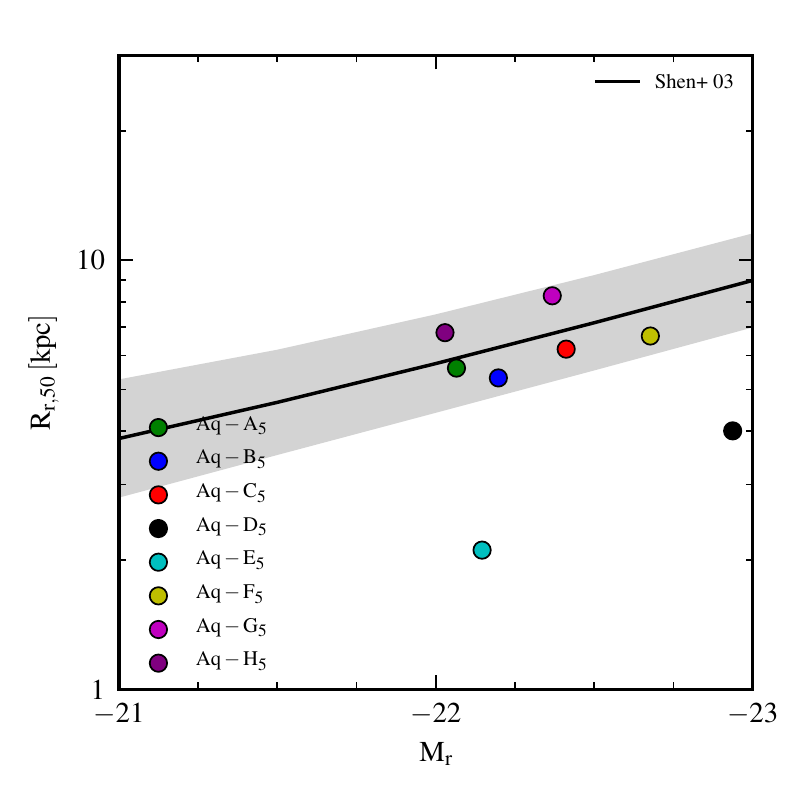}
\caption{\FM{Stellar half-light Petrosian radius versus Petrosian absolute 
  magnitude in the r band} for our simulated galaxies at $z=0$, compared to 
 the median observed relation by \citet{Shen2003} for late-type galaxies in 
 the SDSS. The shaded region represents the observed $1\sigma$ dispersion around 
 the median value.}
\label{fig:halfmassradius}
\end{figure}

In Fig.~\ref{fig:luminositysize}, we compare the scalelength of our
exponential discs, as measured from fits to the \FM{surface brightness
profiles}, with observational data of
the size-luminosity relationship in the $B$ band. The observational
data were taken from the compilation of literature
catalogues\footnote{The references to the catalogues, also given in
  the legend of Fig.~\ref{fig:luminositysize}, are:
  \citet{Mollenhoff2004, MacArthur2003, Graham2001, Graham2002}.}
considered in \cite{Graham2008}, and have been corrected for
inclination and (internal) dust extinction effects by adopting the
prescription described in section 2.2 of their work. We also report
their best-fitting relation (solid black line) and the observed upper
boundary to the disc surface brightnesses (dotted black line).  We
remind that since the disc scalelenghts of the simulated galaxies
were derived from face-on \FM{surface brightness decompositions}, no
correction for inclination is required in this case and dust
extinction is neglected for the determination of the $B$-band
luminosities.  All the simulated galaxies have realistic $B$-band
magnitudes that lie below the (observed) upper boundary to the surface
brightness. It is also interesting to note how the simulated systems
with a substantial bulge (Aq-A, Aq-D and Aq-E) are preferentially
located towards larger discs scale-lenghts \FM{with Aq-E also featuring a 
substantially lower $B$-band luminosity, although within the
observed scatter of the data, with respect to the best-fitting relation.}
Another simple measure of the galaxy
size is obtained by considering the stellar \FM{half-light Petrosian radii in the 
r band $R_{{\rm r}, 50}$}. This can be compared with observational data for the
$R_{{\rm r}, 50}$ -- $M_{\rm r}$ relationship, which is done in
Fig.~\ref{fig:halfmassradius}. We find that the simulated galaxies
fall right into the observational distribution of half-light radii at
comparable stellar \FM{luminosities} found by \citet{Shen2003} for the Sloan
Digital Sky Survey (SDSS), \FM{the most notable exceptions being the Aq-D and Aq-E
haloes that produce galaxies that are too compact given their luminosity}.  
We hence conclude that our galaxies have
sizes for their luminosities that agree well with the observational
data.

\begin{figure*}
\resizebox{17.8cm}{!}{\includegraphics{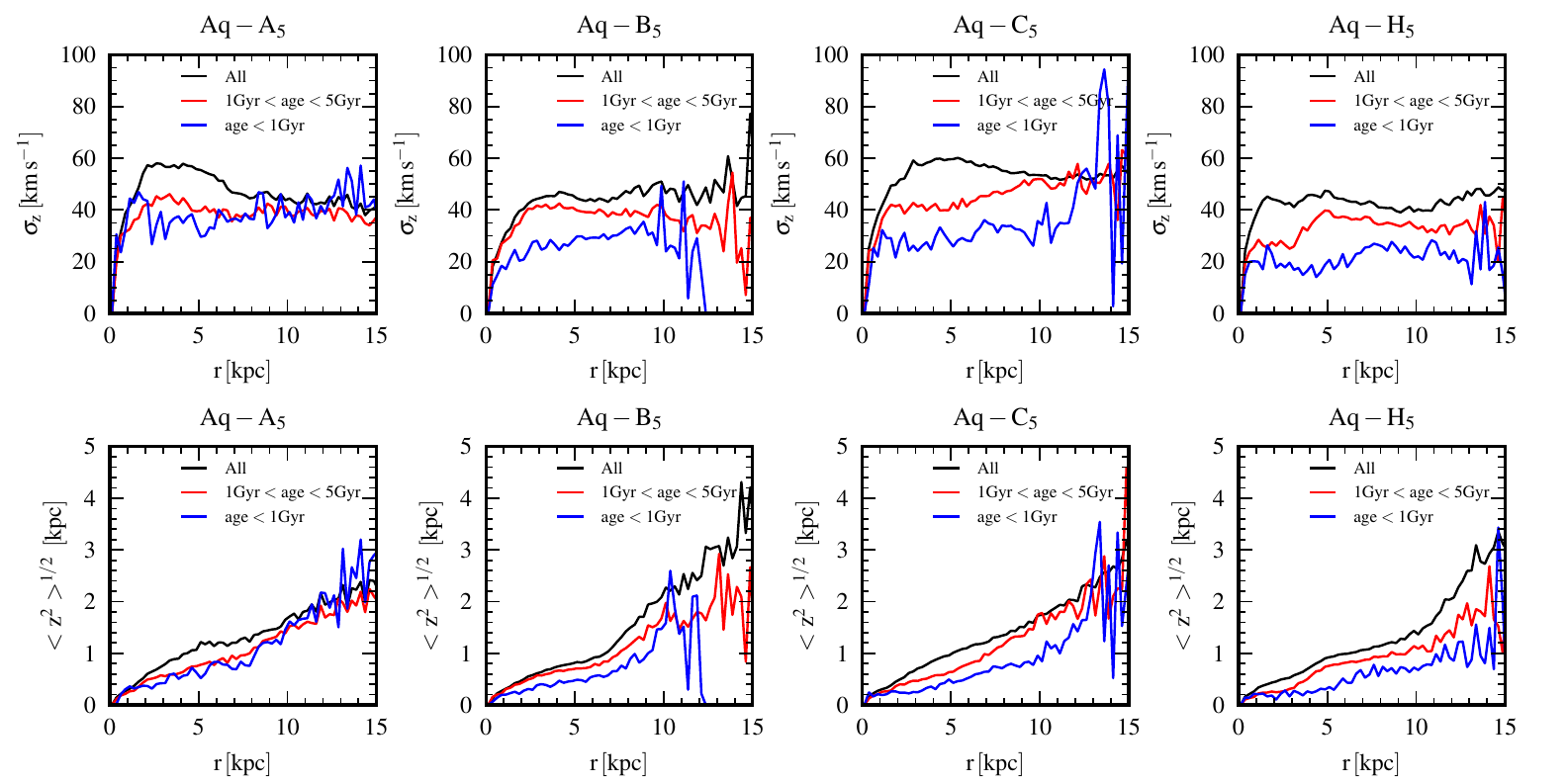}}
\caption{\FM{Vertical velocity dispersion (upper panels) and
scaleheight (lower panels), computed as the mass-weighted second
moment of the $z$-coordinate, of disc stars as a function of
galactocentric radius for a subset of the simulated haloes hosting
prominent disc components. Disc stars are selected as stars
with circularity parameters $\epsilon > 0.7$ and the plots show the
above quantities for three different age cuts: `young' stars (blue
lines), `old' stars (red lines) and the total sample of disc stars
(black lines). Young stars tend to be preferentially colder and hence feature
a thinner vertical distribution with respect to old stars, but
the typical disc scaleheights estimated from this analysis ($\sim
0.5~\kpc$) are too large when compared to observational data.}} 
\label{fig:sigmastars}
\end{figure*}

\FM{Another important property of stellar discs is their vertical
  structure. Numerical simulations of galaxy formation often find it
  difficult to reproduce the cold (i.e. low velocity dispersion) and
  thin stellar disc structures (typical scaleheight $\sim
  100-200~\pc$) observed in late-type spirals and in the Milky Way
  \citep[e.g.][]{Stinson2013b}. While this may partially reflect a
  lack of resolution needed to correctly capture the vertical dynamics
  of the simulated discs -- we remind that even in the case of Aq-C-4,
  our highest resolution simulation, the gravitational softening is
  340 pc -- the treatment of the galaxy's ISM and stellar feedback
  plays an even more important role. In the present simulations, stars
  are born within a gaseous disc approximately in vertical hydrostatic
  equilibrium, where the gas pressure gradient balances the vertical
  gravitational field. The thickness of this gaseous structure is thus
  set by the sound speed of the star-forming phase, which should be
  low enough such that the scaleheight of the gaseous disc is
  comparable with that of the thin stellar disc. This could represent
  a potential problem for the \cite{SFR_paper} ISM model adopted in
  this work, because the effective sound speed needed to stabilize the
  gaseous disc against fragmentation is larger than that expected for
  a stellar disc with $\sim 100~\pc$ thickness.

\begin{figure}
\resizebox{0.45\textwidth}{!}{\includegraphics{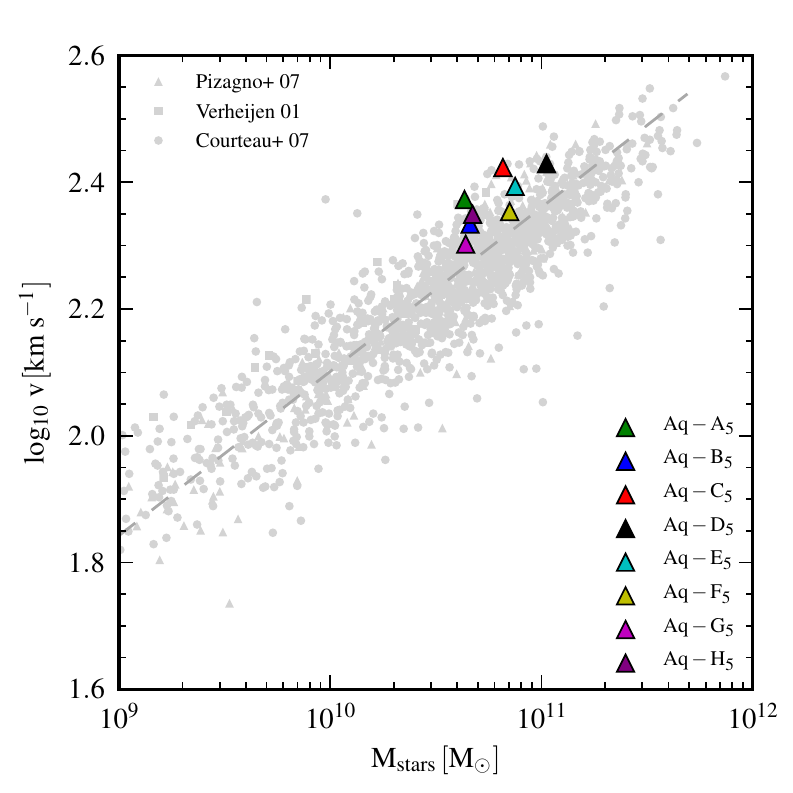}}
\caption{The Tully-Fisher relation for the simulated haloes at $z =
  0$.  Grey symbols represent the observational data sets given in the
  figure legend, while the long dashed line is the best-fitting relation
  derived from the same set of observations by
  \citet{Dutton2011}. Triangular symbols show our simulation
  results. All the simulated galaxies fall comfortably within the
  observed scatter of the relation.  }
\label{fig:tully-fisher}
\end{figure}

In Fig.~\ref{fig:sigmastars}, we present the mass-weighted vertical
velocity dispersion (upper panels) and the mass-weighted second moment
of the $z$-coordinate (a measure of the disc scale-height, lower
panels) as a function of galactocentric radius of disc stars
(i.e. stars with circularity parameter $\epsilon > 0.7$) for a subset
of the simulated haloes hosting prominent discs.  We display these
quantities for three different age cuts: `young' stars with age $<
1~\Gyr$, `old' stars with $1~\Gyr <$ age $< 5~\Gyr$, and the total
sample of stars. The vertical velocity dispersion lies between $20$
and $60~\kms$, with a rapid increase in the innermost kpc and
subsequently with a rather flat trend as a function of radius. These
flat profiles imply strongly flared discs, with scaleheights rapidly
increasing from $\sim 100~\pc$ to a few $\kpc$ in the outer
regions. There is an apparent trend of both velocity dispersion and
scaleheight with stellar age in the sense that young stars have
preferentially a lower velocity dispersion and hence a smaller
scaleheight compared to old stars. Overall, the scaleheight of the
simulated discs are too large with a typical average value of
$\sim0.5~\kpc$ within a radius of $10~\kpc$. This excessive thickness
of the simulated discs is most likely due to our ISM treatment,
although the trend with stellar age suggests that at least part of it
might have a different origin, for example disc heating by
satellites. }

In Fig.~\ref{fig:tully-fisher}, we consider the Tully-Fisher relation
\citep{Tully1977} of the simulated galaxies at $z = 0$. We determine
the stellar masses associated with each galaxy by simply summing the
masses of the stellar particles that lie within $10\%$ of the virial
radius of each halo (the centre of the halo coincides with the
position of its potential minimum), while for the rotation velocity we
take the circular velocity defined as $v_{\rm c}(r)=
\sqrt{{GM(<r)}/{r}}$, where $M(<r)$ is the enclosed total mass at the
radius $r$. Different choices for the radius at which $v_{\rm c}$ is
measured and adopted as characteristic galaxy velocity are possible,
and depending on the detailed shape of $v_{\rm c}(r)$ (see
Fig.~\ref{fig:circvel} and Section \ref{sec:rotation} for a more
detailed discussion of the rotation curves of the simulated
galaxies) somewhat different values of the rotation velocity can be
assigned to each galaxy. In what follows, we use for the Tully-Fisher relation 
the rotation velocity at $R_{\star, 80}$ , corresponding to the radius
within which $80\%$ of the stellar mass is enclosed. At that position
the circular velocity of the galaxy has already reached the flat part
of the rotation curve. We have investigated the range of systematic
uncertainty in assigning a circular velocity to each galaxy by also
considering the peak value of the rotation curves, finding no dramatic
difference, except for the Aq-E halo, which in this case is pushed
slightly above the scatter of the observed Tully-Fisher relation.

In Fig.~\ref{fig:tully-fisher} we include, as grey symbols, a
collection of observed galaxies \citep[from][]{Verheijen2001, 
Courteau2007, Pizagno2007} for which surface photometry and
measurements of rotation velocities were available in the
literature. This data set was already analysed by \cite{Dutton2011} and
we follow their method to convert galaxy luminosities into (MPA/JHU
group) stellar masses, with the appropriate normalization for a
\cite{Chabrier2003} initial mass function. The included best fit to the Tully-Fisher
relation (long dashed line) is also taken from \citet{Dutton2011}. Our
simulated galaxies are close to the predictions of this best-fitting
relation, and even the two cases that are located furthest away
(haloes Aq-A and Aq-C) are comfortably within the observed
scatter. However, the simulations are not evenly distributed around
the best-fitting line, rather they tend to lie preferentially above the
observed relation, with none of the simulated points falling
below. This may indicate still slightly too concentrated galaxies, or
could in part reflect a selection bias in our sample. None the less,
the comparison is encouraging and shows the ability of our
cosmological \arepo\ simulations to form extended disc galaxies with
the correct rotational speed.

\begin{figure*}
\begin{center}
\resizebox{15cm}{!}{\includegraphics{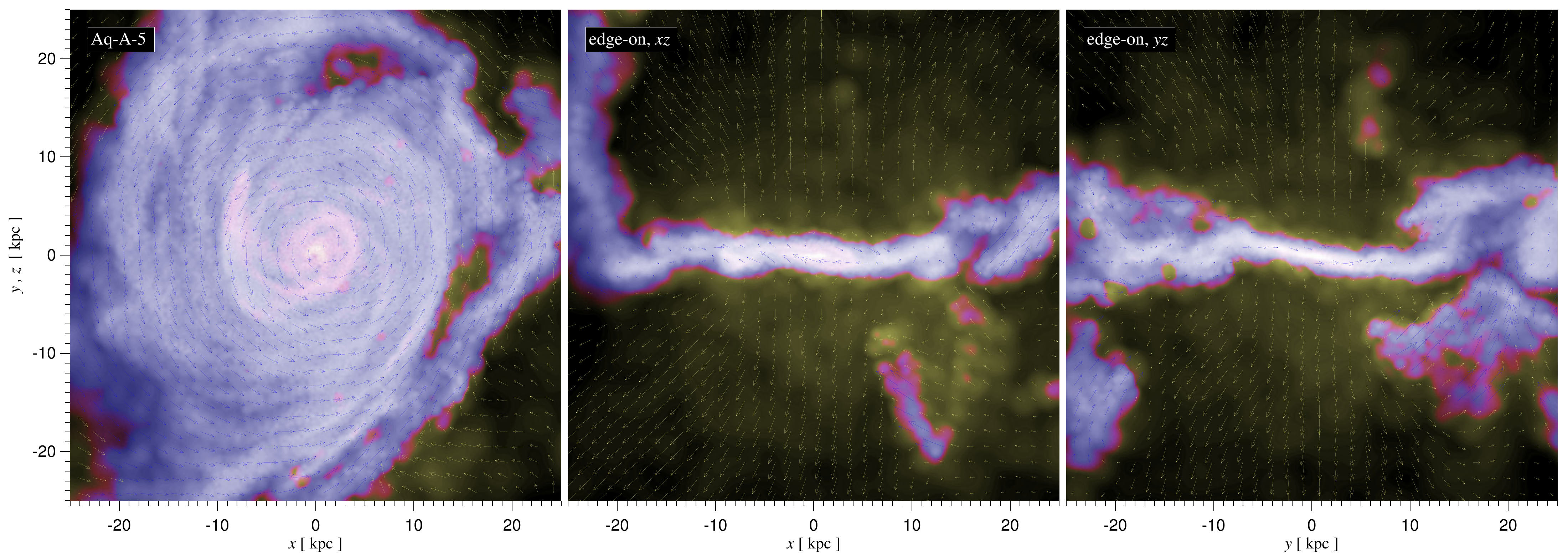}}\\%
\resizebox{15cm}{!}{\includegraphics{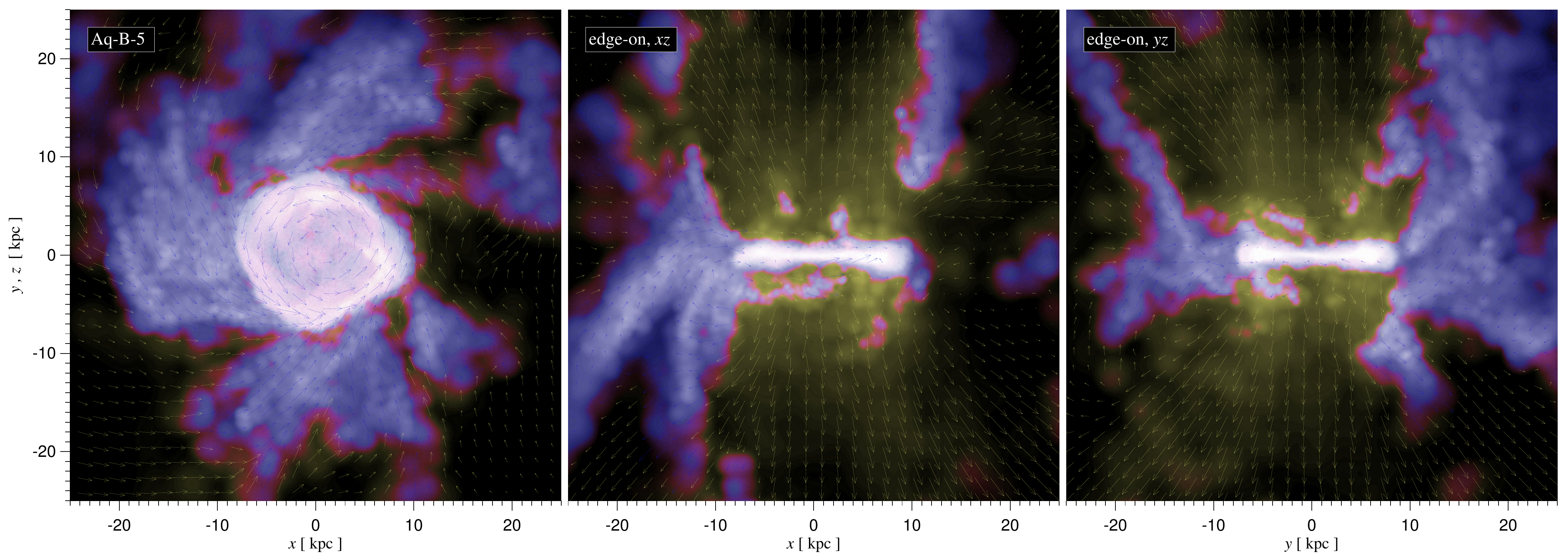}}\\%
\resizebox{15cm}{!}{\includegraphics{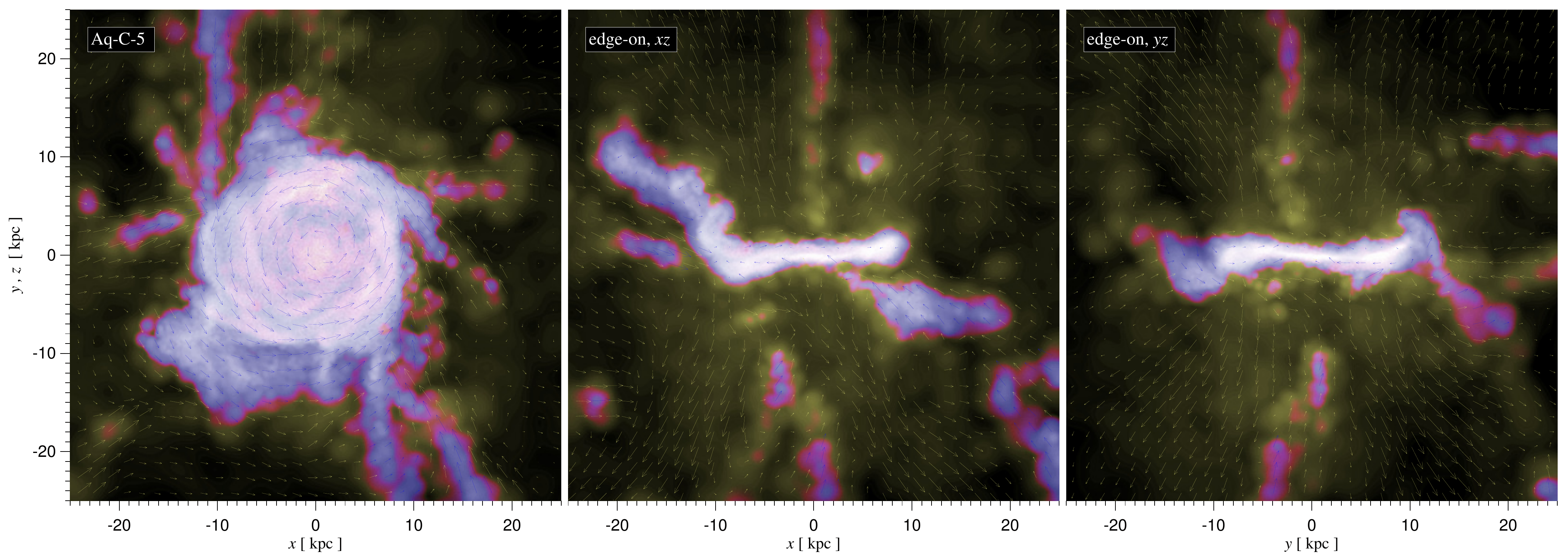}}\\%
\resizebox{15cm}{!}{\includegraphics{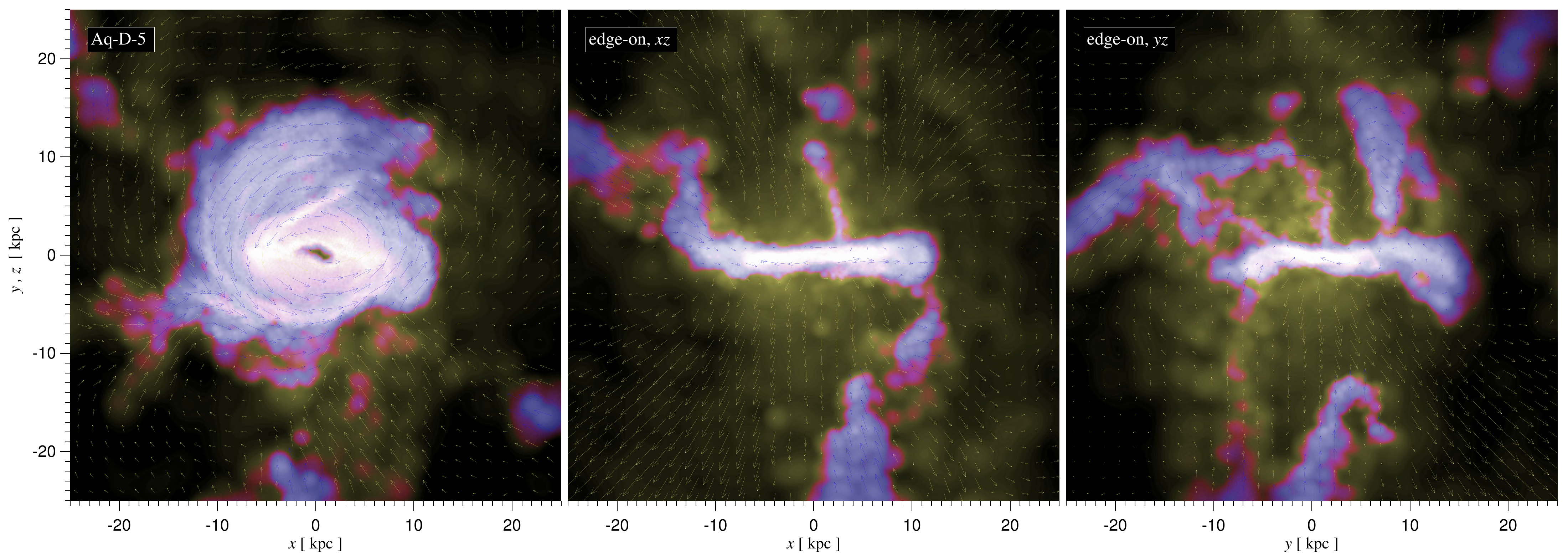}}\\%
\end{center}
\caption{Gas distributions at $z=0$ for the galaxies in haloes Aq-A to
  Aq-D. Each row of panels shows a face-on (left) and two edge-on (centre and
  right) views of the projected gas density for the resolution level
  5. The chosen projection box is $50~\kpc$ on a side and extends for
  a total of $10~\kpc$ in the projection direction, centred on the
  halo potential minimum. The density-weighted gas velocity field is
  overplotted on the density distribution as arrows with lengths
  proportional to the gas velocity. The colour hue encodes the
  density-weighted gas temperature, ranging from blue (cold) to yellow
  (hot). The velocity vectors are drawn either in blue or yellow
  according to the local temperature in order to improve
  visibility. An extended gaseous disc supported by rotation is
  clearly visible in all cases, and in the edge-on panels, a global
  outflow motion of low-density gas, due to our kinetic wind feedback,
  can be seen.}
\label{fig:gasdisk}
\end{figure*}

\begin{figure*}
\begin{center}
\resizebox{15cm}{!}{\includegraphics{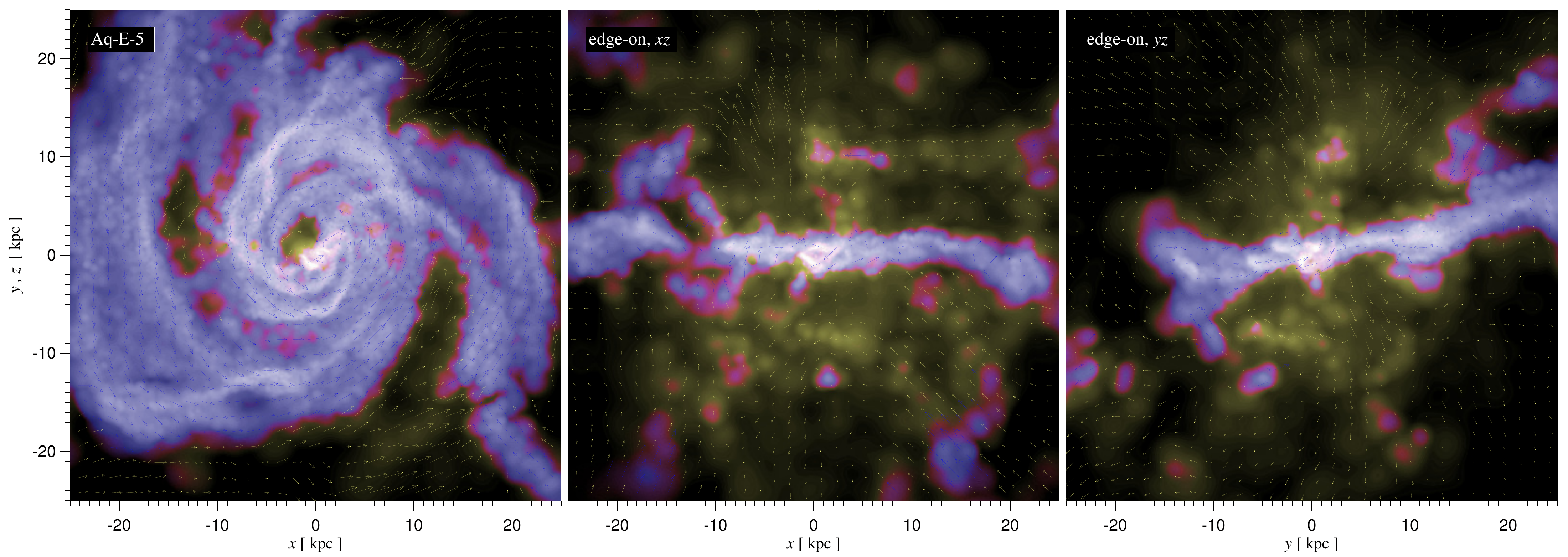}}\\%
\resizebox{15cm}{!}{\includegraphics{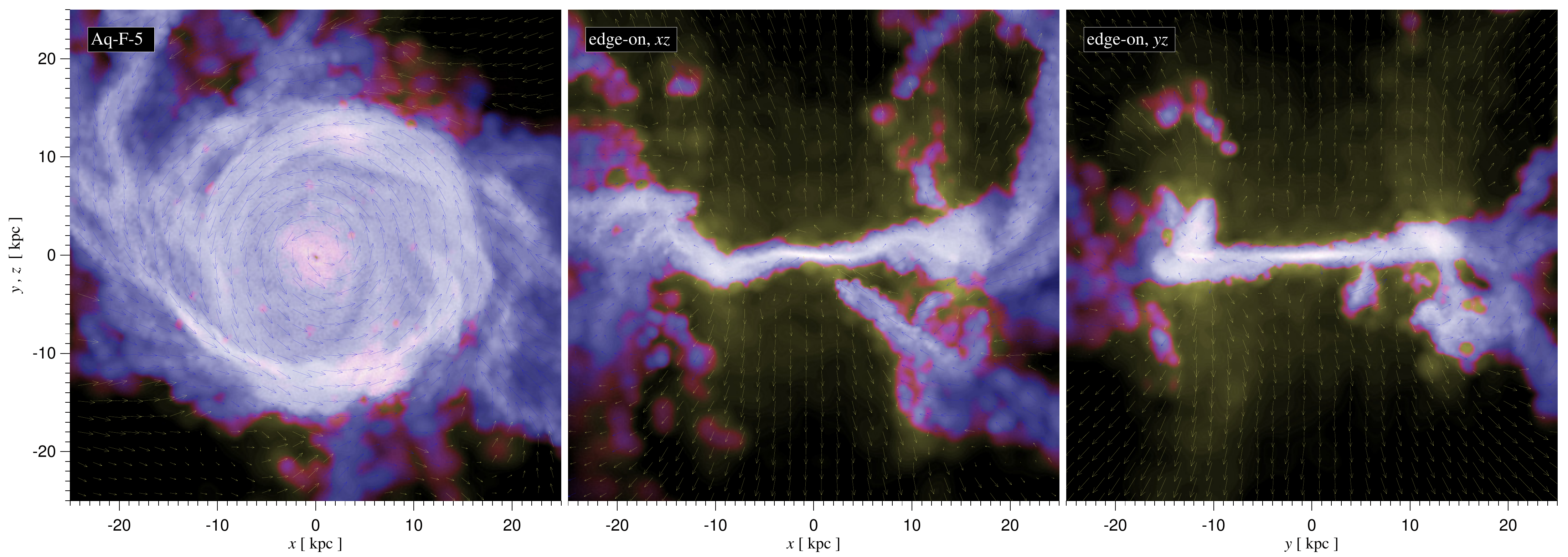}}\\%
\resizebox{15cm}{!}{\includegraphics{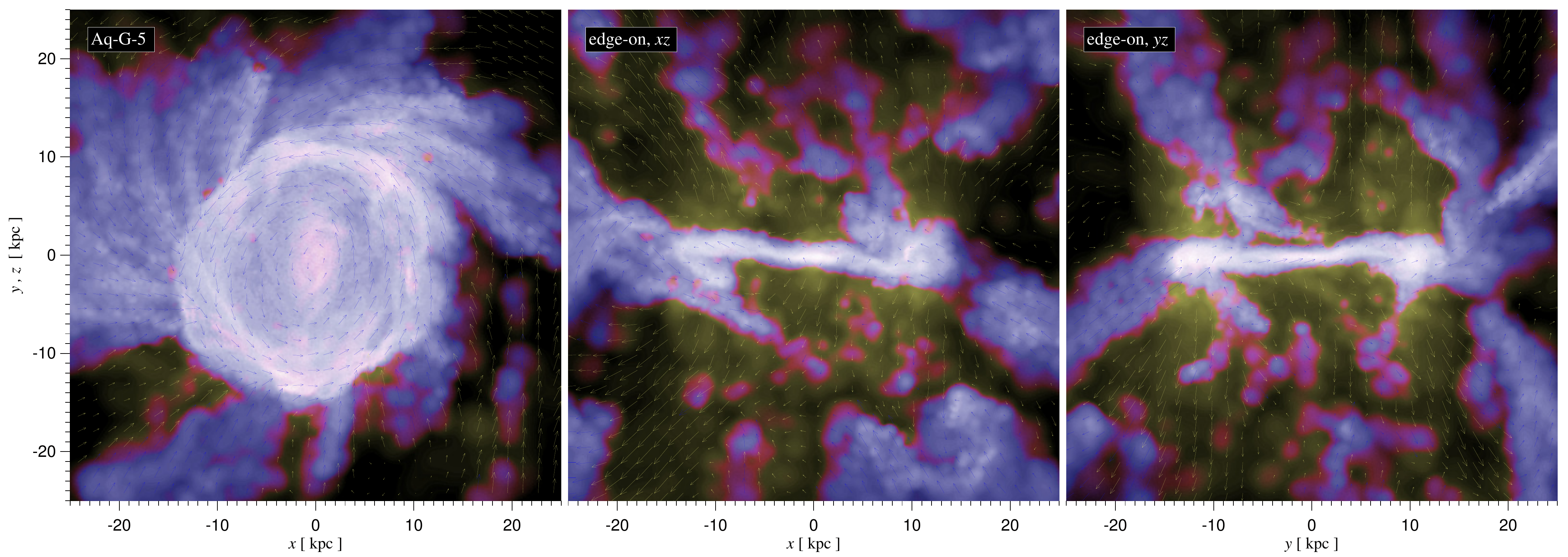}}\\%
\resizebox{15cm}{!}{\includegraphics{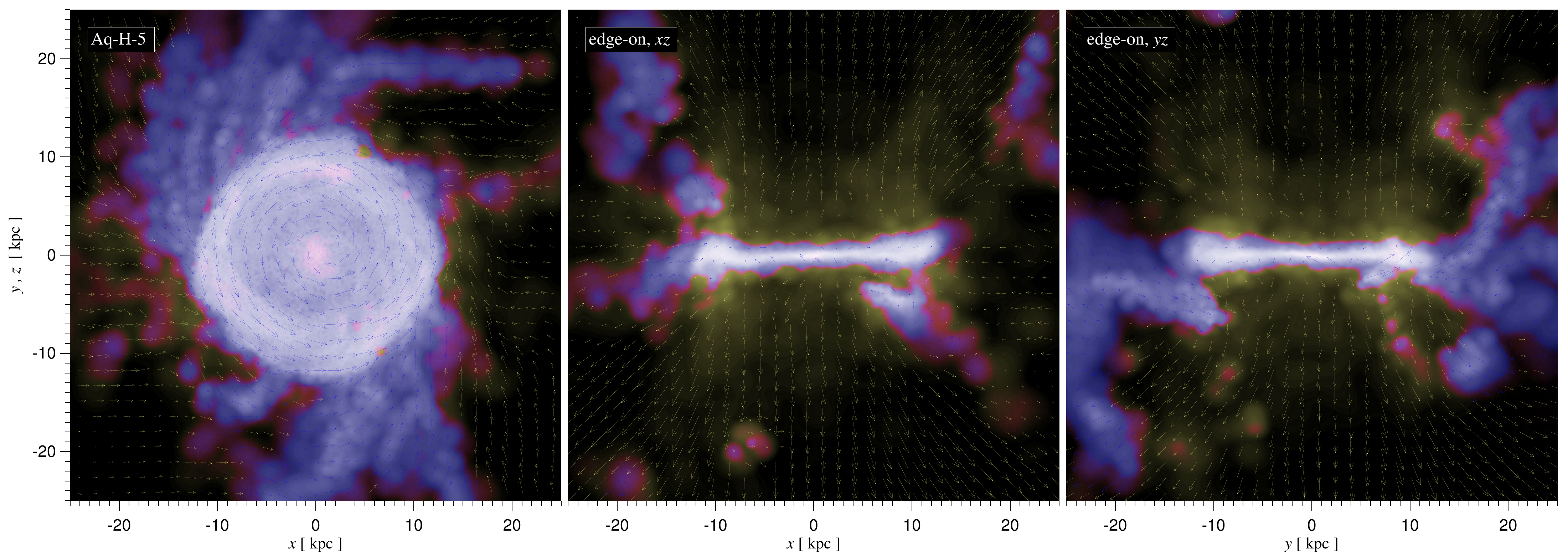}}\\%
\end{center}
\caption{The same as Fig.~\ref{fig:gasdisk}, but for haloes Aq-E to Aq-H.}
\label{fig:gasdisks2}
\end{figure*}

\FM{A more careful comparison with the observed Tully-Fisher relation
  would require us to consider the actual kinematics of the gaseous
  and/or stellar components taken at some characteristic radius,
  usually $2.2$ times the disc scalelength in some photometric
  band. Although this is in principle certainly feasible, we would
  like to note that for the scope of this work we are not interested
  in reproducing this relation in exquisite detail (which would also
  need a larger galaxy sample), but only in showing that our
  simulation methodology is able to produce realistic discs with
  reasonable rotation speed. We also note that if one examines the
  kinematic state of the gaseous and stellar components (see Figs.
  \ref{fig:circvel} and \ref{fig:circvelstars}) more closely, the
  actual rotation velocities show only a slight deviation from the
  theoretical values adopted in the previous analysis. The deviation
  is usually towards lower velocities and this would improve the
  overall agreement between our simulated set of galaxies and the
  observations.}

\subsection{Gaseous discs}

In Figs.~\ref{fig:gasdisk} and \ref{fig:gasdisks2}, we show
projections of the gas density and gas velocity fields for all
eight simulated galaxies at $z=0$. For each galaxy, three panels
arranged in one row are shown, corresponding to a face-on and two
edge-on projections along the principal directions defined for the
$(x'y'z')$ frame of each galaxy.  In each case, the projections show
all the gas in a $50\times 50\times 10~{\rm kpc}$ box around a galaxy's
centre. To enhance the visibility of the dense gas, the projections
are weighted by the local gas density of each cell. The colour map
encodes the gas density as image brightness, and the average gas
temperature as colour hue. In addition, velocity field vectors are
overplotted on a uniform grid of dimension $32 \times 32$. The length
of the velocity vectors is proportional to the magnitude of the
velocity in the projection plane, and either a blue or yellow colour is
used for the arrows depending on the local temperature in order to
enhance visual clarity.

The gas velocity fields in the face-on projections show a high degree
of regular circular motions, especially in the dense neutral gas
phase, which in the denser parts coincides with the star-forming
regions. The surface density profiles of the gas are nearly flat,
similar to the \ramses\ simulations of \citet{Agertz2011}. In a few
cases, one also sees distortions of the gas disc, as expected based on
the stellar distributions. For example, Aq-D-5 shows a bar-like gas
distribution and a central density depression that is presumably
related to AGN feedback.  The lowest gas surface density is seen in
Aq-E-5, which appears red in the stellar projections. Apparently,
comparatively little star formation is ongoing in this system, but
this is still occurring in a disc-like configuration, which is however
roughly orthogonally oriented compared to the spin of the old
stars. The morphology of the gas discs is quite smooth \FM{-- a result primarily
due to the fact that large density fluctuations in the gaseous disc 
are prevented by the effective equation of state used in the treatment of the ISM --}  in stark
contrast to the flocculent gas structures in the SPH simulations of
\citet{Guedes2011} and \citet{Aumer2013b}, that are a \FM{consequence} of their
delayed cooling and/or early stellar feedback models.

Of particular interest in the gas projections are the edge-on views,
which reveal the presence of low-density winds emanating from the
star-forming discs. These outflows are a direct consequence of the
strong kinetic wind feedback realized in our models. Gas streams on to
the discs mostly at large radii and from directions close to the disc
plane.  However, there are also fountain-like inflows of dense gas
scattered over the disc plane. The gas flow in the circumgalactic
medium is clearly quite complicated, reflecting the expected complex
interactions between the galactic winds, the gaseous halo and
cosmological infall. We note that there is no evidence for a
population of small dense gaseous `blobs' in the halo as seen in many
SPH calculations \citep[][]{Kaufmann2006, Guedes2011}.

\begin{figure}
\includegraphics{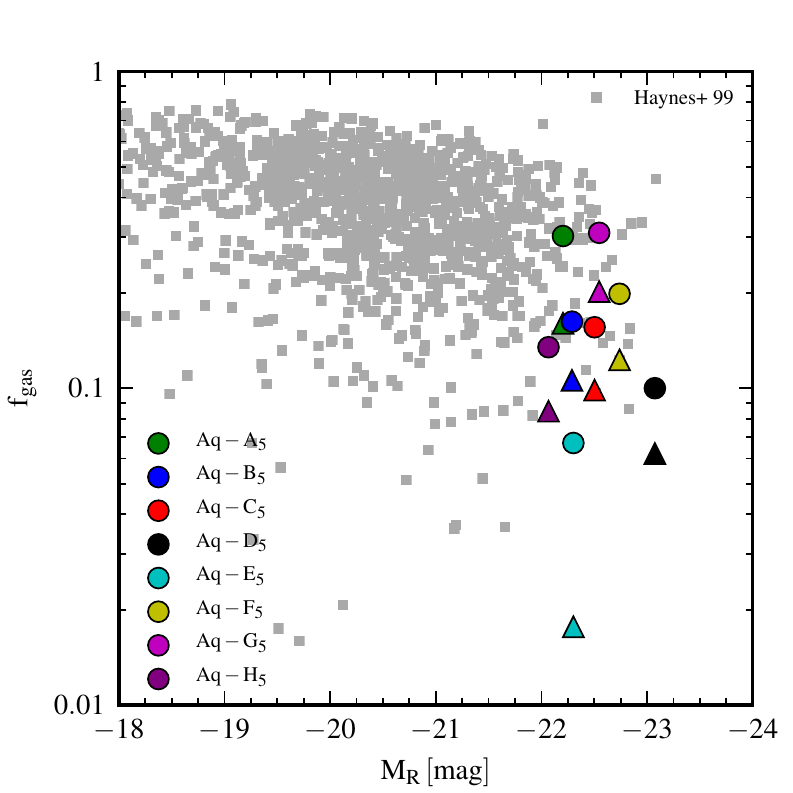}
\caption{Gas mass fraction versus $R$-band magnitude for the eight
  simulated haloes (coloured symbols). \FM{The gas fractions were computed by
  considering all the baryons within $0.1\times R_{\rm vir}$ (circles) or only
  the star-forming gas within the same radius (triangles).} We
  compare the outcome of the simulations with a catalogue of galaxies
  compiled by \citet{HaynesHI1999, HaynesOpt1999}, for \hi\ and
  optical observations, respectively. The observed gas \FM{fractions} have
  been corrected for He abundance. }
\label{fig:gasfraction}
\end{figure}

In Fig.~\ref{fig:gasfraction}, we compare the gas fractions of the
simulated galaxies at the present day relative to observational data
in the $f_{\rm gas}-M_{\rm R}$ plane. The gas fractions are defined
as
\begin{equation}
f_{\rm gas} = \frac{M_{\rm gas}}{M_{\rm gas} + M_{\rm stars}},
\end{equation}
where $M_{\rm stars}$ and $M_{\rm gas}$ are the stellar and the gas
mass within a radius $r = 0.1\times R_{\rm vir}$.  
Stellar $R$-band magnitudes follow from the total stellar luminosity within the same
radius in that passband\footnote{Actually, in the current
  implementation of the stellar photometry $R$-band magnitudes are not
  available. To compute them we converted SDSS r-band magnitudes by
  using eq.~A8 in \cite{Windhorst1991}.}. 
\FM{For the gas, we use two different selection criteria: in one case, 
we measure the total gas fraction (circles) obtained by taking all the gas
contained within the adopted radial cut, whilst in the second (triangles) 
only star-forming gas, again inside $r = 0.1\times R_{\rm vir}$, is considered.}
Again, dust attenuation
has not been taken into account. Data points in
Fig.~\ref{fig:gasfraction} have been obtained by cross-correlating two
samples of nearby galaxies from \citet{HaynesHI1999, HaynesOpt1999}
presenting $21$ cm \hi\ observations and $I$-band photometry,
respectively.  \hi\ integrated line fluxes are converted into gas
masses through the formula \citep[see e.g.][]{Wakker1991}
\begin{equation}
M_{{\rm H}\,{\small\rm I}} = 2.35\times10^6 \left(\frac{D}{\rm Mpc}\right)^2
\left(\frac{S}{\rm Jy\,km~s^{-1}}\right)\,\, {\rm M}_{\odot},
\end{equation}
where $D$ is the distance to the galaxy and $S$ the integrated
\hi\ line flux.  
The resulting masses are then multiplied by a factor of $1.37$ to correct
for helium abundance. $I$-band magnitudes are transformed to $R$ band by
assuming an average $(R-I)$ colour of $0.5$ as in the original paper
by \citet{HaynesOpt1999}. They are also used as an input
to compute the observed stellar masses through a linear fit of the
simulation results in the $M_{\rm star}-M_{\rm R}$ plane. The
\FM{total} gas fractions that we measure for the simulations are
between $10$ and $30\%$, which is reassuringly high and consistent with the
observational data.  Strong feedback models often tend to have
significant difficulties in retaining a sufficiently large amount of
gas available for star formation at late times. Our models do not
appear to suffer from this problem \FM{as can readily be seen from the
  gas fractions of the star-forming phase, which are approximately a
  factor of $2$ smaller than the total gas fractions but still
  sufficiently high to allow star formation to be active at $z = 0$.}
The only exception is the Aq-E halo, which has a gas fraction of $\sim
5\%$ \FM{($\sim 2 \%$ if only star-forming gas is taken into
  account)}. However, this system represents a particular case which
is characterized by a rather strong merger event at $z \sim 1$. This
triggered significant BH growth and associated feedback in our
simulation, similar to the scenario of merger-induced formation of red
ellipticals described in \citet{SpringelMatteoHernquist}.

\section{Formation history}\label{SecHistory}

\begin{figure*}
\begin{center}
\resizebox{16.5cm}{!}{\includegraphics{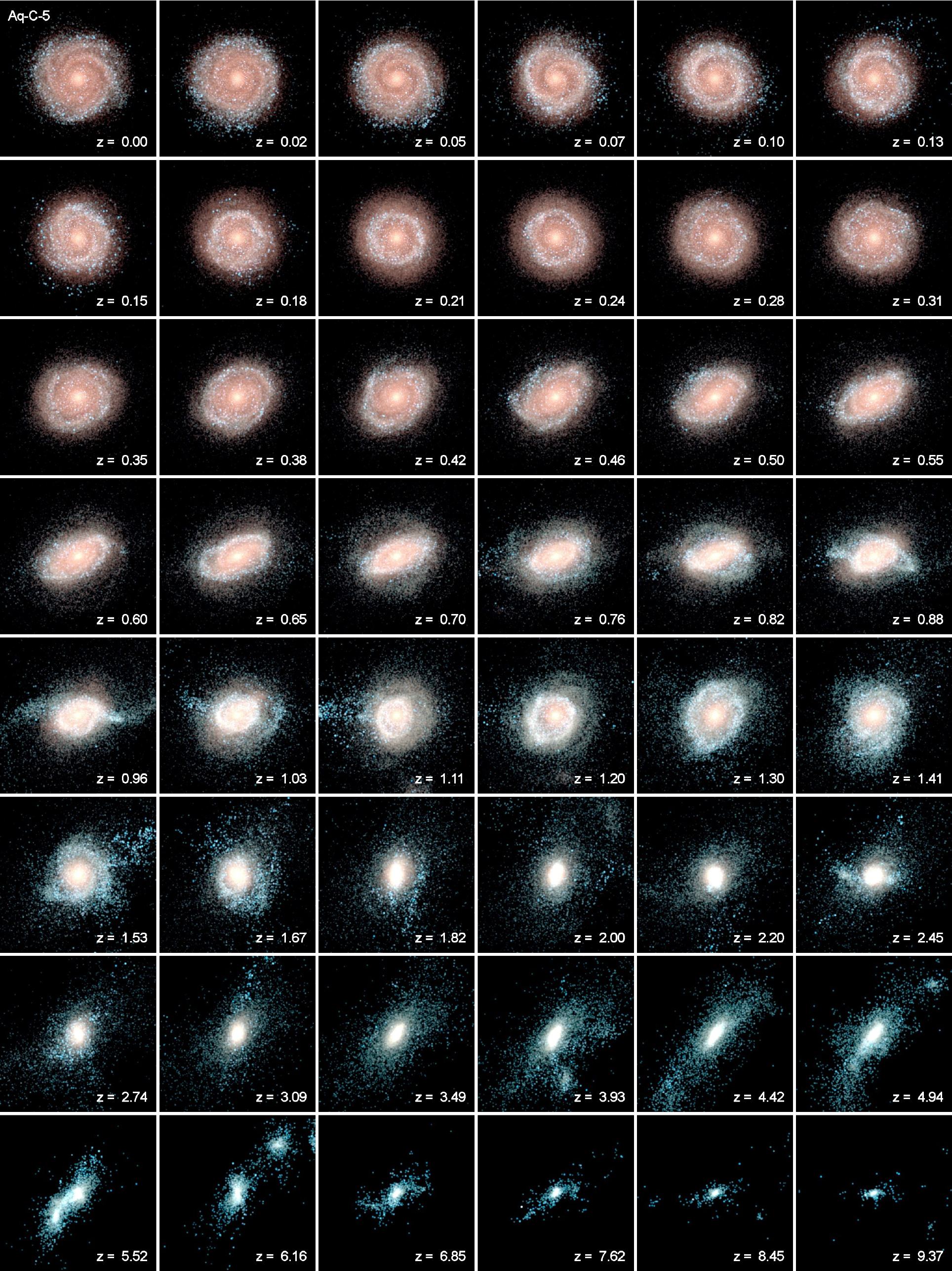}}
\end{center}
\caption{Time evolution sequence of the formation of the Aq-C-5
  galaxy, from $z\sim 10$ to $0$, as labelled. We show all stars in a
  box of length $50\,{\rm kpc}$ (physical) on a side, centred on the
  potential minimum of the main progenitor's dark matter halo. The
  $z=0$ frame has been oriented for a face-on view, and this
  projection direction is kept in all other panels. Rest-frame stellar
  luminosities are mapped to image intensity and colour as in
  Fig.~\ref{fig:stellardisks}.}
\label{fig:timevolvAqC5}
\end{figure*}

\begin{figure*}
\begin{center}
\resizebox{16.5cm}{!}{\includegraphics{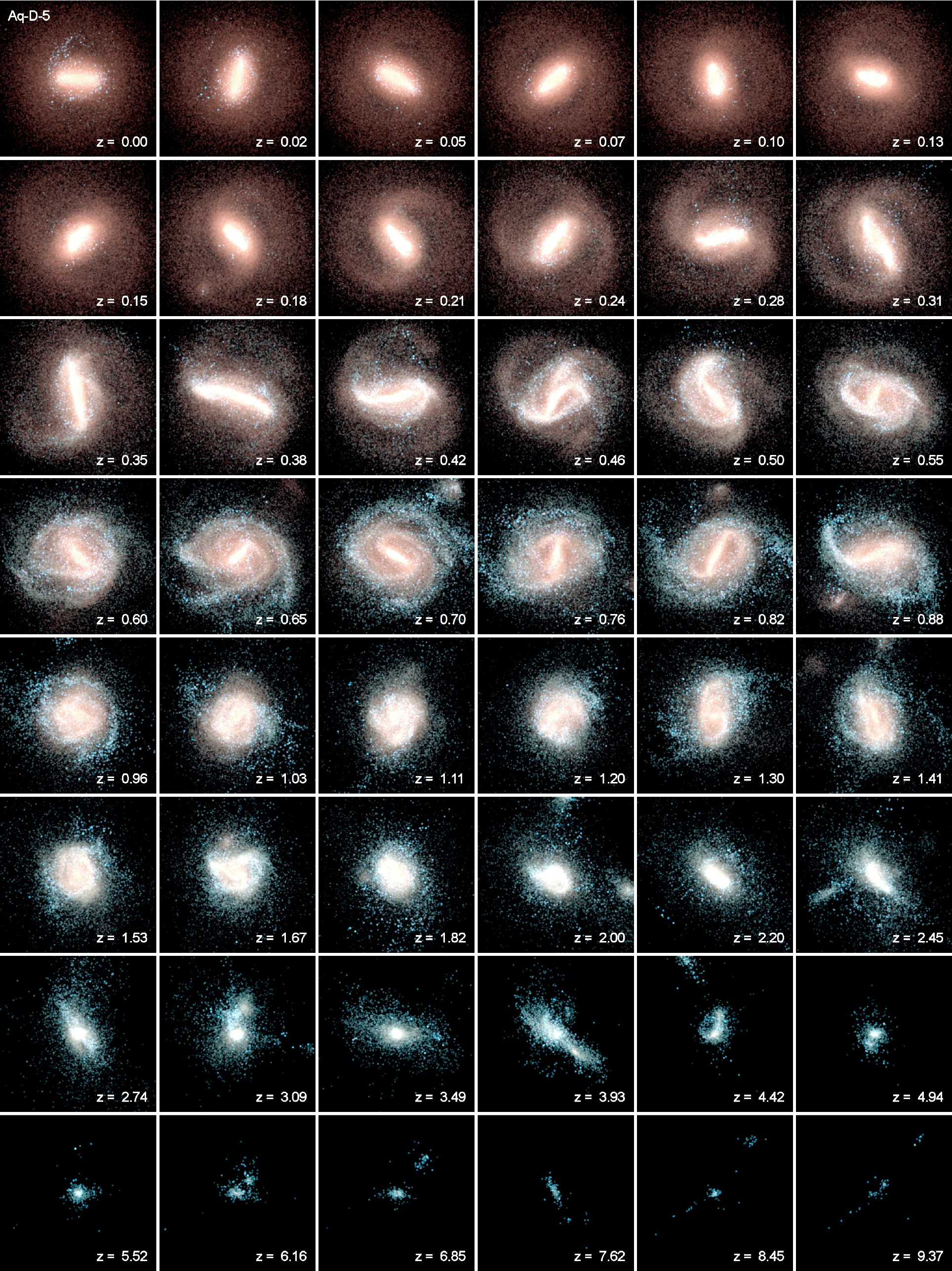}}
\end{center}
\caption{Time evolution sequence of the formation of the Aq-D-5 galaxy
  (as in Fig.~\ref{fig:timevolvAqC5}, but for the Aq-D-5 simulation).}
\label{fig:timevolvAqD5}
\end{figure*}

A first qualitative impression of the formation history of our Milky
Way-sized galaxies can be obtained from time evolution sequences of
the stellar distributions of the simulations.  In
Figs.~\ref{fig:timevolvAqC5} and \ref{fig:timevolvAqD5}, we show
this for Aq-C-5 and Aq-D-5 in an exemplary fashion.  We have selected
Aq-C-5 as a prototypical case making a nice disc at $z=0$ (and because
it is of particular interest due to its prior analysis in the Aquila
project), and Aq-D-5 to show a partially failed disc, in this case
through the formation of a strong bar. All the panels of the
evolutionary sequence show the stars in a region of fixed physical
size ($50\,{\rm kpc}$ on a side centred on the galaxy's most massive
progenitor), so that the images directly reflect the size
evolution. The stars are shown with their luminosities in the
rest frame, with the same mapping to colour and intensity as in
Fig.~\ref{fig:stellardisks}. This mapping is kept the same at all
redshifts and output times, so that the brightness and colour
evolution of the images indicates the variations in the star formation
histories (SFHs) of the galaxies displayed at the individual redshifts. In
particular, we can directly infer from the images that the heyday of
blue disc formation lies in the redshift range $z \sim 0.5-1.5$ for
our simulations. At higher redshift, the galaxies in our sample tend
to be small and blue, whereas they redden progressively towards lower
redshifts.

\subsection{Star formation history}

\begin{figure}
\resizebox{0.45\textwidth}{!}{\includegraphics{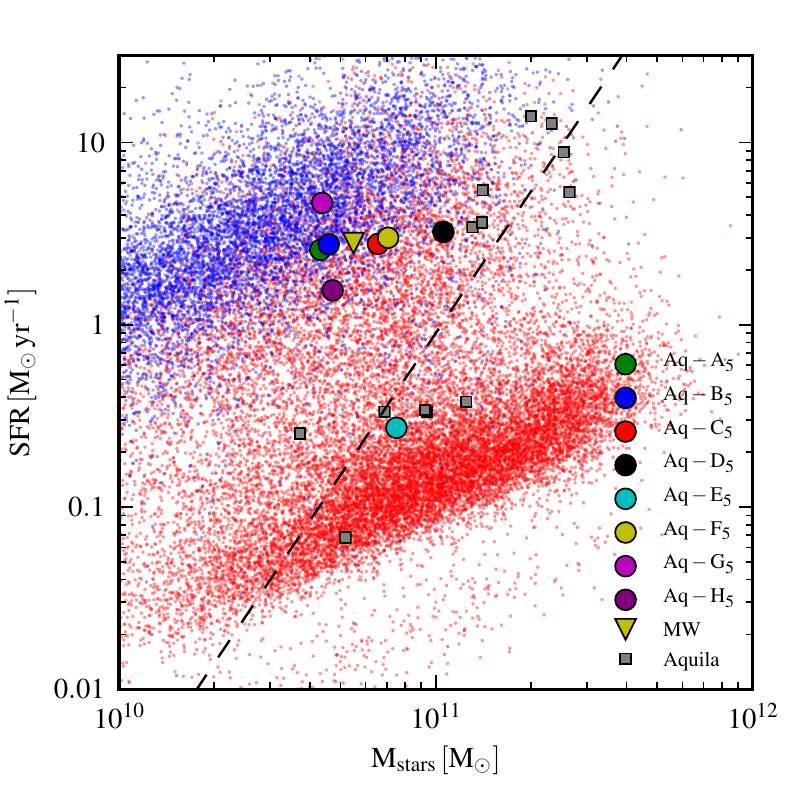}}
\caption{Present-day SFR, averaged over the past 0.5
  Gyr, versus stellar mass. The coloured circles show the results of
  the simulations, while the down-pointing triangle marks the position
  of the Milky Way as given by \citet{Leitner2011}. Background dots
  are a random sub-sample ($10\%$) including only nearby galaxies ($z
  < 0.1$) of the SDSS MPA-JHU DR7.  This subsample is
  divided into a blue cloud and a red sequence by adopting the same
  colour cut used in the Aquila project \citep{Scannapieco2012}. The
  small boxes show the results for the Aq-C halo obtained in the
  Aquila simulations. They tend to lie along the dashed line, which is
  systematically offset relative to the position of the Milky Way;
  once the stellar mass is small enough in the Aquila simulations, the
  present-day SFR is not sufficiently high.  }
\label{fig:sfrvsstarmass}
\end{figure}

The realistic gas fractions measured for our galaxies at the present
epoch support SFRs comparable to expectations for
late-type galaxies at $z=0$. This is seen in
Fig.~\ref{fig:sfrvsstarmass}, where we compare the present-day SFRs 
of our simulations with observational data as a
function of stellar mass. We measure the SFR
time averaged over the past $0.5~\Gyr$ by determining the stellar mass
younger than $0.5~\Gyr$ in a sphere of radius $0.1 \times R_{\rm vir}$
centred on the halo's potential minimum. For measuring the total
stellar mass no age cut is applied. Note that this is the same
procedure as adopted in the Aquila comparison
project~\citep{Scannapieco2012}. The background dots in
Fig.~\ref{fig:sfrvsstarmass} are a random subsample of nearby ($z <
0.1$) galaxies from the SDSS MPA-JHU DR7 data release 7 (DR7)\footnote{http://www.mpa-garching.mpg.de/SDSS/DR7/}, divided 
into the so-called `blue cloud' and `red sequence' on the basis of the
colour condition \citep[see again][]{Scannapieco2012}
\begin{equation}
(g -r) = 0.59 + 0.052 \log\left(\frac{M_{\rm star}}{10^{10}
    M_{\odot}}\right), 
\end{equation}
with the galaxy symbols coloured accordingly. Only $10\%$ of the total
sample is plotted. All the simulated galaxies are actively
star-forming and tend to cluster around the current location of the
Milky Way in the blue cloud, which is marked by the down-pointing
triangle, with properties taken from \citet[table 2]{Leitner2011}.
Again, a clear outlier in our galaxy set is the Aq-E halo, which is
located in the outskirts of the red sequence. This is not surprising:
given the small amount of gas left in the galaxy (see
Fig.~\ref{fig:gasfraction}) a low SFR is expected. We
also include in Fig.~\ref{fig:sfrvsstarmass} the results of the Aquila
comparison project at resolution level 5, for {\em all} the employed
codes and feedback models (grey boxes). Interestingly, these
simulations appear to systematically miss the location of the Milky
Way and cluster around the dashed line. Those simulations that manage
to reduce the SFR sufficiently end up being too red,
while those with enough current star formation are too massive. This
highlights the difficulty to arrive at simulation models that suppress
star formation strongly especially at high redshift, but allow a high
SFR at low redshift. Our new code combined with the
implemented feedback physics achieves this considerably better than
the models studied in the Aquila project.


\begin{figure*}
\hspace*{-0.4cm}\resizebox{18.8cm}{!}{\includegraphics{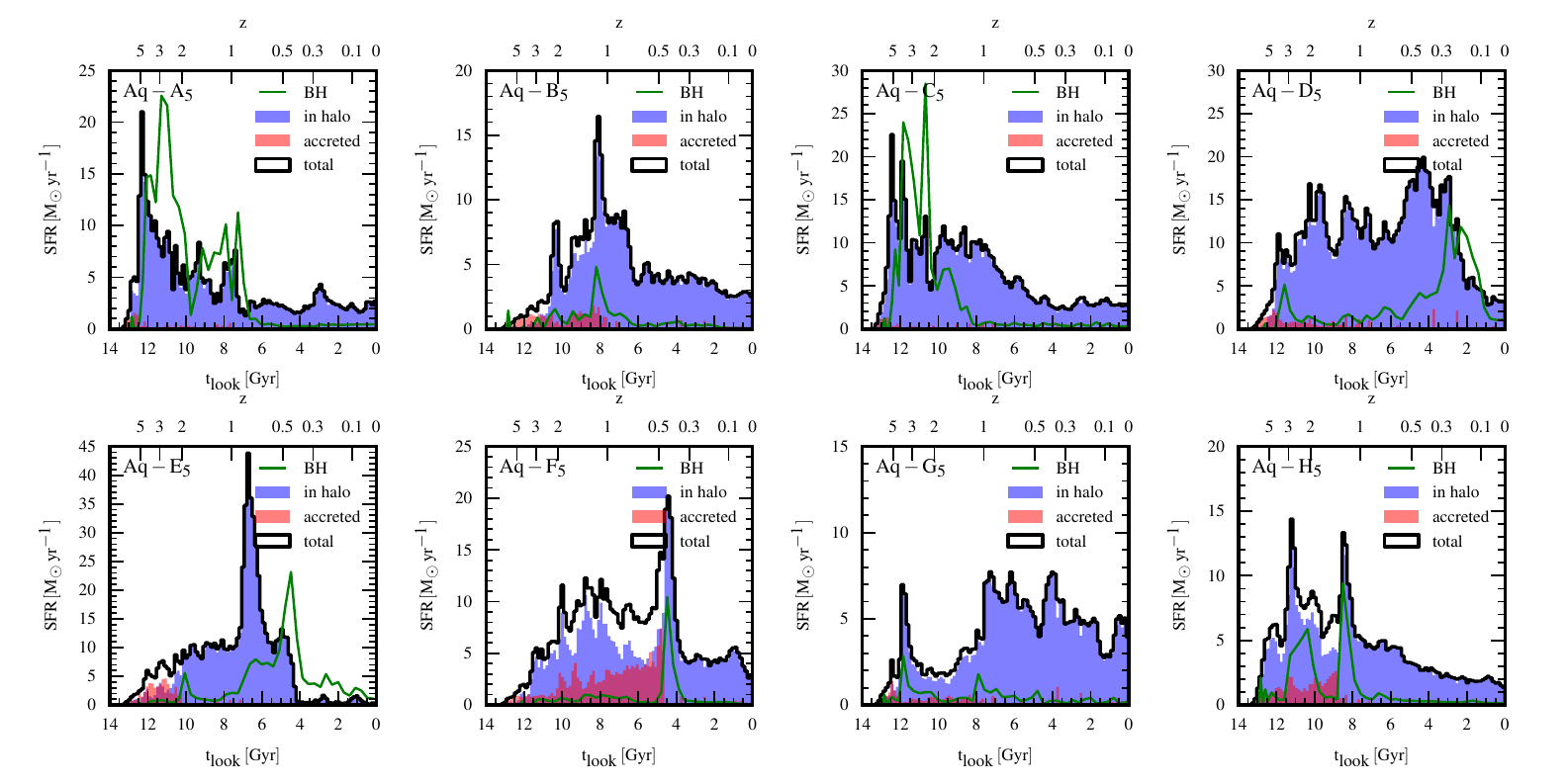}}
\caption{SFHs, as a function of look-back time and
  redshift, for the eight Aquarius haloes. Only stars within $r <
  0.1~r_{200}$ at $z = 0$ were considered in the measurement. For each
  halo, the total SFR is split between the
  contributions of stars born in the halo's most massive progenitor
  (blue histograms) and those that were accreted from other haloes or
  substructures within the halo (red histograms). The latter component
  is usually a small fraction of the total stellar mass, i.e.~most
  stars of the galaxies form in situ. We also overplot the growth
  history of the supermassive BH residing at the centre of
  each galaxy (green lines). The growth rate is the average growth
  rate between two simulation outputs, scaled by a factor of $1/250$ to
  give them a comparable size as the SFRs (in solar
  masses per year).}
\label{fig:sfr}
\end{figure*}

In Fig.~\ref{fig:sfr}, we show plots of the formation history of the
stars contained in each galaxy. This is based on the age histogram of
stars that end up in the galaxies at the present time (again selected
as all the stellar particles in a sphere of radius $0.1\times R_{\rm
  vir}$ centred on the halo potential minimum). SFRs are then 
  computed by dividing the stellar mass in each age bin
by the width of the bin itself, so that they effectively represent the
average SFR over the temporal bin width. Since we model the stellar
mass return to the ISM in our simulations, the initial mass of each
star particle (i.e., the mass that the particle has when it is
created) is employed for the derivation of the SFRs. We used 100 bins
of $140~\Myr$ for a total time span of $14~\Gyr$. We additionally mark
the formation rate of stars that have formed in situ in the main
progenitor (blue histograms), or have been accreted from substructures
(red histograms). The accreted stars are usually a small fraction,
amounting to $\lsim 10\%$ of the total stellar mass, with the notable
exception of the Aq-F halo for which this fraction increases up to
$23\%$. We also overplot the accretion rate of the central BH (green
lines), determined as the average accretion rate between two
simulation outputs and rescaled by a factor of $1/250$ to make it visible
in the plots, to compare its formation history to that of stars. The
general picture that emerges from this comparison is that of a
connection between the two components, which show similar features in
their formation histories at roughly the same look-back times. We
address this point in more detail in Section~\ref{sec:bhgrowth}.

From the figure it can be seen that SFHs
come in a variety of shapes, that reflect the underlying mass assembly
history of the haloes. The general trend is that of a peak of the SFR
at early times ($z \sim 3-5$), with typical maximum SFRs of the 
order of $20\,{\rm M}_\odot{\rm yr}^{-1}$, followed by a
decline and a rather constant SFR at $z \lsim 1$, with most of our
galaxies ending up with SFRs of $\sim 2-3\,{\rm
  M}_\odot {\rm yr}^{-1}$ at $z=0$. However, there are many exceptions
to this rule, especially in systems where a large bulge is found at
present times. In these cases the SFR tends to peak below $z\sim 1$
and then to drop quite abruptly (see haloes Aq-D, Aq-E and Aq-F). This
can either be a signature of a strong merger event, a significant
tidal perturbation or the result of secular evolution that triggers a
bar instability. Another interesting case is that of the halo Aq-G,
which features unusually little star formation at high redshift, a
period of quite low star formation in the redshift interval $1-3$
followed by quite intense star formation from $z=1$ to the present,
and finally a merger even right at the present time that triggers a
$z\simeq 0$ increase of the SFR. This history implies
a quite young age of this galaxy.

\begin{figure}
\includegraphics{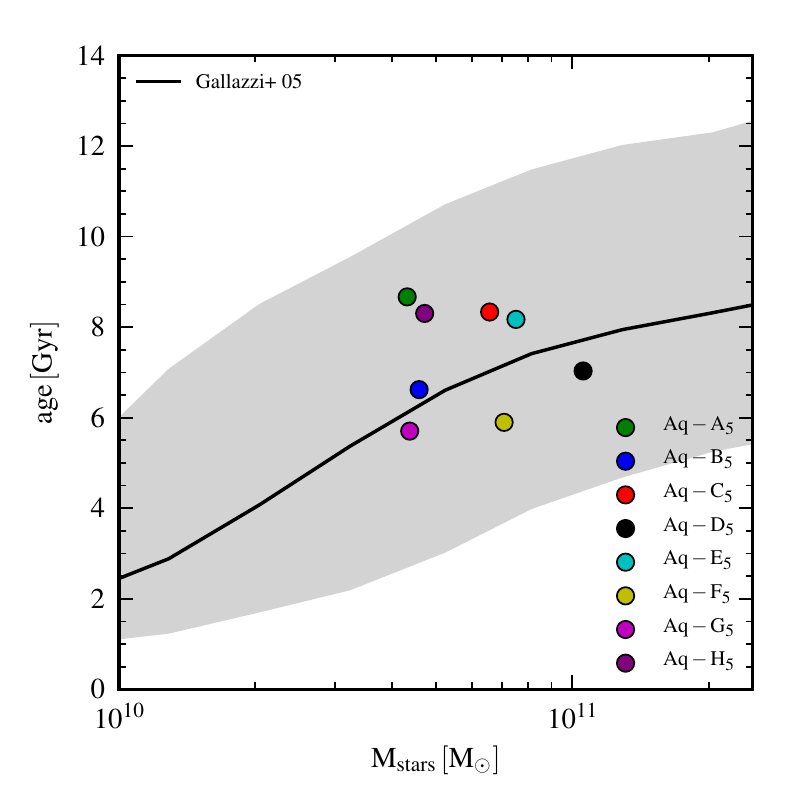}
\caption{Mass-weighted stellar age versus stellar mass for our
  simulated galaxies at $z=0$, compared to the median observational
  relation by \citet{Gallazzi2005}. The grey band indicates the region
  between the 16th and 84th percentiles of the observed age
  distribution as a function of the stellar mass.}
\label{fig:agevsmass}
\end{figure}

This expectation is borne out in Fig.~\ref{fig:agevsmass}, which
compares the mean mass-weighted stellar ages of our galaxies with SDSS
data, as a function of stellar mass. Here, the galaxy Aq-G has indeed
the youngest stellar population. It also has one of the smallest
stellar masses among our set, making it in fact agree rather well with
the observational determination of the age-stellar mass relation of
late-type galaxies in the SDSS carried out by \cite{Gallazzi2005}. Our
other galaxies also agree well with these measurements, but our
dynamic range in galaxy mass is too small to say whether we also
reproduce the significant decline in age towards smaller stellar
masses seen in the observations.

Based on the kinematic D/B decomposition carried out in
Section \ref{sec:stellardics}, we can also determine the mean stellar
ages of these components individually. As expected, the bulge
components tend to be old; they have mean mass-weighted ages of $\sim
7-8$ Gyr, while the stellar disc components are on average younger
with mean ages of the order of $\sim 4-5$ Gyr. Given also the general
shape of the SFH curves, this suggests that bulge
formation is associated with the early peak in the SFR while discs are
built in a more gradual way at later times, consistent with the
inside-out picture of~\citet{Fall1980}.

\begin{figure}
\includegraphics{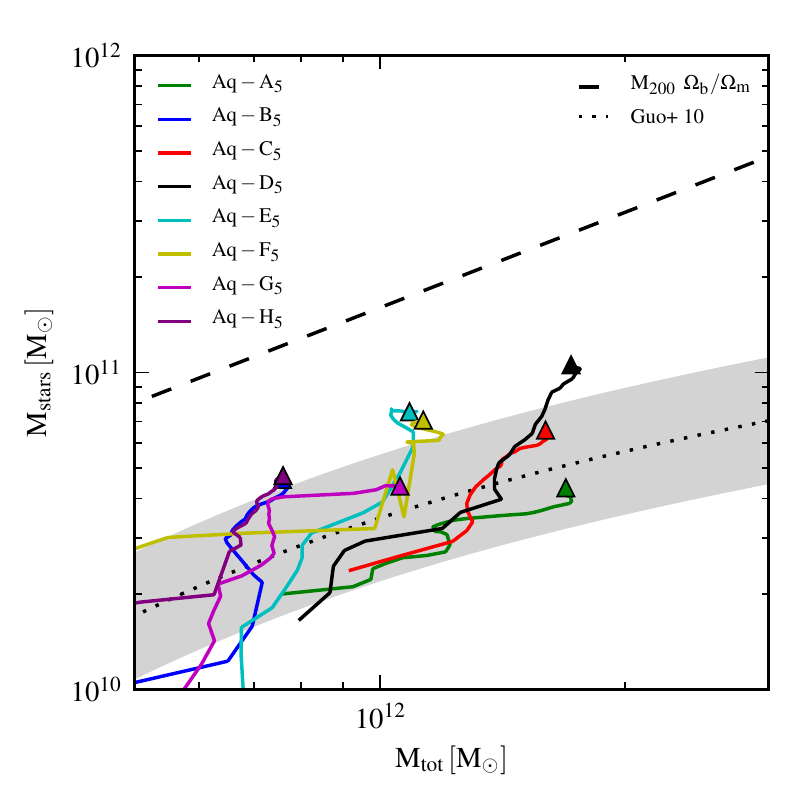}
\caption{Stellar mass versus halo mass for the eight Aquarius
  haloes. The evolutionary tracks are plotted from $z = 2$ to the
  present time. The dashed line represents the baryonic mass
  associated with the halo if the universal baryon fraction is
  assumed. The dotted line shows the results obtained by abundance
  matching \citep{Guo2010} and the grey region encloses an uncertainty
  of $\pm 0.2~{\rm dex}$ around the mean value predicted by abundance
  matching. The evolutionary tracks of the haloes are in good
  agreement with abundance matching expectations, although there is a
  tendency of a slight overproduction of stars in a subset of our
  systems at $z = 0$.}
\label{fig:stellarvstotalmass}
\end{figure}

Of particular interest is how the stellar masses compare to the total
virial masses of haloes. This is seen in
Fig.~\ref{fig:stellarvstotalmass}, where we compare all of our
simulations against the stellar mass-halo mass relation (dotted
line) derived by \cite{Guo2010} from abundance matching arguments. The
shaded region around the expected relation is determined such that its
upper and lower boundaries are at $\pm 0.2~{\rm dex}$ from the
fiducial values, while the dashed line shows the baryonic content of a
halo of a given virial mass if the universal baryon fraction is
assumed. For each system, we include the results at a range of output
times, from redshift $z=2$ to $0$, such that a continuous track is
formed. As the abundance matching results only weakly depend on
redshift \citep{Behroozi2010, Moster2010}, this then also gives a
useful test to see whether our galaxies evolve consistently with this
expected relation. We note that extending this test to still higher
redshift would require taking into account the residual redshift
dependence of the abundance matching results and the different growth
histories of our haloes.

Reassuringly, the stellar masses of our systems are indeed small
enough to be consistent with abundance matching arguments, although a
subset of our galaxies is at the upper end of what is acceptable
within the expected scatter.  A good match is even obtained for two of
the more massive galaxies of the sample, namely Aq-A and Aq-C, which
have virial masses of almost $2\times 10^{12}\,\mo$ but still manage
to have a sufficiently small stellar mass and to form an extended
disc. Matching the stellar mass-halo mass relation has not been
achieved in the Aquila comparison project, except for runs G3-TO,
G3-CR and R-AGN, which however did not form realistic disc systems
with the employed implementations of strong feedback. Also, as pointed
out by \cite{Guo2011} and \cite{Sawala2011}, previous simulations --
even the most successful ones among them -- had generally failed to
match this relation. Only very recently, a few studies
\citep{Guedes2011,  Aumer2013b, Stinson2013a} have reported successful
simulations of individual galaxies conforming with the stellar mass-halo 
mass relation. Similar to our work, these successes have
ultimately become possible due to an increased feedback strength and a
more efficient coupling of the feedback energy to the gas
phase. However, we note that most of these works employed halo masses
considerably less massive than used here, typically around $\sim 0.7
\times 10^{12}\,{\rm M}_\odot$. This lower mass makes it easier to
suppress star formation.  A case in point is provided by the high-mass
systems in \citet{Aumer2013b}, which show overly high low-redshift
star formation, whereas their lower mass systems do not have this
problem. To the best of our knowledge, our runs are the first examples
of the successful formation of late-time spirals in haloes in the mass
range $1-2\times 10^{12}\,\mo$, which is the size of the Milky Way.

\subsection{Central black hole growth}\label{sec:bhgrowth}

\begin{figure*}
\hspace*{-0.4cm}\resizebox{18.7cm}{!}{\includegraphics{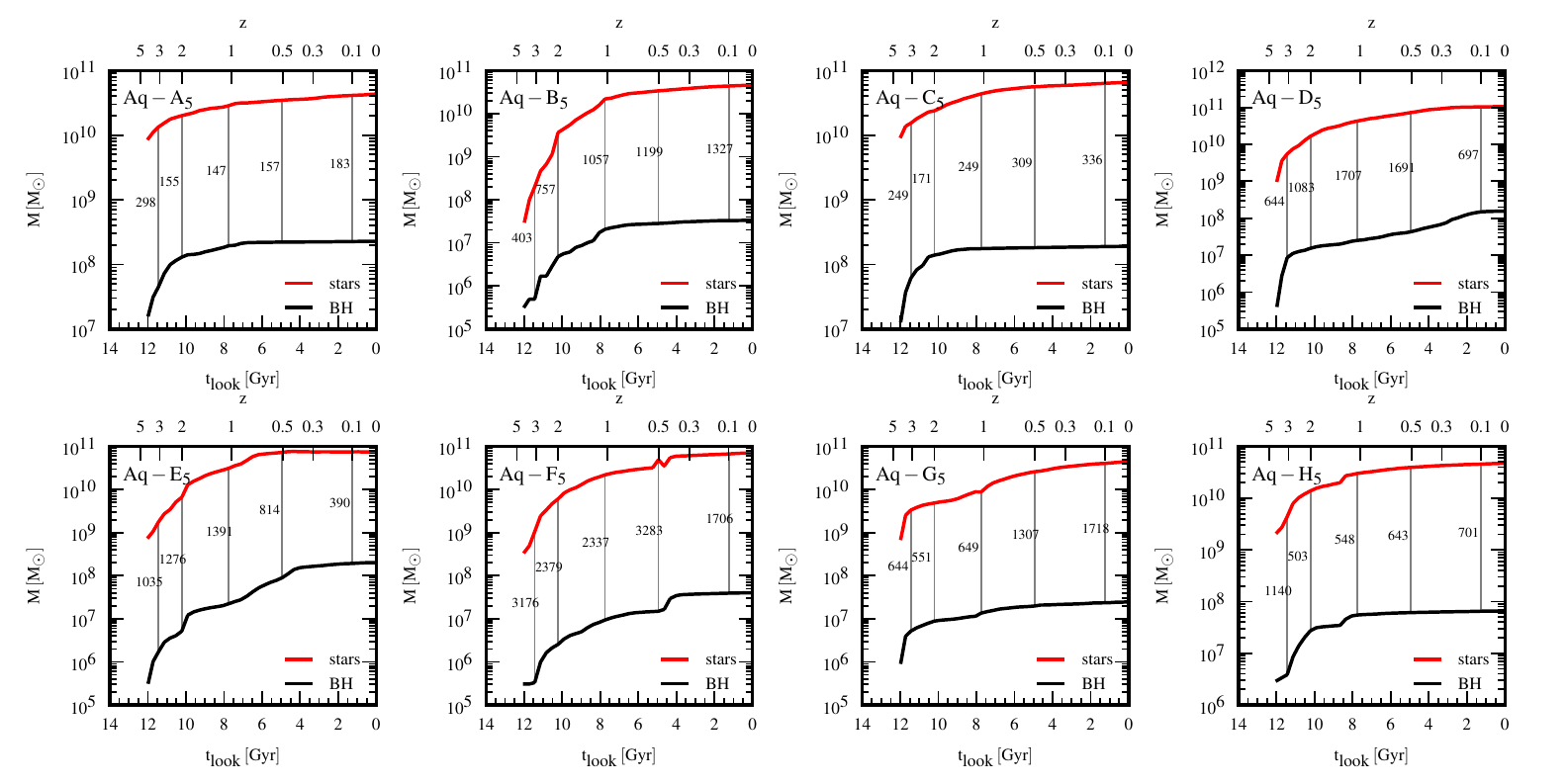}}
\caption{Evolution of the galaxy's stellar (red lines) and central
  BH (black lines) masses, as a function of look-back time and
  redshift, for the eight Aquarius haloes. The stellar masses are
  computed as the sum of the mass of each star particle within $10\%$
  of the halo virial radius at a given time, while for the BH
  we plot for any selected time the most massive BH present in
  the halo's most massive progenitor.  The evolutionary trends show
  that both the stellar content and the BH mass of the
  simulated galaxies steadily grow with time, although the rate at
  which the mass is acquired usually declines at late times (see also
  Fig.~\ref{fig:sfr}).  For each halo the mass assembly history of the
  two components shows similar features, supporting the notion of a
  (tight) link between them. The vertical lines and the associated
  numbers give the mass ratio of stars to central BH at a
  number of selected redshifts.}
\label{fig:stellarbhmass}
\end{figure*}

In Fig.~\ref{fig:stellarbhmass}, we compare the BH mass growth to
the stellar mass growth as a function of look-back time and
redshift. The stellar mass growth curves are obtained by summing the
masses of all the stellar particles within $10\%$ of the halo virial
radius at any given look-back time, while the BH masses are those of
the most massive BH in the halo progenitor again at any given
time. For all the haloes the two curves exhibit similar features. Both
components show a steady growth over cosmic time, starting with a
rapid growth rate at high redshift that subsequently flattens at low
redshift ($z \lsim 1$). These similarities in the general behaviour
are indicative of a co-evolution between the stellar content of a
galaxy and its central BH, consistent with the observed relation
between the masses of the two components \citep[e.g.][]{Magorrian1998,
  Haering2004}.

This is not entirely unexpected since the growth of both components
relies on the supply of the same fuel (i.e.~gas), but it is not
immediately obvious why two processes, such as star formation and gas
accretion on to a BH, occurring on vastly different physical scales
(the star-forming disc and the circumnuclear regions of a galaxy)
should be so tightly related \citep[for a recent review
  see][]{Kormendy2013}. In fact, different conjectures have been made
about the cause for the observed apparent co-evolution and tight
galaxy -- BH relations, ranging from feedback-regulated BH growth to
statistical averaging as a result of successive mergers. We recall
that our simulations are based on a local model for feedback-regulated
growth of BHs through gas accretion. This produces a quite close
co-evolution of the BH mass and the stellar mass in galaxies, as seen
in Fig.~\ref{fig:stellarbhmass}.

A closer inspection of the figure reveals however a difference in the
slope of the evolutionary curves, especially at late times, where the
BH mass in the majority of cases grows more slowly than the stellar
mass, as can be inferred from the mass ratios between stellar mass and
BH mass indicated at selected redshifts in the individual panels. This
implies that our simulations typically predict a higher BH
mass to stellar mass ratio at high redshift than seen in the local
Universe. There is tentative observational evidence that this may
indeed be the case; unusually large BH masses have for example been
discovered in galaxies in the local Universe that are structurally
similar to galaxies at much earlier times \citep{vandenBosch2012}.
Another feature that can be seen in the evolution of the BH masses is
the occurrence of merger events. They appear as sudden increases
(almost in a step-like fashion) in the mass assembly history of the
BHs, caused by brief phases of exponential Eddington-limited BH
growth. Similar features are present, but less visible, in the stellar
mass evolution curves as well.

All of the BHs in our galaxies end up with relatively high masses
around $10^8~\mo$, which are however still consistent with the BH mass-stellar 
mass relation of \cite{Haering2004}. The fact that we get
reasonable BH masses together with realistic disc properties in
self-consistent cosmological simulations represents an important
achievement of our galaxy formation model. This is especially
noteworthy as other BH feedback schemes are known to negatively impact
galaxy morphologies. For example, in \citet{Haas2012} the gaseous
discs in $z=2$ galaxies were lost once AGN feedback was
activated. Similarly, in the Aquila comparison project
\citep{Scannapieco2012}, the nice (but overly massive) disc formed by
\ramses\ was lost once BH feedback was activated.

\section{Halo mass structure}  \label{SecBaryonImpact}

\begin{figure*}
\hspace*{-0.3cm}\resizebox{18.6cm}{!}{\includegraphics{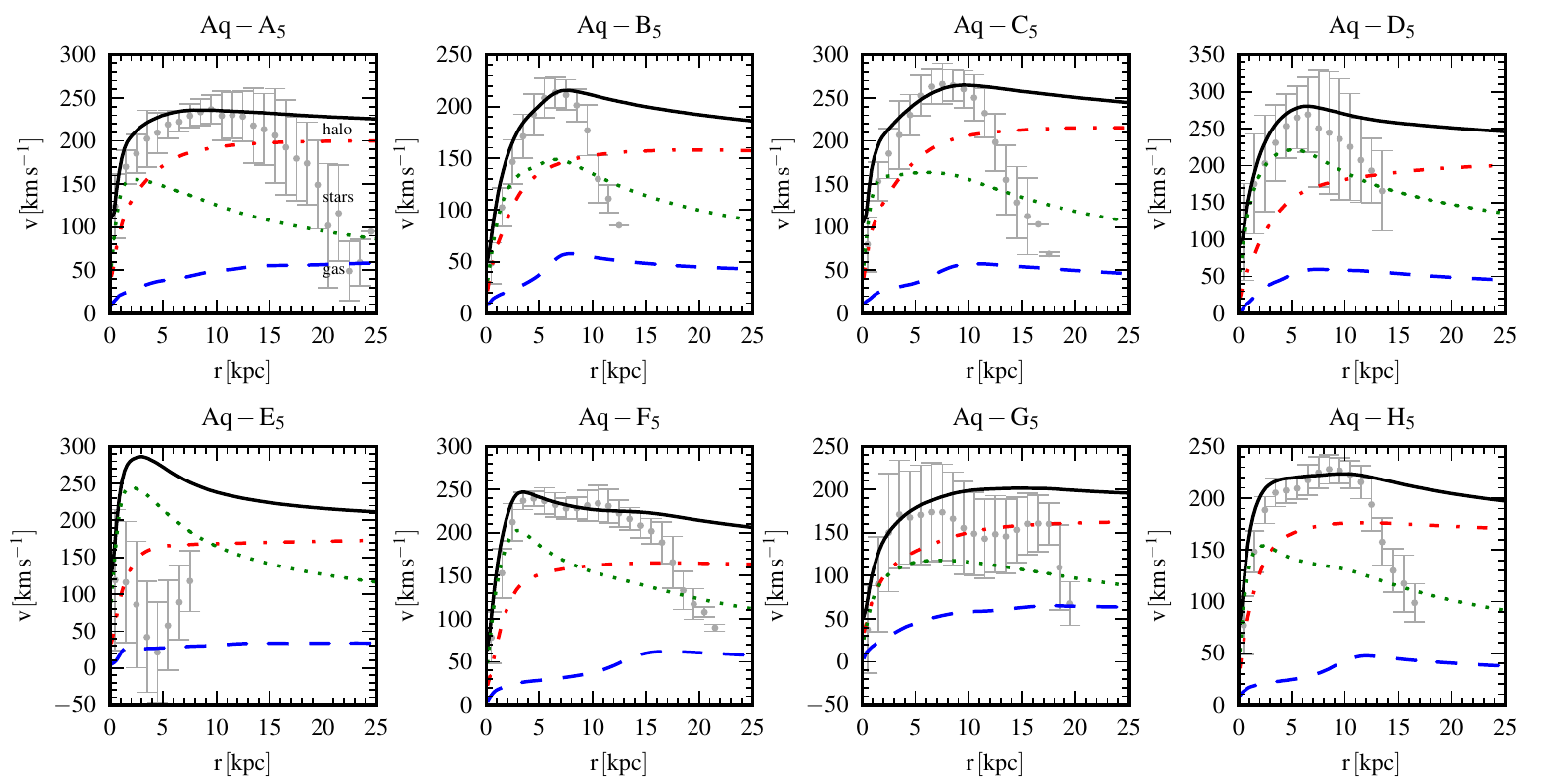}}
\caption{Rotation curves ($v_{\rm c}(r) = \sqrt{GM(<r)/r}$) for the
  eight Aquarius haloes at $z = 0$. Different line types give the
  contributions of various mass components to the total circular
  velocity: stars (dotted lines), dark matter (dot-dashed lines)
  and gas (dashed lines).  The total rotation curves are given by
  solid lines, \FM{while points with error bars show the rotation velocity
  of the star-forming gas within $0.1\times R_{\rm vir}$.}  
  It is apparent that in most of the cases quite flat
  rotation curves are present. Some haloes (Aq-D, Aq-E and to a lesser
  degree Aq-F) however show that a spheroidal stellar structure (i.e.,
  a bulge) still provides the dominant contribution to the circular
  velocity in the centre. \FM{The position of the sudden drop in the star-forming gas rotation velocity
  that can be observed in some of the haloes is set by the extension of this component.} 
  }
\label{fig:circvel}
\end{figure*}

\begin{figure*}
\hspace*{-0.3cm}\resizebox{18.6cm}{!}{\includegraphics{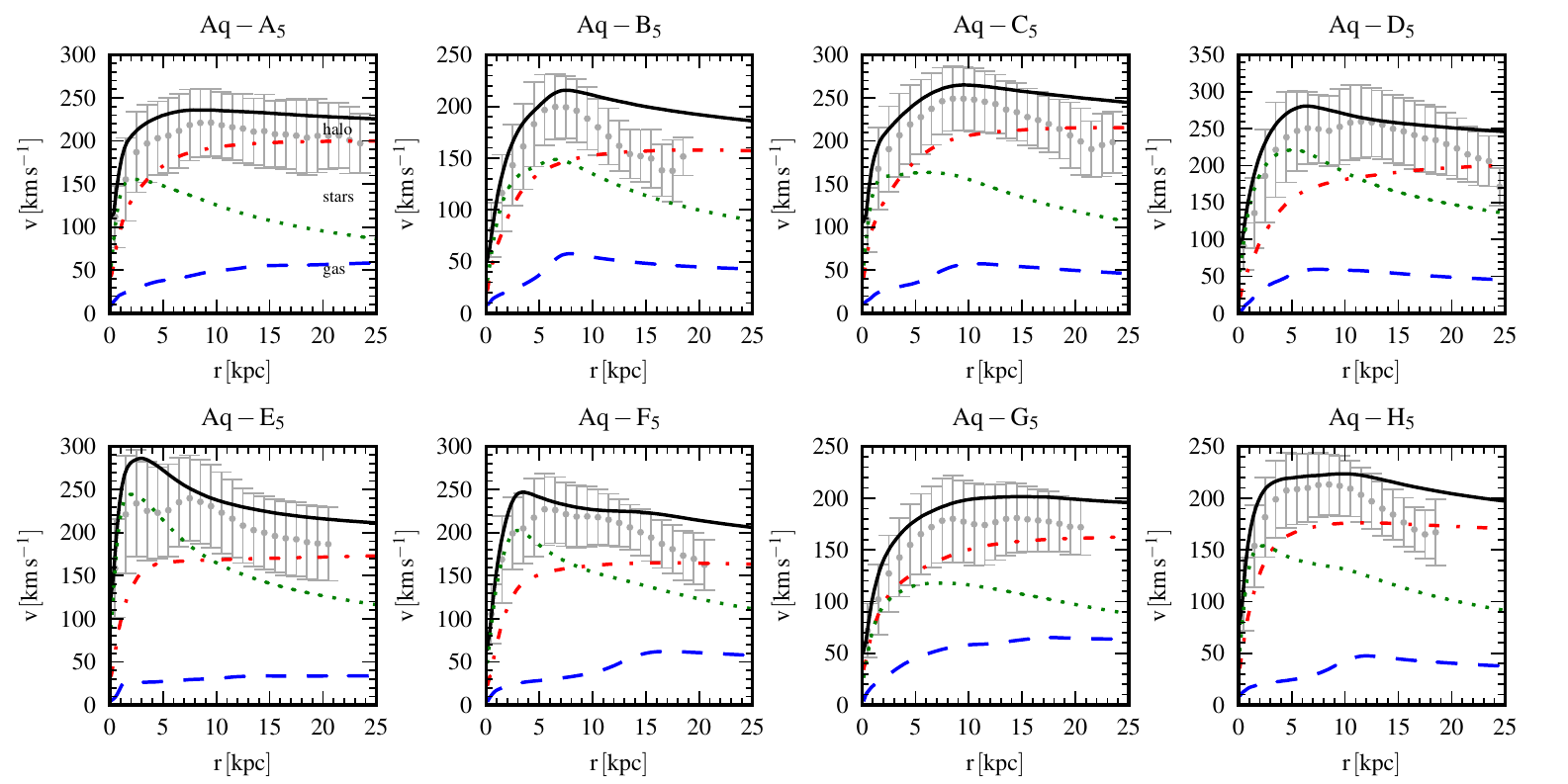}}
\caption{\FM{The same as in Fig.~\ref{fig:circvel} but points with
error bars represent the rotation velocity of disc stars selected by
considering only stars within $0.1\times R_{\rm vir}$ and circularity 
parameter $\epsilon > 0.7$. The figure shows that disc stars are 
indeed rotationally supported.}}
\label{fig:circvelstars}
\end{figure*}

In this section, we discuss the overall mass structure of our
galaxies, as reflected for example in their rotation curves. This is
interesting for at least two reasons. On the one hand, the inner shape
of the rotation curves is arguably one of the two primary areas where
significant `small-scale tensions' between $\Lambda$CDM and
observational data may be present (the other contentious area is the
abundance, central structure and spatial distribution of satellites,
a topic beyond the scope of this paper). On the other hand,
baryonic effects have recently been claimed to substantially affect
the central dark matter structure, even in large spiral galaxies
\citep{Maccio2012}. The claimed effect of strong, repetitive outflows
originating in supernova feedback is that of introducing a dark matter
core, thereby overcoming the adiabatic contraction of the halo that is
usually expected as result of baryons cooling out in the centre. Given
that our galaxies experience strong outflows, including repeated
`explosive' ones from strong quasar feedback, it is interesting to
check to what extent we can confirm this finding in our simulations.

\subsection{Rotation curves}\label{sec:rotation}

The detailed shape of the rotation curve of a galaxy encodes key
information about the mass distribution within the system. For
instance, a pronounced peak of the rotation velocity in the innermost
regions followed by a rapid decline is indicative of the presence of a
massive and compact structure \citep[often associated with a large,
  dominant stellar bulge in earlier simulation work of galaxy
  formation, see also][]{Scannapieco2012}, whilst late-type spirals
are characterized by an almost flat profile of the rotational velocity
which in the outer parts requires dark matter if the ordinary laws of
gravity hold.

In Fig.~\ref{fig:circvel}, we present the rotation curves of our
simulated galaxies. Contributions to the total rotation velocity
(solid lines) have also been separately computed via $v_{c}(r) =
\sqrt{GM(<r) / r)} $ for the primary mass components that constitute
each galaxy: stars (dotted lines), gas (dashed lines) and dark matter
(dot-dashed lines). \FM{Points with error bars show instead the
  (mass-weighted) rotation velocity of cold star-forming gas within $0.1\times R_{\rm vir}$
  around the galaxy symmetry axis.}  With the exception of Aq-E, all
galaxies have approximately flat rotation curves that show a rapid
rise in the centre followed by a slowly declining trend after the
maximum velocity has been reached. Baryons (in the form of stars) tend
to dominate only in the innermost couple of kpc.  However, for systems
Aq-D, Aq-E and Aq-F this behaviour extends out to about 10 kpc and is
responsible for the appearance of a more pronounced peak in the
rotation profiles. These galaxies have \FM{pronounced} central bulges
(and also the largest stellar masses), with Aq-E leading the set in
this respect. Aq-G has the least massive stellar distribution in the
inner parts and is dark matter dominated everywhere except for the
innermost kpc.

\begin{figure*}
\hspace*{-0.2cm}\resizebox{18.5cm}{!}{\includegraphics{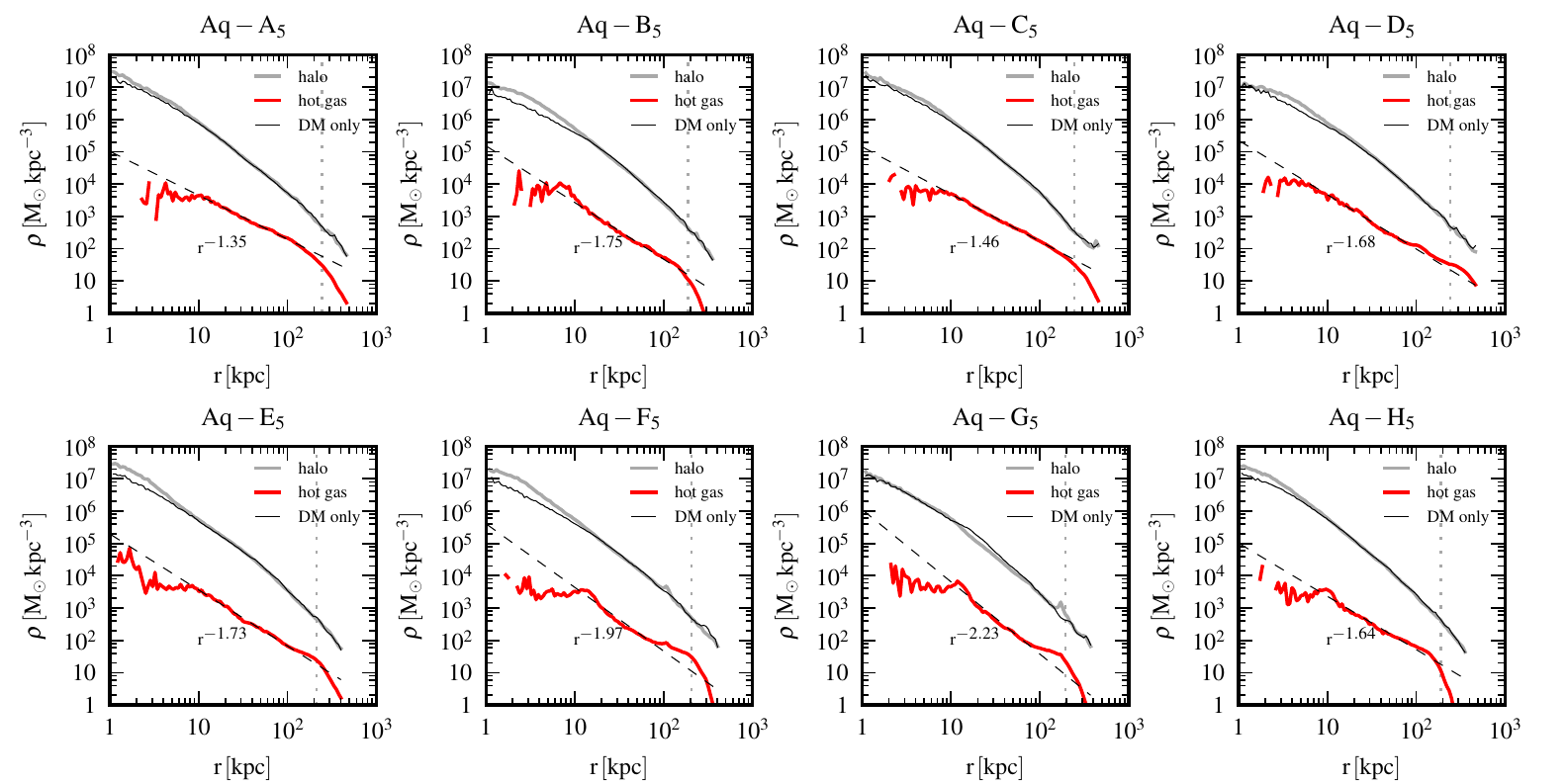}}
\caption{Spherically averaged dark matter (grey lines) and hot gas ($T
  > 3.0\times 10^5~\K$, red lines) density profiles of the eight
  simulated haloes. The density profiles obtained in the pure dark
  matter simulations (thin black lines) of the same haloes, rescaled
  by the factor $1 - \Omega_{\rm b} / \Omega_{\rm m}$, are also shown
  for comparison. The dotted vertical lines indicate the position of
  the virial radius of each halo. The contraction of the dark matter
  haloes due to the cooling and the infall of baryons in the central
  regions can be clearly detected in the majority of the simulated
  objects, whereas there is no sign of dark matter core formation in
  any of the galaxies. In the radial range between $\sim 10\,{\rm
    kpc}$ and the virial radius, the hot gas follows to good
  approximation a power-law profile (dashed lines), $\rho_{\rm hot}(r)
  \propto r^{\alpha}$, with $\alpha$ varying between $-1.35$ and $-2.2$
  in the different systems.}
\label{fig:darkstructure}
\end{figure*}

For what concerns the gas component, its contribution to the
\FM{circular} velocity is quite subdominant but not completely
negligible either, since velocities of up to $\approx 50~\kms$ are
reached. Again, the exception is represented by the gas-poor system
Aq-E, where only a maximum velocity of $\approx 35~\kms$ is
attained. It is also interesting to note that the shapes of the gas
rotation curves are rather different with respect to those of the
other components, especially in systems Aq-B, Aq-C, Aq-F and Aq-H.
This likely reflects a depletion of (cold) gas in the central regions
due to the efficient star formation and associated feedback processes
operating there. \FM{The actual kinematics of the gas closely follows 
the computed rotation velocity profiles, showing that the star-forming 
gas phase settles into a rotationally supported gaseous disc. In some 
of the haloes a sudden drop in the gas rotation velocity can be observed, and
the radius at which this break occurs indicates the radial extension of
the cold gas component. 

Fig.~\ref{fig:circvelstars} is complementary to
Fig.~\ref{fig:circvel} in the sense that it presents again the
rotation curves of our simulated galaxies but this time the points
with error bars show the rotation velocity of disc stars, that is
stars with $r < 0.1\times R_{\rm vir}$ and circularity parameter $\epsilon > 0.7$. 
Again, it is readily apparent that disc stars chosen according to this kinematic
criterion comprise a rotationally supported structure.}

\subsection{Baryonic physics impact on dark matter}  

The modification of the dark matter structure of a Milky Way-sized
halo due to baryonic effects is an extremely interesting, and so far
unsettled question.  While dark matter only simulations of Milky
Way-sized haloes can be viewed as quite reliable these days, with
several groups reaching consistent results independently and with
different methods \citep[e.g.][]{Springel2008, Stadel2009}, the
situation with respect to baryons is far less clear.
\citet{Navarro1996} were the first to demonstrate with $N$-body
simulations that a sudden loss of a large fraction of the baryonic
component from a dwarf galaxy halo could imprint a (small) dark matter
core.  \citet{Gnedin2002} however showed that even for assumptions of
maximum feedback the central halo cusp could not be destroyed by
outflows, with the inner density lowered only by a moderate factor. A
similar conclusion was reached by \citet{Ogiya2011}.
\citet{Machenko2006} argued that a larger impact is possible if there
are strong random bulk motions of baryons in dark matter potentials in
the early Universe, an idea that has recently gained
popularity. Several studies have argued that baryonic effects can
induce through this process dark matter cores in dwarf galaxies
\citep{Governato2010, Governato2012, Pontzen2012}, or even in large
Milky Way size haloes or galaxy clusters \citep{Maccio2012,
  Martizzi2012}. However, the recent work by
\citet{GarrisonKimmel2013} questions these results.  There is simply
not enough energy available in supernovae to achieve the reported dark
matter flattenings. Also, cyclic blow outs are not necessarily more
effective than a single large burst.

\begin{figure*}
\begin{center}
\resizebox{14cm}{!}{\includegraphics{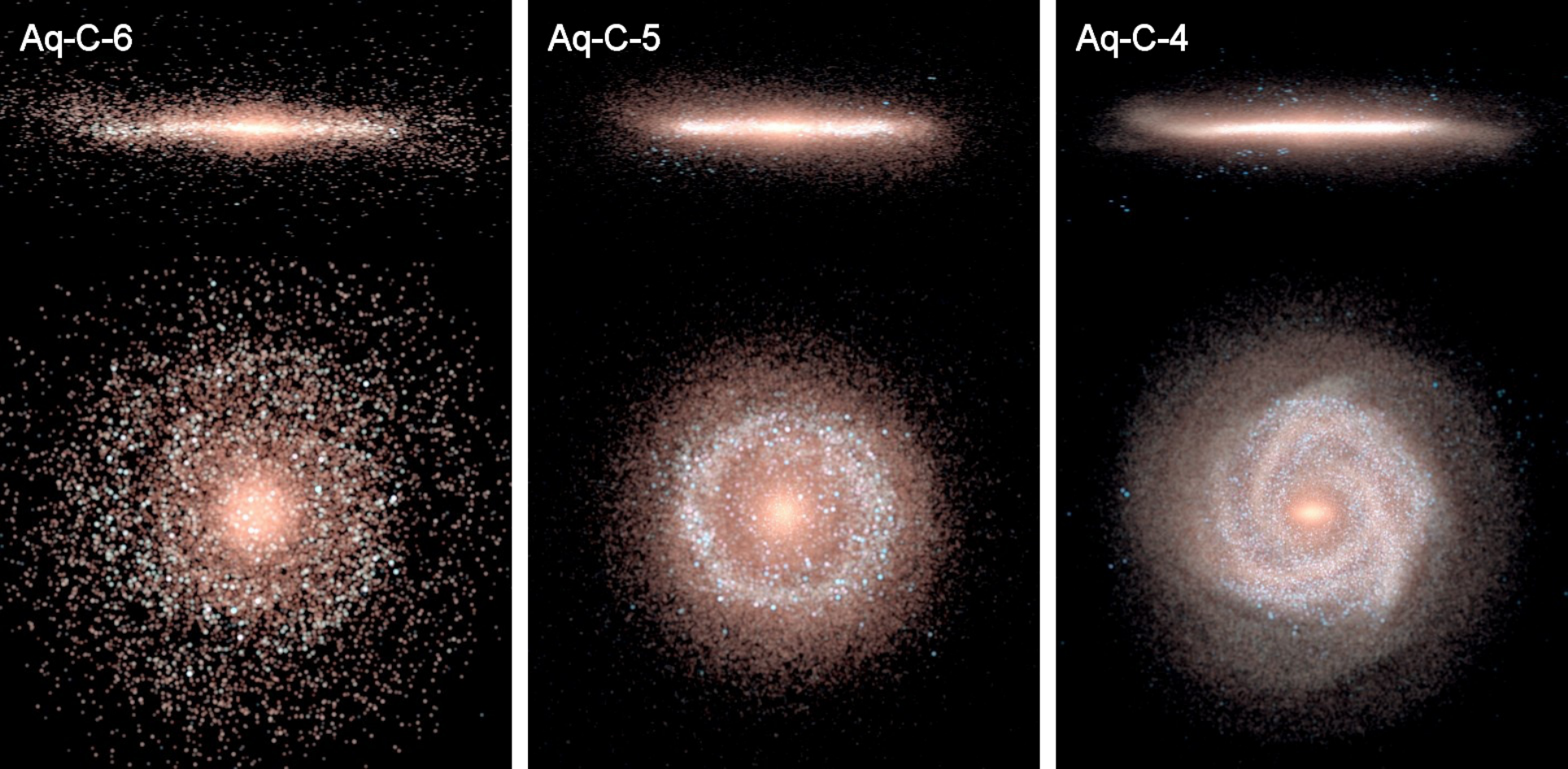}}
\end{center}
\caption{Stellar disc morphologies for the simulations Aq-C-6, Aq-C-5
  and Aq-C-4 (from left to right, respectively). As in
  Fig.~\ref{fig:stellardisks}, the three panels feature an edge-on
  (top) and a face-on (bottom) view of the projected stellar density
  to allow a visual comparison of the disc structure and its
  composition in terms of stellar age as a function of the
  resolution.}
\label{fig:resolutionmorphology}
\end{figure*}

We provide here a simple first analysis of this important issue, which
has also significant bearings on dark matter indirect detection in the
Galaxy \citep[e.g.][]{Yang2013}. In Fig.~\ref{fig:darkstructure}, we
show spherically averaged dark matter density profiles for our eight
different haloes (grey solid lines). We compare them to corresponding
runs carried out with the same initial conditions but using dark
matter alone (black thin lines). In these comparison simulations the
baryons behave effectively as if they were dark matter as well. To
make the comparison of the dark matter profiles more direct, we
rescale the measured dark matter density of the latter runs by the
factor $1 - \Omega_{\rm b} / \Omega_{\rm m}$, where $\Omega_{\rm b}$
and $\Omega_{\rm m}$ are the density parameters of baryons and dark
matter in our hydrodynamical models, respectively. In this way, any
difference in the recovered dark matter density profiles in the two
sets of runs can be ascribed to the influence of baryonic physics
alone. We see that in all cases the central dark matter density of our
simulations that include baryons is {\em increased}, showing the
expected effect for a mild adiabatic contraction. There is no trace of
core formation in our results, despite the fact that we have
substantial early feedback from winds and repeated quasar-driven
outflows. These are apparently not sufficiently strong to trigger the
alleged process of core formation, a finding consistent with the
recent analysis of idealized test simulation by
\citet{GarrisonKimmel2013}.

In Fig.~\ref{fig:darkstructure}, we also include measurements of the
spherically averaged density profiles of the hot ($T > 3\times
10^5~\K$) gas component in our hydrodynamical simulations. This gas
component forms a pressure-supported atmosphere, known as corona,
which is nearly in hydrostatic equilibrium with the dark matter
potential. The hot corona is rather extended, reaching distances up
the halo virial radius (the vertical dotted lines in the figure) and
beyond. The slope of its density profile is in general shallower than
that of the dark matter, with a tendency of forming an approximately
constant density state in the central regions ($r \lsim 10~\kpc$)
where the galaxy is located.  Compared to the Eris simulation
\citep{Guedes2011}, our hot gas profiles have considerably steeper
profiles; none is as strongly flattened as Eris, which shows a
power-law profile with slope $-1.13$ over a similar radial range as
our simulations.

\section{Resolution study}  \label{SecResolution}

It is interesting to examine how robust our results are to changes in
numerical resolution when {\em the same halo} is simulated at
drastically different mass resolution. Experience shows that the high
non-linearity of the feedback loops needed to tame star formation in
Milky Way-sized haloes make the numerical results often less robust
than is desirable when the resolution is changed. The Aquarius systems
are ideal for examining this issue as the high quality of the initial
conditions has been validated to high accuracy with dark matter only
simulations.

\begin{figure*}
\begin{center}
\resizebox{18cm}{!}{\includegraphics{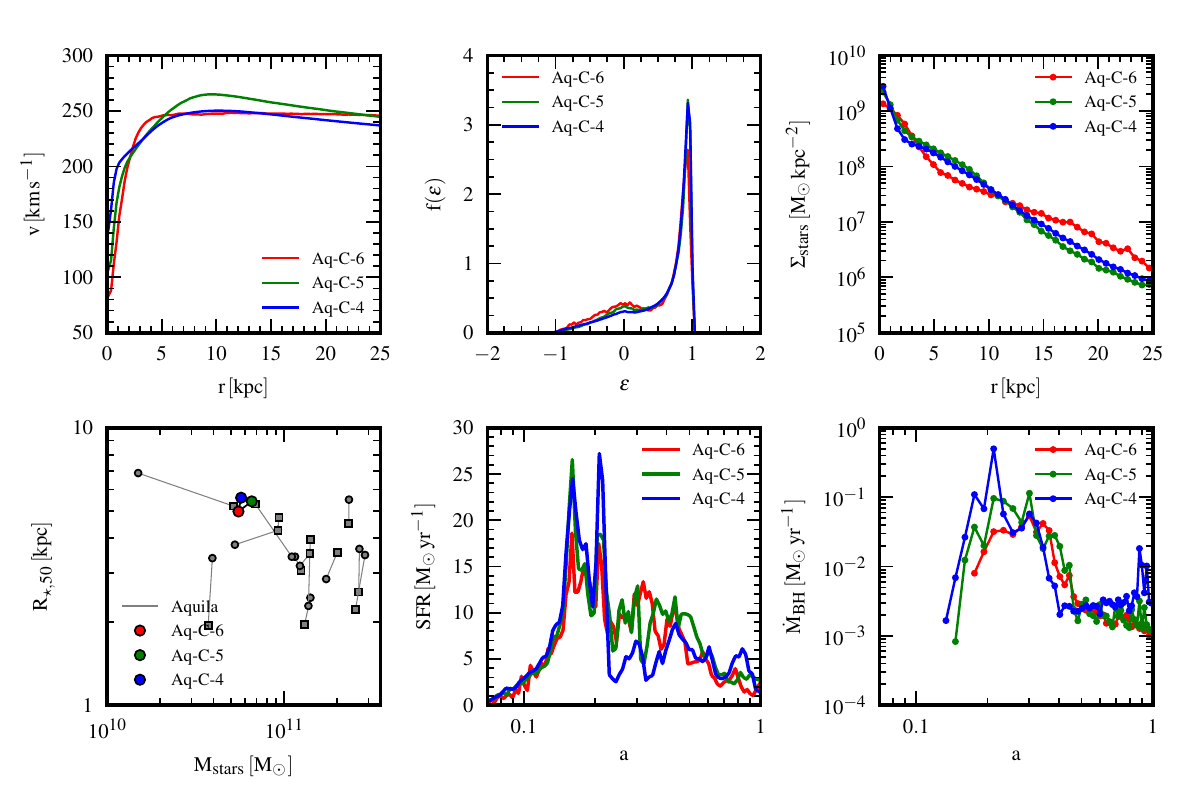}}
\end{center}
\caption{Comparison of the results obtained for the Aq-C halo at three
  different levels of numerical resolution, spanning a dynamic range
  of 64 in mass: Aq-C-6 (red), Aq-C-5 (green) and Aq-C-4 (blue). The
  six panels show: the circular velocity profiles (top left), the
  circularity distributions of stellar orbits (top middle), the
  stellar surface density profiles (top right), the relation between
  stellar half-mass radius and stellar mass (bottom left), the SFR 
  history (bottom middle) and finally, the growth rate
  of the central supermassive BH (bottom right). In the
  comparison of $R_{\star,50}$ with $M_{\rm stars}$, we include the
  simulation results of the Aquila comparison project for the same
  Aq-C halo, and simulations carried out at different numerical
  resolution (indicated by squares for level 5 and circles for level 6) 
  but with the same code, and physical models are connected by thin lines.}
\label{fig:resolutionstudy}
\end{figure*}

To allow a straightforward comparison with the Aquila code comparison
project we focus on the Aq-C system, which we have simulated both at
eight times higher (Aq-C-4) and eight times lower (Aq-C-6) mass
resolution than used in our default set up. In
Fig.~\ref{fig:resolutionmorphology}, we compare the stellar
morphology we obtain for these three cases. The images are constructed
in the same way as those shown in Fig.~\ref{fig:stellardisks}, and use
the same mapping of stellar luminosity to image intensity and
colour. We have opted to show the $z=0.2$ output instead of $z=0$,
because the Aq-C-4 run experiences a close tidal encounter with a
massive satellite at around this time, which subsequently affects the
ongoing star formation in Aq-C-4 and temporarily restricts it to a
narrow inner ring.  This changes the appearance of new young blue
stars significantly, reducing the good visual correspondence of the
systems Aq-C-5 and Aq-C-4 right at the particular output time of
$z=0$, which is a bit misleading given the overall good agreement of
the runs.  We note that also in the Aquila comparison project, a late
time merger in the Aq-C system complicates the comparison of the
different simulations as the exact timing and orbit of this incoming
substructure depends sensitively on simulation details.

More quantitatively, in Fig.~\ref{fig:resolutionstudy} we compare
several different basic quantities and examine how they compare with
each other in our Aq-C-4, Aq-C-5 and Aq-C-6 simulations. We consider
the rotation curves, eccentricity distributions, surface mass density
profiles, the relation between stellar half-mass radius and stellar
mass, SFR history and BH growth history. We
find that the eccentricity distribution converges particularly well,
which is reassuring. This suggests that the overall morphology is
quite robustly predicted by our simulation methodology, even at the
comparatively low resolution of Aq-C-6. Also, the stellar surface mass
density profiles converge quite well, especially between the higher
resolution simulations Aq-C-5 and Aq-C-4. The low-resolution run shows
still a quite similar disc profile, which is however slightly
flatter. Similarly, the rotation curves exhibit very similar shape
with only small deviations around the bulge and peak regions. Finally,
the growth histories of the stars and the galaxy's supermassive 
BH show good overall agreement, with well-aligned patterns for the
most important star bursts and merger events. This demonstrates that
our simulations reliably and reproducibly track the same formation
history of the Aq-C galaxy despite a variation of the mass resolution
over a factor of 64, and despite the presence of very strong negative
(winds and BH growth) and positive (enhanced cooling due to
metal enrichment) feedback.

The lower-left panel in Fig.~\ref{fig:resolutionstudy} shows our three
different resolution simulations as symbols in the half-mass --
stellar mass plane. The three runs of the Aq-C object are connected by
thin lines, forming a small triangle. For comparison, we also include
in the figure the results of all the Aq-C simulations carried out in
the Aquila comparison project, which employed a large set of other
simulation codes and feedback models. In Aquila, the two lower
resolutions corresponding to Aq-C-6 and Aq-C-5 were considered, and we
include these results as symbols (circles for level 6 and squares for
level 5), connecting each pair of runs carried out with the same code
by a thin line. Note that the sides of the small triangle we obtained
for our \arepo\ runs are shorter than any of the connecting lines for
other codes, highlighting the good convergence properties of our
simulation methodology relative to other implementations. Especially,
the galaxy sizes appear to be more robust in our code than in other
implementations of the galaxy formation physics. This good convergence
for galaxy sizes can be interpreted as circumstantial evidence that
angular momentum conservation, which is not manifest in mesh codes, is
sufficiently accurate in our simulations. Interestingly, two
simulations from the Aquila project reached almost the same
combination of stellar mass and galaxy size as found in our
simulations. These are the \gadget 3 run with BHs (GR-BH) and
the \ramses\ run with AGN-feedback (R-AGN), both at resolution
level 5. However, these simulations formed a dominant spheroid with at
best a feeble disc component.  The majority of the Aquila galaxies was
however simply too massive, by up to a factor of $\sim 5$, and also
too compact given their stellar mass.

\section{Discussion}  \label{SecDiscussion}

\subsection{Comparison with previous studies}

Previous studies, in particular the Aquila comparison project
\citep{Scannapieco2012}, have shown that the formation of disc
galaxies hinges strongly on the relative level of star formation at
early and late times.  In fact, there is a clear correlation in the
sense that the more successful models are those which manage to
suppress high-redshift star formation while still allowing efficient
star formation at late times. The difference this makes is
particularly evident in a comparison of the results of
\citet{Scannapieco2009} and \citet{Aumer2013b} which partially overlap
in the set of objects studied and use largely the same numerical
techniques, except that \citet{Aumer2013b} invoke an additional strong
feedback channel in order to account for `early stellar feedback'. The
difference in the results they obtain is striking. While in
\citet{Scannapieco2009} none of the galaxies contain more than 20 per
cent of the total stellar mass in their discs, \citet{Aumer2013b}
achieve the highest D/B ratios reported in the literature
thus far, with values for D/T as large as $\sim 0.6$. This is
comparable to what we reach here.

The disc fractions reported by \citet{Guedes2011} and
\citet{Agertz2011} are similarly high but are based on photometric
decompositions, which is known to produce inflated disc fractions
compared to kinematic decompositions \citep[e.g.][]{Scannapieco2011}.
We note that the Eris simulation did not include high temperature
metal cooling, an effect that would have significantly boosted
cooling, given that outflows are strong in Eris.  The galaxy studied
by \citet{Agertz2011} has accounted fully for metal-line cooling (as
we do) and promptly obtained rotation curves peaks that are too high,
indicating still too efficient cooling and a stellar content
inconsistent with abundance matching results. A similar, albeit more
severe problem is seen in the recent simulations reported by
\citet{Few2012} based on the \ramses\ code, which feature quite well-defined 
discs but also rotation curves rising to the centre as a
result of overly luminous bulges.

It is worth stressing that our simulations study the same initial
conditions as in \citet{Scannapieco2009} and (in part) in
\citet{Aumer2013b}, but use a fundamentally different numerical method
and a radically different treatment of feedback processes.  In
particular, we do not invoke early stellar feedback, rather we resort
to supernova-driven outflows, modelled as energy-driven winds whose
velocity is tied to the characteristic velocity of haloes. The
efficiency of this process is assumed to be close to the upper
envelope of what is energetically plausible. We do not implement the
feedback through a `delayed cooling' approach that has become very
popular in recent times in cosmological simulations of galaxy
formation \citep{Stinson2006, Guedes2011}. As has been pointed by
\citet{Agertz2012}, the present implementations of these schemes
appear to considerably exaggerate the impact of the Type II supernova energy input
they are meant to mimic. It also needs to be seen whether the
optimistic assumptions made about the efficiency of radiative pressure
feedback from early stars are plausible. Serious doubts about this
seem justified \citep{Krumholz2013}.

We note that we do not use the high star formation threshold advocated
by \citet{Governato2010} and \citet{Guedes2011}. Hence, this is not a
unique requirement to form realistic disc galaxies, even though it may
still be essential within a certain numerical framework for treating
hydrodynamics, star formation and feedback. Our feedback model follows
the philosophy of an explicit subgrid model that does not require
model adjustments when the numerical resolution is changed. While
physical processes that are subgrid (such as the formation of
individual stars or the launching of the galactic wind) remain
unresolved when the resolution is increased, this approach allows
numerically and physically well-posed simulation schemes with results
that do not strongly depend on numerical resolution. This is highly
desirable to obtain meaningful results in cosmological structure
formation, where invariably galaxies of widely different mass form
concurrently.

Another interesting point to observe is that the haloes of most of our
galaxies are considerably more massive than those of the best
successful disc galaxies reported in the literature thus far.  When
accounting for the different virial mass definition adopted by Eris
(which gives $\sim 15\%$ higher values than $M_{200}$), in fact {\em
  all} of our haloes are more massive than Eris and the g1536 halo of
\citet{Stinson2013a}. It becomes more difficult to control excessive
star formation in such larger haloes, and we consider it to be a
success of our model that this is possible in haloes above
$10^{12}\,{\rm M}_\odot$.  Only \citet{Aumer2013b} have reported good
discs in such more massive haloes, but their simulations tend to show
too high SFRs at low redshift in the corresponding
galaxies, a problem that our simulations do not have, in part due to the
inclusion of AGN feedback.  Also their high mass galaxies at $z=0$
appear to be too small, and the trend between stellar mass and size is
too flat.

\subsection{Conclusions}

Our simulations are probably the most successful hydrodynamical
simulations of the formation of Milky Way-like galaxies within a full
cosmological context published thus far. The models Aq-C and Aq-F are
the best Milky Way look alikes among our set. They feature galaxies
with high D/T ratio that form in sufficiently massive dark matter
haloes to qualify as Milky Way systems.  At the same time, these
galaxies have a stellar mass, a current SFR, a
rotation curve, a size and a present-day gas fraction in reasonable
agreement with properties of the Milky Way.  
Our models are also the first successful disc formation runs that simultaneously 
grow a supermassive BH of reasonable size even though back-reaction
on the forming galaxies through strong AGN feedback is included.  The
final BH masses we obtain tend to be significantly higher than
the mass of Sagittarius A$^\star$ in the centre of the Milky Way. 
\FM{This disagreement may not be very profound since it is well known that
Sagittarius A$^\star$ is undermassive with respect to what is expected from 
the local BH mass scaling relations \citep{Haering2004}, which our BH model 
is designed to match. The discrepancy} could be rectified by assuming a 
higher coupling efficiency for the BH feedback. The self-regulated nature of 
the BH growth would then still inject the same energy -- and leave the galaxy 
properties to first order unchanged -- but achieve this with a smaller BH mass
growth. However, the good match to the BH mass scaling relations
may then be lost.

As shown by \citet{Vogelsberger2013} and \citet{Torrey2013}, the
simulation methodology used here also produces very promising results
for the galaxy population as a whole.  This suggests that
hydrodynamical studies of galaxy formation in large volumes, and at
extremely high resolution in individual haloes, are finally turning
into serious competition for semi-analytic models of galaxy formation
in terms of their ability to predict galaxies in reasonable agreement
with observational data. The higher physical fidelity of hydrodynamic
simulations, especially when it comes to the dynamics of the diffuse
gas phase, make them ultimately the superior tool for exploring and
understanding galaxy formation physics.  We note that our methods also
represent a significant advance in computational efficiency compared
to previous generations of hydrodynamical galaxy formation
simulations, despite the fact that we employ a complicated
unstructured moving-mesh code. Our simulation of Aq-C-4 has consumed
less than $10^5$ CPU hours on a cluster with Intel Sandy Bridge 2.7
GHz CPUs (2.7 GHz) and 512 MPI tasks, including all group finding and
\subfind\ post-processing. We also recall the good numerical
convergence of our approach, which to our knowledge is presently
unmatched by other simulation methodologies in the field.

The near-term future of cosmological hydrodynamic simulations appears
truly exciting. There are now several groups and codes that have made
great strides towards forming realistic galaxies, both in cosmological
simulations like the ones discussed here, and in simulations of
isolated galaxies \citep[e.g.][]{Hopkins2012a}. This will be of
enormous help for advancing our theoretical understanding of galaxy
formation and for properly interpreting the wealth of observational
data. However, the devil is very much in the details of feedback, and
for the time being this remains a major headache which compromises
some of the predictive power of simulations. A `first principles'
understanding of how galaxies form will ultimately require simulations
that fully account for the physics of the ISM and the cosmological
context at the same time, a truly formidable task for the future.

\section*{Acknowledgements}

\FM{We thank an anonymous referee for a constructive report. We are
grateful to} Mark Vogelsberger, Shy Genel, Lars Hernquist, Debora Sijacki,
Paul Torrey, Christoph Pfrommer, Christopher Hayward and Ewald
Puchwein for useful discussions.  FM and VS acknowledge support by
the DFG Research Centre SFB-881 `The Milky Way System' through project
A1.  This work has also been supported by the European Research
Council under ERC-StG grant EXAGAL-308037 and by the Klaus Tschira
Foundation.

\bibliographystyle{mn2e}
\bibliography{paper}

\begin{thebibliography}{}

\bibitem[\protect\citeauthoryear{{Abadi}, {Navarro}, {Steinmetz} \&
  {Eke}}{{Abadi} et~al.}{2003}]{Abadi2}
{Abadi} M.~G.,  {Navarro} J.~F.,  {Steinmetz} M.,    {Eke} V.~R.,  2003, ApJ,
  597, 21

\bibitem[\protect\citeauthoryear{{Agertz}, {Kravtsov}, {Leitner} \&
  {Gnedin}}{{Agertz} et~al.}{2013}]{Agertz2012}
{Agertz} O.,  {Kravtsov} A.~V.,  {Leitner} S.~N.,    {Gnedin} N.~Y.,  2013,
  \apj, 770, 25

\bibitem[\protect\citeauthoryear{{Agertz}, {Teyssier} \& {Moore}}{{Agertz}
  et~al.}{2011}]{Agertz2011}
{Agertz} O.,  {Teyssier} R.,    {Moore} B.,  2011, \mnras, 410, 1391

\bibitem[\protect\citeauthoryear{{Aumer}, {White}, {Naab} \&
  {Scannapieco}}{{Aumer} et~al.}{2013}]{Aumer2013b}
{Aumer} M.,  {White} S.~D.~M.,  {Naab} T.,    {Scannapieco} C.,  2013, \mnras,
  434, 3142

\bibitem[\protect\citeauthoryear{{Balogh}, {Pearce}, {Bower} \& {Kay}}{{Balogh}
  et~al.}{2001}]{Balogh2001}
{Balogh} M.~L.,  {Pearce} F.~R.,  {Bower} R.~G.,    {Kay} S.~T.,  2001, \mnras,
  326, 1228

\bibitem[\protect\citeauthoryear{{Battaglia} et~al.,}{{Battaglia}
  et~al.}{2005}]{Battaglia2005}
{Battaglia} G.,  et~al., 2005, \mnras, 364, 433

\bibitem[\protect\citeauthoryear{{Bauer} \& {Springel}}{{Bauer} \&
  {Springel}}{2012}]{Bauer2012}
{Bauer} A.,  {Springel} V.,  2012, \mnras, 423, 2558

\bibitem[\protect\citeauthoryear{{Behroozi}, {Conroy} \& {Wechsler}}{{Behroozi}
  et~al.}{2010}]{Behroozi2010}
{Behroozi} P.~S.,  {Conroy} C.,    {Wechsler} R.~H.,  2010, \apj, 717, 379

\bibitem[\protect\citeauthoryear{{Boylan-Kolchin}, {Springel}, {White},
  {Jenkins} \& {Lemson}}{{Boylan-Kolchin} et~al.}{2009}]{MilleniumII}
{Boylan-Kolchin} M.,  {Springel} V.,  {White} S.~D.~M.,  {Jenkins} A.,
  {Lemson} G.,  2009, \mnras, 398, 1150

\bibitem[\protect\citeauthoryear{{Boylan-Kolchin}, {Springel}, {White} \&
  {Jenkins}}{{Boylan-Kolchin} et~al.}{2010}]{Boylan-Kolchin2010}
{Boylan-Kolchin} M.,  {Springel} V.,  {White} S.~D.~M.,    {Jenkins} A.,  2010,
  \mnras, 406, 896

\bibitem[\protect\citeauthoryear{{Boylan-Kolchin}, {Bullock} \&
  {Kaplinghat}}{{Boylan-Kolchin} et~al.}{2011}]{Boylan-Kolchin2011}
{Boylan-Kolchin} M.,  {Bullock} J.~S.,    {Kaplinghat} M.,  2011, \mnras, 415,
  L40

\bibitem[\protect\citeauthoryear{{Boylan-Kolchin}, {Bullock} \&
  {Kaplinghat}}{{Boylan-Kolchin} et~al.}{2012}]{Boylan-Kolchin2012}
{Boylan-Kolchin} M.,  {Bullock} J.~S.,    {Kaplinghat} M.,  2012, \mnras, 422,
  1203

\bibitem[\protect\citeauthoryear{{Boylan-Kolchin}, {Bullock}, {Sohn}, {Besla}
  \& {van der Marel}}{{Boylan-Kolchin} et~al.}{2013}]{Boylan-Kolchin2012b}
{Boylan-Kolchin} M.,  {Bullock} J.~S.,  {Sohn} S.~T.,  {Besla} G.,    {van der
  Marel} R.~P.,  2013, \apj, 768, 140

\bibitem[\protect\citeauthoryear{{Brooks} et~al.,}{{Brooks}
  et~al.}{2011}]{Brooks2011}
{Brooks} A.~M.,  et~al., 2011, \apj, 728, 51

\bibitem[\protect\citeauthoryear{{Chabrier}}{{Chabrier}}{2003}]{Chabrier2003}
{Chabrier} G.,  2003, \apjl, 586, L133

\bibitem[\protect\citeauthoryear{{Courteau}, {Dutton}, {van den Bosch},
  {MacArthur}, {Dekel}, {McIntosh} \& {Dale}}{{Courteau}
  et~al.}{2007}]{Courteau2007}
{Courteau} S.,  {Dutton} A.~A.,  {van den Bosch} F.~C.,  {MacArthur} L.~A.,
  {Dekel} A.,  {McIntosh} D.~H.,    {Dale} D.~A.,  2007, \apj, 671, 203

\bibitem[\protect\citeauthoryear{{Davis}, {Efstathiou}, {Frenk} \&
  {White}}{{Davis} et~al.}{1985}]{Davis1985}
{Davis} M.,  {Efstathiou} G.,  {Frenk} C.~S.,    {White} S.~D.~M.,  1985, \apj,
  292, 371

\bibitem[\protect\citeauthoryear{{Dehnen}, {McLaughlin} \& {Sachania}}{{Dehnen}
  et~al.}{2006}]{Dehnen2006}
{Dehnen} W.,  {McLaughlin} D.~E.,    {Sachania} J.,  2006, \mnras, 369, 1688

\bibitem[\protect\citeauthoryear{{Dutton} et~al.,}{{Dutton}
  et~al.}{2011}]{Dutton2011}
{Dutton} A.~A.,  et~al., 2011, \mnras, 416, 322

\bibitem[\protect\citeauthoryear{{Fall} \& {Efstathiou}}{{Fall} \&
  {Efstathiou}}{1980}]{Fall1980}
{Fall} S.~M.,  {Efstathiou} G.,  1980, \mnras, 193, 189

\bibitem[\protect\citeauthoryear{{Faucher-Gigu{\`e}re}, {Lidz}, {Zaldarriaga}
  \& {Hernquist}}{{Faucher-Gigu{\`e}re} et~al.}{2009}]{FaucherGiguere2009}
{Faucher-Gigu{\`e}re} C.-A.,  {Lidz} A.,  {Zaldarriaga} M.,    {Hernquist} L.,
  2009, \apj, 703, 1416

\bibitem[\protect\citeauthoryear{{Few}, {Gibson}, {Courty}, {Michel-Dansac},
  {Brook} \& {Stinson}}{{Few} et~al.}{2012}]{Few2012}
{Few} C.~G.,  {Gibson} B.~K.,  {Courty} S.,  {Michel-Dansac} L.,  {Brook}
  C.~B.,    {Stinson} G.~S.,  2012, \aap, 547, A63

\bibitem[\protect\citeauthoryear{{Gallazzi}, {Charlot}, {Brinchmann}, {White}
  \& {Tremonti}}{{Gallazzi} et~al.}{2005}]{Gallazzi2005}
{Gallazzi} A.,  {Charlot} S.,  {Brinchmann} J.,  {White} S.~D.~M.,
  {Tremonti} C.~A.,  2005, \mnras, 362, 41

\bibitem[\protect\citeauthoryear{{Garrison-Kimmel}, {Rocha}, {Boylan-Kolchin},
  {Bullock} \& {Lally}}{{Garrison-Kimmel} et~al.}{2013}]{GarrisonKimmel2013}
{Garrison-Kimmel} S.,  {Rocha} M.,  {Boylan-Kolchin} M.,  {Bullock} J.~S.,
  {Lally} J.,  2013, \mnras, 433, 3539

\bibitem[\protect\citeauthoryear{{Genel}, {Vogelsberger}, {Nelson}, {Sijacki},
  {Springel} \& {Hernquist}}{{Genel} et~al.}{2013}]{Genel2013}
{Genel} S.,  {Vogelsberger} M.,  {Nelson} D.,  {Sijacki} D.,  {Springel} V.,
  {Hernquist} L.,  2013, \mnras, 435, 1426

\bibitem[\protect\citeauthoryear{{Gnedin} \& {Zhao}}{{Gnedin} \&
  {Zhao}}{2002}]{Gnedin2002}
{Gnedin} O.~Y.,  {Zhao} H.,  2002, \mnras, 333, 299

\bibitem[\protect\citeauthoryear{{Governato} et~al.,}{{Governato}
  et~al.}{2004}]{Governato2004}
{Governato} F.,  et~al., 2004, \apj, 607, 688

\bibitem[\protect\citeauthoryear{{Governato}, {Willman}, {Mayer}, {Brooks},
  {Stinson}, {Valenzuela}, {Wadsley} \& {Quinn}}{{Governato}
  et~al.}{2007}]{Governato2007}
{Governato} F.,  {Willman} B.,  {Mayer} L.,  {Brooks} A.,  {Stinson} G.,
  {Valenzuela} O.,  {Wadsley} J.,    {Quinn} T.,  2007, \mnras, 374, 1479

\bibitem[\protect\citeauthoryear{{Governato} et~al.,}{{Governato}
  et~al.}{2010}]{Governato2010}
{Governato} F.,  et~al., 2010, \nat, 463, 203

\bibitem[\protect\citeauthoryear{{Governato} et~al.,}{{Governato}
  et~al.}{2012}]{Governato2012}
{Governato} F.,  et~al., 2012, \mnras, 422, 1231


\bibitem[\protect\citeauthoryear{{Graham}}{{Graham}}{2001}]{Graham2001}
{Graham} A.~W.,  2001, \aj, 121, 820

\bibitem[\protect\citeauthoryear{{Graham}}{{Graham}}{2002}]{Graham2002}
{Graham} A.~W.,  2002, \mnras, 334, 721

\bibitem[\protect\citeauthoryear{{Graham} \& {Worley}}{{Graham} \&
  {Worley}}{2008}]{Graham2008}
{Graham} A.~W.,  {Worley} C.~C.,  2008, \mnras, 388, 1708

\bibitem[\protect\citeauthoryear{{Guedes}, {Callegari}, {Madau} \&
  {Mayer}}{{Guedes} et~al.}{2011}]{Guedes2011}
{Guedes} J.,  {Callegari} S.,  {Madau} P.,    {Mayer} L.,  2011, \apj, 742, 76

\bibitem[\protect\citeauthoryear{{Guo}, {White}, {Li} \&
  {Boylan-Kolchin}}{{Guo} et~al.}{2010}]{Guo2010}
{Guo} Q.,  {White} S.,  {Li} C.,    {Boylan-Kolchin} M.,  2010, \mnras, 404,
  1111

  \bibitem[\protect\citeauthoryear{{Guo} et~al.,}{{Guo}  et~al.}{2011}]{Guo2011}
{Guo} Q.,  et~al., 2011, \mnras, 413, 101

\bibitem[\protect\citeauthoryear{{Haas}, {Schaye}, {Booth}, {Dalla Vecchia},
  {Springel}, {Theuns} \& {Wiersma}}{{Haas} et~al.}{2013}]{Haas2012}
{Haas} M.~R.,  {Schaye} J.,  {Booth} C.~M.,  {Dalla Vecchia} C.,  {Springel}
  V.,  {Theuns} T.,    {Wiersma} R.~P.~C.,  2013, \mnras, 435, 2955

\bibitem[\protect\citeauthoryear{{H{\"a}ring} \& {Rix}}{{H{\"a}ring} \&
  {Rix}}{2004}]{Haering2004}
{H{\"a}ring} N.,  {Rix} H.-W.,  2004, \apjl, 604, L89

\bibitem[\protect\citeauthoryear{{Haynes}, {Giovanelli}, {Salzer}, {Wegner},
  {Freudling}, {da Costa}, {Herter} \& {Vogt}}{{Haynes}
  et~al.}{1999a}]{HaynesOpt1999}
{Haynes} M.~P.,  {Giovanelli} R.,  {Salzer} J.~J.,  {Wegner} G.,  {Freudling}
  W.,  {da Costa} L.~N.,  {Herter} T.,    {Vogt} N.~P.,  1999a, \aj, 117, 1668

\bibitem[\protect\citeauthoryear{{Haynes}, {Giovanelli}, {Chamaraux}, {da
  Costa}, {Freudling}, {Salzer} \& {Wegner}}{{Haynes}
  et~al.}{1999b}]{HaynesHI1999}
{Haynes} M.~P.,  {Giovanelli} R.,  {Chamaraux} P.,  {da Costa} L.~N.,
  {Freudling} W.,  {Salzer} J.~J.,    {Wegner} G.,  1999b, \aj, 117, 2039

\bibitem[\protect\citeauthoryear{{Hopkins}, {Quataert} \& {Murray}}{{Hopkins}
  et~al.}{2012}]{Hopkins2012a}
{Hopkins} P.~F.,  {Quataert} E.,    {Murray} N.,  2012, \mnras, 421, 3488

\bibitem[\protect\citeauthoryear{{Hummels} \& {Bryan}}{{Hummels} \&
  {Bryan}}{2012}]{Hummels2012}
{Hummels} C.~B.,  {Bryan} G.~L.,  2012, \apj, 749, 140

\bibitem[\protect\citeauthoryear{{Katz} \& {Gunn}}{{Katz} \&
  {Gunn}}{1991}]{Katz1991}
{Katz} N.,  {Gunn} J.~E.,  1991, \apj, 377, 365

\bibitem[\protect\citeauthoryear{{Kaufmann}, {Mayer}, {Wadsley}, {Stadel} \&
  {Moore}}{{Kaufmann} et~al.}{2006}]{Kaufmann2006}
{Kaufmann} T.,  {Mayer} L.,  {Wadsley} J.,  {Stadel} J.,    {Moore} B.,  2006,
  \mnras, 370, 1612

\bibitem[\protect\citeauthoryear{{Kaufmann}, {Mayer}, {Wadsley}, {Stadel} \&
  {Moore}}{{Kaufmann} et~al.}{2007}]{Kaufmann2007}
{Kaufmann} T.,  {Mayer} L.,  {Wadsley} J.,  {Stadel} J.,    {Moore} B.,  2007,
  \mnras, 375, 53

\bibitem[\protect\citeauthoryear{{Kere{\v s}}, {Vogelsberger}, {Sijacki},
  {Springel} \& {Hernquist}}{{Kere{\v s}} et~al.}{2012}]{Keres2012}
{Kere{\v s}} D.,  {Vogelsberger} M.,  {Sijacki} D.,  {Springel} V.,
  {Hernquist} L.,  2012, \mnras, 425, 2027

\bibitem[\protect\citeauthoryear{{Kormendy} \& {Ho}}{{Kormendy} \&
  {Ho}}{2013}]{Kormendy2013}
{Kormendy} J.,  {Ho} L.~C.,  2013, \araa, 51, 511

\bibitem[\protect\citeauthoryear{{Krumholz} \& {Thompson}}{{Krumholz} \&
  {Thompson}}{2013}]{Krumholz2013}
{Krumholz} M.~R.,  {Thompson} T.~A.,  2013, \mnras, 434, 2329

\bibitem[\protect\citeauthoryear{{Leitner} \& {Kravtsov}}{{Leitner} \&
  {Kravtsov}}{2011}]{Leitner2011}
{Leitner} S.~N.,  {Kravtsov} A.~V.,  2011, \apj, 734, 48

\bibitem[\protect\citeauthoryear{{Li} \& {White}}{{Li} \&
  {White}}{2008}]{Li2008}
{Li} Y.-S.,  {White} S.~D.~M.,  2008, \mnras, 384, 1459

\bibitem[\protect\citeauthoryear{{MacArthur}, {Courteau} \&
  {Holtzman}}{{MacArthur} et~al.}{2003}]{MacArthur2003}
{MacArthur} L.~A.,  {Courteau} S.,    {Holtzman} J.~A.,  2003, \apj, 582, 689

\bibitem[\protect\citeauthoryear{{Macci{\`o}}, {Stinson}, {Brook}, {Wadsley},
  {Couchman}, {Shen}, {Gibson} \& {Quinn}}{{Macci{\`o}}
  et~al.}{2012}]{Maccio2012}
{Macci{\`o}} A.~V.,  {Stinson} G.,  {Brook} C.~B.,  {Wadsley} J.,  {Couchman}
  H.~M.~P.,  {Shen} S.,  {Gibson} B.~K.,    {Quinn} T.,  2012, \apjl, 744, L9

\bibitem[\protect\citeauthoryear{{Magorrian} et~al.,}{{Magorrian}
  et~al.}{1998}]{Magorrian1998}
{Magorrian} J.,  et~al., 1998, \aj, 115, 2285

\bibitem[\protect\citeauthoryear{{Martig}, {Bournaud}, {Croton}, {Dekel} \&
  {Teyssier}}{{Martig} et~al.}{2012}]{Martig2012}
{Martig} M.,  {Bournaud} F.,  {Croton} D.~J.,  {Dekel} A.,    {Teyssier} R.,
  2012, \apj, 756, 26

\bibitem[\protect\citeauthoryear{{Martizzi}, {Teyssier} \& {Moore}}{{Martizzi}
  et~al.}{2013}]{Martizzi2012}
{Martizzi} D.,  {Teyssier} R.,    {Moore} B.,  2013, \mnras, 432, 1947

\bibitem[\protect\citeauthoryear{{Mashchenko}, {Couchman} \&
  {Wadsley}}{{Mashchenko} et~al.}{2006}]{Machenko2006}
{Mashchenko} S.,  {Couchman} H.~M.~P.,    {Wadsley} J.,  2006, \nat, 442, 539

\bibitem[\protect\citeauthoryear{{M{\"o}llenhoff}}{{M{\"o}llenhoff}}{2004}]{Mo%
llenhoff2004}
{M{\"o}llenhoff} C.,  2004, \aap, 415, 63

\bibitem[\protect\citeauthoryear{{Moster}, {Somerville}, {Maulbetsch}, {van den
  Bosch}, {Macci{\`o}}, {Naab} \& {Oser}}{{Moster} et~al.}{2010}]{Moster2010}
{Moster} B.~P.,  {Somerville} R.~S.,  {Maulbetsch} C.,  {van den Bosch} F.~C.,
  {Macci{\`o}} A.~V.,  {Naab} T.,    {Oser} L.,  2010, \apj, 710, 903

\bibitem[\protect\citeauthoryear{{Navarro}, {Eke} \& {Frenk}}{{Navarro}
  et~al.}{1996}]{Navarro1996}
{Navarro} J.~F.,  {Eke} V.~R.,    {Frenk} C.~S.,  1996, \mnras, 283, L72

\bibitem[\protect\citeauthoryear{{Navarro} \& {Steinmetz}}{{Navarro} \&
  {Steinmetz}}{2000}]{Navarro2000}
{Navarro} J.~F.,  {Steinmetz} M.,  2000, \apj, 538, 477

\bibitem[\protect\citeauthoryear{{Nelson}, {Vogelsberger}, {Genel}, {Sijacki},
  {Kere{\v s}}, {Springel} \& {Hernquist}}{{Nelson} et~al.}{2013}]{Nelson2013}
{Nelson} D.,  {Vogelsberger} M.,  {Genel} S.,  {Sijacki} D.,  {Kere{\v s}} D.,
  {Springel} V.,    {Hernquist} L.,  2013, \mnras, 429, 3353

\bibitem[\protect\citeauthoryear{{Ogiya} \& {Mori}}{{Ogiya} \&
  {Mori}}{2011}]{Ogiya2011}
{Ogiya} G.,  {Mori} M.,  2011, \apjl, 736, L2

\bibitem[\protect\citeauthoryear{{Okamoto}, {Eke}, {Frenk} \&
  {Jenkins}}{{Okamoto} et~al.}{2005}]{Okamoto2005}
{Okamoto} T.,  {Eke} V.~R.,  {Frenk} C.~S.,    {Jenkins} A.,  2005, \mnras,
  363, 1299

\bibitem[\protect\citeauthoryear{{Piontek} \& {Steinmetz}}{{Piontek} \&
  {Steinmetz}}{2011}]{Piontek2011}
{Piontek} F.,  {Steinmetz} M.,  2011, \mnras, 410, 2625

\bibitem[\protect\citeauthoryear{{Pizagno} et~al.,}{{Pizagno}
  et~al.}{2007}]{Pizagno2007}
{Pizagno} J.,  et~al., 2007, \aj, 134, 945

\bibitem[\protect\citeauthoryear{{Pontzen} \& {Governato}}{{Pontzen} \&
  {Governato}}{2012}]{Pontzen2012}
{Pontzen} A.,  {Governato} F.,  2012, \mnras, 421, 3464

\bibitem[\protect\citeauthoryear{{Power}, {Navarro}, {Jenkins}, {Frenk},
  {White}, {Springel}, {Stadel} \& {Quinn}}{{Power} et~al.}{2003}]{Power2003}
{Power} C.,  {Navarro} J.~F.,  {Jenkins} A.,  {Frenk} C.~S.,  {White} S.~D.~M.,
   {Springel} V.,  {Stadel} J.,    {Quinn} T.,  2003, \mnras, 338, 14

\bibitem[\protect\citeauthoryear{{Puchwein} \& {Springel}}{{Puchwein} \&
  {Springel}}{2013}]{Puchwein2013}
{Puchwein} E.,  {Springel} V.,  2013, \mnras, 428, 2966

\bibitem[\protect\citeauthoryear{{Robertson}, {Yoshida}, {Springel} \&
  {Hernquist}}{{Robertson} et~al.}{2004}]{Robertson2004}
{Robertson} B.,  {Yoshida} N.,  {Springel} V.,    {Hernquist} L.,  2004, \apj,
  606, 32

\bibitem[\protect\citeauthoryear{{Sakamoto}, {Chiba} \& {Beers}}{{Sakamoto}
  et~al.}{2003}]{Sakamoto2003}
{Sakamoto} T.,  {Chiba} M.,    {Beers} T.~C.,  2003, \aap, 397, 899

\bibitem[\protect\citeauthoryear{{Sales}, {Navarro}, {Schaye}, {Dalla Vecchia},
  {Springel} \& {Booth}}{{Sales} et~al.}{2010}]{Sales2010}
{Sales} L.~V.,  {Navarro} J.~F.,  {Schaye} J.,  {Dalla Vecchia} C.,  {Springel}
  V.,    {Booth} C.~M.,  2010, \mnras, 409, 1541

\bibitem[\protect\citeauthoryear{{Sales}, {Navarro}, {Schaye}, {Dalla Vecchia},
  {Springel}, {Haas} \& {Helmi}}{{Sales} et~al.}{2009}]{Sales2009}
{Sales} L.~V.,  {Navarro} J.~F.,  {Schaye} J.,  {Dalla Vecchia} C.,  {Springel}
  V.,  {Haas} M.~R.,    {Helmi} A.,  2009, \mnras, 399, L64

\bibitem[\protect\citeauthoryear{{Sawala}, {Guo}, {Scannapieco}, {Jenkins} \&
  {White}}{{Sawala} et~al.}{2011}]{Sawala2011}
{Sawala} T.,  {Guo} Q.,  {Scannapieco} C.,  {Jenkins} A.,    {White} S.,  2011,
  \mnras, 413, 659

\bibitem[\protect\citeauthoryear{{Scannapieco}, {Tissera}, {White} \&
  {Springel}}{{Scannapieco} et~al.}{2008}]{Scannapieco2008}
{Scannapieco} C.,  {Tissera} P.~B.,  {White} S.~D.~M.,    {Springel} V.,  2008,
  MNRAS, 389, 1137

\bibitem[\protect\citeauthoryear{{Scannapieco}, {White}, {Springel} \&
  {Tissera}}{{Scannapieco} et~al.}{2009}]{Scannapieco2009}
{Scannapieco} C.,  {White} S.~D.~M.,  {Springel} V.,    {Tissera} P.~B.,  2009,
  \mnras, 396, 696

\bibitem[\protect\citeauthoryear{{Scannapieco}, {White}, {Springel} \&
  {Tissera}}{{Scannapieco} et~al.}{2011}]{Scannapieco2011}
{Scannapieco} C.,  {White} S.~D.~M.,  {Springel} V.,    {Tissera} P.~B.,  2011,
  \mnras, 417, 154

\bibitem[\protect\citeauthoryear{{Scannapieco} et~al.,}{{Scannapieco}
  et~al.}{2012}]{Scannapieco2012}
{Scannapieco} C.,  et~al., 2012, \mnras, 423, 1726

\bibitem[\protect\citeauthoryear{{S{\'e}rsic}}{{S{\'e}rsic}}{1963}]{Sersic1963}
{S{\'e}rsic} J.~L.,  1963, Bol. Asociacion Argentina de Astron. La Plata
  Argentina, 6, 41

\bibitem[\protect\citeauthoryear{{Shen}, {Mo}, {White}, {Blanton}, {Kauffmann},
  {Voges}, {Brinkmann} \& {Csabai}}{{Shen} et~al.}{2003}]{Shen2003}
{Shen} S.,  {Mo} H.~J.,  {White} S.~D.~M.,  {Blanton} M.~R.,  {Kauffmann} G.,
  {Voges} W.,  {Brinkmann} J.,    {Csabai} I.,  2003, \mnras, 343, 978

\bibitem[\protect\citeauthoryear{{Sijacki}, {Vogelsberger}, {Kere{\v s}},
  {Springel} \& {Hernquist}}{{Sijacki} et~al.}{2012}]{Sijacki2012}
{Sijacki} D.,  {Vogelsberger} M.,  {Kere{\v s}} D.,  {Springel} V.,
  {Hernquist} L.,  2012, \mnras, 424, 2999

\bibitem[\protect\citeauthoryear{{Sommer-Larsen}, {G{\"o}tz} \&
  {Portinari}}{{Sommer-Larsen} et~al.}{2003}]{Sommer2003}
{Sommer-Larsen} J.,  {G{\"o}tz} M.,    {Portinari} L.,  2003, \apj, 596, 47

\bibitem[\protect\citeauthoryear{{Springel}}{{Springel}}{2005}]{Springel2005b}
{Springel} V.,  2005, MNRAS, 364, 1105

\bibitem[\protect\citeauthoryear{{Springel}}{{Springel}}{2010}]{Arepo}
{Springel} V.,  2010, \mnras, 401, 791

\bibitem[\protect\citeauthoryear{{Springel} \& {Hernquist}}{{Springel} \&
  {Hernquist}}{2003}]{SFR_paper}
{Springel} V.,  {Hernquist} L.,  2003, MNRAS, 339, 289

\bibitem[\protect\citeauthoryear{{Springel}, {Di Matteo} \&
  {Hernquist}}{{Springel} et~al.}{2005a}]{Springel2005a}
{Springel} V.,  {Di Matteo} T.,    {Hernquist} L.,  2005a, \mnras, 361, 776

\bibitem[\protect\citeauthoryear{{Springel} et~al.,}{{Springel}
  et~al.}{2005b}]{Millenium}
{Springel} V.,  et~al., 2005b, \nat, 435, 629

\bibitem[\protect\citeauthoryear{{Springel}, {Di Matteo} \&
  {Hernquist}}{{Springel} et~al.}{2005c}]{SpringelMatteoHernquist}
{Springel} V.,  {Di Matteo} T.,    {Hernquist} L.,  2005c, \apjl, 620, L79

\bibitem[\protect\citeauthoryear{{Springel}, {Frenk} \& {White}}{{Springel}
  et~al.}{2006}]{Springel2006}
{Springel} V.,  {Frenk} C.~S.,    {White} S.~D.~M.,  2006, \nat, 440, 1137

\bibitem[\protect\citeauthoryear{{Springel} et~al.,}{{Springel}
  et~al.}{2008}]{Springel2008}
{Springel} V.,  et~al., 2008, MNRAS, 391, 1685

\bibitem[\protect\citeauthoryear{{Stadel}, {Potter}, {Moore}, {Diemand},
  {Madau}, {Zemp}, {Kuhlen} \& {Quilis}}{{Stadel} et~al.}{2009}]{Stadel2009}
{Stadel} J.,  {Potter} D.,  {Moore} B.,  {Diemand} J.,  {Madau} P.,  {Zemp} M.,
   {Kuhlen} M.,    {Quilis} V.,  2009, \mnras, 398, L21

\bibitem[\protect\citeauthoryear{{Stinson}, {Seth}, {Katz}, {Wadsley},
  {Governato} \& {Quinn}}{{Stinson} et~al.}{2006}]{Stinson2006}
{Stinson} G.,  {Seth} A.,  {Katz} N.,  {Wadsley} J.,  {Governato} F.,
  {Quinn} T.,  2006, \mnras, 373, 1074

\bibitem[\protect\citeauthoryear{{Stinson}, {Bailin}, {Couchman}, {Wadsley},
  {Shen}, {Nickerson}, {Brook} \& {Quinn}}{{Stinson}
  et~al.}{2010}]{Stinson2010}
{Stinson} G.~S.,  {Bailin} J.,  {Couchman} H.,  {Wadsley} J.,  {Shen} S.,
  {Nickerson} S.,  {Brook} C.,    {Quinn} T.,  2010, \mnras, 408, 812

\bibitem[\protect\citeauthoryear{{Stinson}, {Brook}, {Macci{\`o}}, {Wadsley},
  {Quinn} \& {Couchman}}{{Stinson} et~al.}{2013a}]{Stinson2013a}
{Stinson} G.~S.,  {Brook} C.,  {Macci{\`o}} A.~V.,  {Wadsley} J.,  {Quinn}
  T.~R.,    {Couchman} H.~M.~P.,  2013a, \mnras, 428, 129

\bibitem[\protect\citeauthoryear{{Stinson} et~al.,}{{Stinson}
  et~al.}{2013b}]{Stinson2013b}
{Stinson} G.~S.,  et~al., 2013b, \mnras, preprint (arXiv:1301.5318)

\bibitem[\protect\citeauthoryear{{Torrey}, {Vogelsberger}, {Sijacki},
  {Springel} \& {Hernquist}}{{Torrey} et~al.}{2012}]{Torrey2012}
{Torrey} P.,  {Vogelsberger} M.,  {Sijacki} D.,  {Springel} V.,    {Hernquist}
  L.,  2012, \mnras, 427, 2224

\bibitem[\protect\citeauthoryear{{Torrey}, {Vogelsberger}, {Genel}, {Sijacki},
  {Springel} \& {Hernquist}}{{Torrey} et~al.}{2013}]{Torrey2013}
{Torrey} P.,  {Vogelsberger} M.,  {Genel} S.,  {Sijacki} D.,  {Springel} V.,
  {Hernquist} L.,  2013, submitted to MNRAS, preprint (arXiv:1305.4931)

\bibitem[\protect\citeauthoryear{{Tully} \& {Fisher}}{{Tully} \&
  {Fisher}}{1977}]{Tully1977}
{Tully} R.~B.,  {Fisher} J.~R.,  1977, \aap, 54, 661

\bibitem[\protect\citeauthoryear{{Uhlig}, {Pfrommer}, {Sharma}, {Nath},
  {En{\ss}lin} \& {Springel}}{{Uhlig} et~al.}{2012}]{Uhlig2012}
{Uhlig} M.,  {Pfrommer} C.,  {Sharma} M.,  {Nath} B.~B.,  {En{\ss}lin} T.~A.,
   {Springel} V.,  2012, \mnras, 423, 2374

\bibitem[\protect\citeauthoryear{{van den Bosch}, {Gebhardt}, {G{\"u}ltekin},
  {van de Ven}, {van der Wel} \& {Walsh}}{{van den Bosch}
  et~al.}{2012}]{vandenBosch2012}
{van den Bosch} R.~C.~E.,  {Gebhardt} K.,  {G{\"u}ltekin} K.,  {van de Ven} G.,
   {van der Wel} A.,    {Walsh} J.~L.,  2012, \nat, 491, 729

\bibitem[\protect\citeauthoryear{{Verheijen}}{{Verheijen}}{2001}]{Verheijen200%
1}
{Verheijen} M.~A.~W.,  2001, \apj, 563, 694

\bibitem[\protect\citeauthoryear{{Vogelsberger}, {Sijacki}, {Kere{\v s}},
  {Springel} \& {Hernquist}}{{Vogelsberger} et~al.}{2012}]{Vogelsberger2012}
{Vogelsberger} M.,  {Sijacki} D.,  {Kere{\v s}} D.,  {Springel} V.,
  {Hernquist} L.,  2012, \mnras, 425, 3024

\bibitem[\protect\citeauthoryear{{Vogelsberger}, {Genel}, {Sijacki}, {Torrey},
  {Springel} \& {Hernquist}}{{Vogelsberger} et~al.}{2013}]{Vogelsberger2013}
{Vogelsberger} M.,  {Genel} S.,  {Sijacki} D.,  {Torrey} P.,  {Springel} V.,
  {Hernquist} L.,  2013, MNRAS, preprint (arXiv:1305.2913)

\bibitem[\protect\citeauthoryear{{Wakker} \& {van Woerden}}{{Wakker} \& {van
  Woerden}}{1991}]{Wakker1991}
{Wakker} B.~P.,  {van Woerden} H.,  1991, \aap, 250, 509

\bibitem[\protect\citeauthoryear{{Wang}, {Frenk}, {Navarro}, {Gao} \&
  {Sawala}}{{Wang} et~al.}{2012}]{Wang2012}
{Wang} J.,  {Frenk} C.~S.,  {Navarro} J.~F.,  {Gao} L.,    {Sawala} T.,  2012,
  \mnras, 424, 2715

\bibitem[\protect\citeauthoryear{{Wilkinson} \& {Evans}}{{Wilkinson} \&
  {Evans}}{1999}]{Wilkinson1999}
{Wilkinson} M.~I.,  {Evans} N.~W.,  1999, \mnras, 310, 645

\bibitem[\protect\citeauthoryear{{Windhorst} et~al.,}{{Windhorst}
  et~al.}{1991}]{Windhorst1991}
{Windhorst} R.~A.,  et~al., 1991, \apj, 380, 362

\bibitem[\protect\citeauthoryear{{Xue} et~al.,}{{Xue}  et~al.}{2008}]{Xue2008}
{Xue} X.~X.,  et~al., 2008, \apj, 684, 1143

\bibitem[\protect\citeauthoryear{{Yang}, {Feng}, {Li} \& {Fan}}{{Yang}
  et~al.}{2013}]{Yang2013}
{Yang} R.-Z.,  {Feng} L.,  {Li} X.,    {Fan} Y.-Z.,  2013, \apj, 770, 127

\end{thebibliography}

\label{lastpage}

\end{document}